\documentclass{aa}  
\usepackage{natbib}
\bibpunct{(}{)}{;}{a}{}{,} 
\usepackage{graphicx}
\usepackage[varg]{txfonts}
\usepackage[colorlinks=true, linkcolor=blue]{hyperref}
\hypersetup{colorlinks, allcolors = {blue}}
\usepackage{booktabs} 
\usepackage{placeins}
\begin{document} 
    
    \title{Impacts of radiative cooling on the images of a black hole shadow and extended jets in two-temperature GRMHD simulations}
    \titlerunning{Impacts of radiative cooling on the images of black hole shadow and jets}

    \author{Mingyuan Zhang\inst{1,2,3,}\textcolor{blue}{\thanks{Corresponding authors: \email{mzhang22@sjtu.edu.cn;\\ mizuno@sjtu.edu.cn}}}\orcid{0009-0002-8053-643X}
        \and
            Yosuke Mizuno\inst{1,2,3,4,}\textcolor{blue}{$^\star$}\orcid{0000-0002-8131-6730}
        \and
            Indu K. Dihingia\inst{5,1}\orcid{0000-0002-4064-0446}
        \and
            Christian M. Fromm\inst{6,4}\orcid{0000-0002-1827-1656}
        \and
            Ziri Younsi\inst{7}\orcid{0000-0001-9283-1191} 
        \and
            \\Hai Yang\inst{8,1}\orcid{0000-0003-1220-3422} 
        \and
            Alejandro Cruz-Osorio\inst{9}\orcid{0000-0002-3945-6342}
            }
    \authorrunning{Zhang et al.}
    
    \institute{Tsung-Dao Lee Institute, Shanghai Jiao Tong University, Shanghai 201210, PR China
        \and
            School of Physics and Astronomy, Shanghai Jiao Tong University, Shanghai 200240, PR China
        \and
            Key Laboratory for Particle Astrophysics and Cosmology (MOE) and Shanghai Key Laboratory for Particle Physics and Cosmology, Shanghai Jiao Tong University, Shanghai 200240, PR China
        \and
            Institut f\"{u}r Theoretische Physik, Goethe Universit\"{a}t, Max-von-Laue-Str. 1, D-60438 Frankfurt, Germany
        \and         
            Institute of Fundamental Physics and Quantum Technology, School of Physical Science and Technology, Ningbo University, Ningbo, Zhejiang 315211, PR China
        \and
           Institut f\"ur Theoretische Physik und Astrophysik, Universit\"at W\"urzburg, Emil-Fischer-Str. 31, D-97074 W\"urzburg, Germany
        \and
            Mullard Space Science Laboratory, University College London, Holmbury St. Mary, Dorking, Surrey, RH5 6NT, UK
        \and
            Nicolaus Copernicus Astronomical Center, Polish Academy of Sciences, Bartycka 18, PL-00-716 Warszawa, Poland
        \and
            Instituto de Astronom\'{\i}a, Universidad Nacional Aut\'onoma de M\'exico, AP 70-264, Ciudad de M\'exico 04510, M\'exico
            }


 
    \abstract
    {The recent $230$~GHz observations from the Event Horizon Telescope (EHT) collaboration have successfully imaged the supermassive black hole shadow of the M87 galaxy. However, the relatively high radiative efficiency observed in the hot accretion flow suggests that radiative cooling is nonnegligible and should be considered when calculating the electron temperature.}
    {In this study, we compared accretion models without and with radiative cooling across a range of mass accretion rates, aiming to assess the impact of cooling on the disk structure, electron temperature distribution, black hole shadow morphology, broadband spectral energy distributions (SEDs), and flux variability.}
    {We performed general relativistic radiative transfer (GRRT) calculations on two-temperature, radiative, general relativistic magnetohydrodynamic simulations, employing different electron heating prescriptions across a range of mass accretion rates, $\dot{M}_{\mathrm{BH}} = (1.0 - 10) \times 10^{-6}\,\dot{M}_{\mathrm{Edd}}$. Our models incorporate the nonthermal electron distribution function, analyzing the radiation transfer due to synchrotron emission at $230$~GHz with a fixed viewing inclination angle of $163^\circ$. These simulations were targeted toward M87$^{*}$.
    } 
    {By comparing density profiles, electron temperature distributions, GRRT images, SEDs, and time variability between models without and with radiative cooling across different mass accretion rates, we find that the radiative cooling sharply decreases the electron temperature in the inner disk around the equatorial plane ($r\lesssim 10\,r_\mathrm{g}$), where the density is highest. In contrast, the temperature in the jet sheath decreases slightly. Radiative cooling leads to a dimmer disk, more extended and brighter jets, and an overall reduction in total flux. For a given accretion rate, cooling reduces the high-frequency flux compared to models without radiative losses. We also find that the time variability mainly originates from the midplane region for both non-cooling and cooling cases. With increasing mass accretion rate, time variability decreases in both non-cooling and cooling cases.
    Although currently below the dynamic range of EHT observations, the features identified in this study can be resolved by next-generation arrays such as the ngEHT.} 
    {}

   \maketitle
   \nolinenumbers
%



\section{Introduction}
\label{introduction}

The nearby radio galaxy M87 is an ideal target for studying black hole accretion and jet formation \citep{2014ARA&A..52..529Y, 2019ARA&A..57..467B}. It is believed to have a low accretion rate consisting of a geometrically thick, optically thin accretion disk in a radiatively inefficient accretion flow state \citep[RIAF;][]{1995ApJ...444..231N, 2014ARA&A..52..529Y}. The emission at millimeter wavelengths of such low-luminosity accretion flows mainly comes from the synchrotron radiation of electrons \citep{1977ApJ...214..840I, 1994ApJ...428L..13N, 2014ARA&A..52..529Y}. The Event Horizon Telescope (EHT) collaboration, utilizing very-long-baseline interferometry (VLBI) techniques, successfully observed horizon-scale shadow images of M87$^{*}$ at millimeter wavelengths \citep{2019ApJ...875L...4E, 2024A&A...681A..79E}.

General relativistic magnetohydrodynamic (GRMHD) simulation is a valuable tool for improving our understanding of accretion physics, and it can be used to simulate shadow images on horizon scales \citep[e.g.,][]{2003ApJ...589..444G, 2003ApJ...589..458D, 2007CQGra..24S.259N, 2009ApJ...706..497M, 2014A&A...570A...7M, 2016A&A...586A..38M, 2010ApJ...717.1092D, 2012ApJ...755..133S, 2018A&A...612A..34D, 2019A&A...632A...2D, 2018NatAs...2..585M, 2025nfgs.book..327D}. 
A ring-like structure is the most direct observational signature of a black hole shadow, as seen in synthetic images of the GRMHD simulations \citep{2019ApJ...875L...5E, 2022ApJ...930L..16E, 2025A&A...693A.265E}. Two distinct types of accretion flows have been identified. The first is the standard and normal evolution \citep[SANE; e.g.,][]{2012MNRAS.426.3241N,2013MNRAS.436.3856S} state, in which the magnetic field strength is weak, the angular momentum is transferred through magnetorotational instability (MRI), and the accreting gas does not become a magnetically arrested situation during the simulation. Therefore, the accretion process is smoother compared to the second type. The second type of accretion flow is the magnetically arrested disk \citep[MAD; e.g.,][]{2003PASJ...55L..69N, 2011MNRAS.418L..79T}. In the MAD state, the magnetic flux accumulating near the horizon reaches the upper limits, and the accretion flow becomes magnetically arrested. However, this accumulated magnetic flux can be released substantially due to the magnetic flux eruptions, and the reduction of magnetic flux in the inner disk results in a temporary increase in the accretion rate until sufficient flux is advected again \citep[e.g.,][]{2012MNRAS.423.3083M,2014MNRAS.439..503S}. The intermediate state (INSANE) is also a possible third type of accretion flow. For instance, it is suggested to explain the transitions observed in X-ray binaries by \cite{2026MNRAS.546ag148R}.

General relativistic radiative transfer (GRRT) calculations compute black hole shadow images utilizing GRMHD simulation data. The important parameters to model the electromagnetic emission are the electron distribution function (eDF) and the electron temperature, which are related synchrotron emission power. In their M87 GRRT calculation, the EHT collaboration assumed that the eDF is the Maxwell-J$\Ddot{\mathrm{u}}$ttner (thermal) distribution \citep{2019ApJ...875L...5E}. However, the electron distribution can be affected by energy dissipation, particle acceleration, and thermalization \citep{2014ARA&A..52..529Y}. 
Considering only thermal distribution ignores key processes such as magnetic reconnection and turbulent dissipation, which drive electrons toward a nonthermal power-law distribution \citep[][]{2010ApJ...708.1545D, 2013ApJ...773..118H}. In addition, some features of M87 caused by electron acceleration have been observed in the near-infrared and optical bands \citep{2016MNRAS.457.3801P}. Recently, the application of the nonthermal $\kappa$ distribution to MAD models has reproduced the wide opening angle jet morphology at $86$~GHz and fit the broadband spectrum of M87 from the radio to the near-infrared bands \citep{2022NatAs...6..103C}. 
Compared to SANE models, \cite{2022A&A...660A.107F} indicate that the combined constraint from broadband spectrum and jet collimation profile favors MAD model coupling with a $\kappa$ distribution.
Furthermore, \cite{2023ApJ...959L...3D} show the propagation of waves along the shear layer of the jet wind using the $\kappa$ distribution in the MAD regime, which provides a possible source to accelerate the electrons through turbulence or reconnection if the observation could confirm the imprints of such waves.
In addition, \cite{2024A&A...687A..88Z} find that the jet emits more when a $\kappa$ distribution is considered in the MAD model. They suggest that future multifrequency observations, which simultaneously resolve horizon-scale structure and the jet base, could be used to investigate the existence of nonthermal electrons. Recently, \cite{2025ApJ...984...35T} find that anisotropic nonthermal distribution functions can help produce sufficiently bright and limb-brightened jets.
Therefore, nonthermal electron impacts on the near-horizon behavior of accretion flows around M87$^{*}$ play a crucial role and should be examined closely.

Traditionally, the electron temperature is estimated using the gas temperature to calculate the emission using a single-fluid GRMHD simulation. One common parametric prescription is the so-called ``$R-\beta$'' model \citep{2016A&A...586A..38M}, which estimates electron temperature using plasma-$\beta$ and two parameters, $R_{\mathrm{l}}$ and $R_{\mathrm{h}}$, respectively. Although this treatment is much more flexible in allowing a wide exploration of parameter space, it requires extensive parameter searching, and the physical processes related to the optimal parameters are difficult to interpret. In addition, the electron temperature is not only determined by ion temperature but also depends on the microscopic physical processes, such as heating, cooling, and the advection of the electrons \citep{2021ApJ...910L..13E}. Thus, the physically driven two-temperature model is needed to explore electron thermodynamics self-consistently. This approach further eliminates the dependence on hyperparameters $R_{\mathrm{l}}$ and $R_{\mathrm{h}}$ when calculating electron temperature. 
Several studies have shown that the self-consistent two-temperature model is well matched with the parameterized $R-\beta$ model \citep{2021MNRAS.506..741M, 2024A&A...687A..88Z, 2025ApJ...981..145M}. Apart from that, \cite{2026ApJ..1001..227C} recently used the electron temperature obtained through the simulations of turbulent collisionless plasmas on a microscopic scale, providing a better depiction of the jet in morphology and width at 86~GHz than using the parameterized $R-\beta$ model.

\cite{2015MNRAS.454.1848R} introduced the two-temperature model to effectively evolve electron thermodynamics separately from ions by extending the equations in GRMHD simulations, while the energy-momentum and particle number conservation equations still assumed a single fluid. This approach has been applied to model M87$^{*}$ \citep{2018ApJ...864..126R, 2019MNRAS.486.2873C} and Sgr\,A$^{*}$ \citep{2017MNRAS.467.3604R, 2018MNRAS.478.5209C, 2020MNRAS.494.4168D, 2020MNRAS.499.3178Y}. 

While our understanding of accretion flows has improved considerably in the last few decades, neglecting radiative cooling can lead to incomplete models or inconsistent predictions.
For instance, radiative cooling may result in the electron temperature being lower than the ion temperature even in the high magnetization regions where the electron is heated efficiently \citep[e.g.,][]{2011ApJ...735....9M,2018ApJ...864..126R,2019MNRAS.486.2873C}. Moreover, radiative cooling affects the disk structure via the pressure balance inside the disk \citep[e.g.,][]{2012MNRAS.426.1928D,2020MNRAS.499.3178Y,2025ApJ...981L..11S, 2026arXiv260509326S}. Recently, \cite{2025MNRAS.538..698S} found that the two-temperature model with radiative cooling better matches the historical observations in flux variability at $230$~GHz for Sgr\,A$^{*}$. This is achieved by decreasing the total flux and its fluctuations. However, most of the aforementioned studies adopted thermal eDF, and radiative cooling impacts on black hole shadows and extended jets have not yet been fully investigated. This is the purpose of our study. In addition, the measured radiative efficiency of M87$^{*}$ is relatively high for a hot accretion flow model \citep{2021ApJ...910L..13E}. 
These indicate that considering the two-temperature model with radiative cooling using nonthermal eDF may be crucial for accurately modeling M87$^{*}$'s black hole shadow and large-scale jet structure.

We organize this paper as follows: Sect. \ref{introduction} provides a brief overview of the background. In Sect. \ref{setup}, we describe the dissipation and cooling processes included in the GRMHD simulations as well as the nonthermal eDF used in the GRRT calculation. The results are presented and discussed in detail in Sect. \ref{results}. In Sect. \ref{discussion}, we discuss the implications of our findings, address the limitations of the study, and outline possible directions for future work. 

Throughout this paper, we adopt units in which the speed of light is $c = 1$ and the gravitational constant is $G = 1$. We absorb a factor of $\sqrt{4\pi}$ into the definition of the magnetic field four-vector, $b^{\mu}$.

\section{Numerical setup}
\label{setup}

\subsection{General relativistic magnetohydrodynamic simulations} 
\label{2.1}

We performed GRMHD simulations of magnetized accretion flows onto a rotating black hole, considering Coulomb interaction and radiative cooling in a two-temperature framework following \cite{2023MNRAS.518..405D}. Here, we considered radiative cooling as a source term without solving the radiation fields explicitly. 
We utilized the \texttt{BHAC} code \citep{2017ComAC...4....1P, 2019A&A...629A..61O} for this study. 
The metric adopted in the simulation consists of spherical modified Kerr-Schild (MKS) coordinates.
The torus is initialized by a Fishbone-Moncrief hydrodynamic equilibrium solution \citep{1976ApJ...207..962F} with $r_{\mathrm{in}} = 20\,r_{\mathrm{g}}$ and $r_{\mathrm{max}} = 40\,r_{\mathrm{g}}$, where $r_{\mathrm{g}} = GM_{\mathrm{BH}}/c^2$ and $M_{\mathrm{BH}}$ is the black hole mass. An ideal-gas equation of state with a constant relativistic adiabatic index of $\gamma = 4/3$ was used \citep{2013rehy.book.....R}. We note that a value close to $5/3$ is also a considerable choice \citep{2025MNRAS.537.2496C,2025ApJ...980..193G}, and it is also possible to update the adiabatic index self-consistently \citep{2017MNRAS.466..705S,2025MNRAS.538..698S}.
We put a weak single magnetic field loop in this equilibrium torus defined by the vector potential with only one nonzero component, $A_\phi \propto \max(q-0.2, 0)$,
where
\begin{equation}
q=\frac{\rho}{\rho_{\rm max}} \left( \frac{r}{r_{\rm in}}\right)^3 \sin^3 \theta \exp \left( \frac{-r}{400}\right).
\end{equation}
Here, $\rho$ is the fluid rest-mass density, and $\rho_{\rm max}$ is maximum density in the torus.
This field configuration supplies enough magnetic flux onto the black hole to reach the MAD state \citep[e.g.,][]{2003PASJ...55L..69N,2011MNRAS.418L..79T}. To excite the MRI inside the torus, a 4 percent random perturbation was applied to the gas pressure within the torus. 

The electron temperature time evolution in the two-temperature GRMHD simulations is based on solving the electron entropy equation \citep{2015MNRAS.454.1848R,2021MNRAS.506..741M}.
Electron heating is provided by grid-scale dissipation models \citep[e.g.,][]{2015MNRAS.454.1848R}. 
The physical processes include turbulent heating, magnetic reconnection, shock waves, and Ohmic dissipation. 
Our tests used two heating prescriptions: turbulence \citep{2019PNAS..116..771K} and magnetic reconnection \citep{2017ApJ...850...29R}. Apart from that, the energy transfer from protons to electrons through Coulomb interaction \citep{1965pfig.book.....S,1984ApJ...280..319C} was also considered {for radiative cooling cases}. Electron energy loss was considered through radiative cooling processes, namely bremsstrahlung, cyclo-synchrotron radiation of the thermal electrons \citep{1996ApJ...465..312E}, and multiple inverse Compton scattering of the cyclo-synchrotron photons by the thermal electrons \citep{1995ApJ...452..710N}. Note that the radiative cooling processes due to nonthermal electrons in the plasma were ignored for simplicity throughout the simulations. The detailed initial setup about Coulomb interaction and radiative cooling is described in \cite{2023MNRAS.518..405D,2025nfgs.book..327D}.

The outer radial boundary is located at $r = 2\,500\,r_{\mathrm{g}}$. The inner radial position of the simulation domain is well inside the black hole horizon.
The simulation domain was discretized using an effective grid resolution of $384 \times 192 \times 192$ with three layers of static mesh refinement. In particular, \cite{2025ApJ...981..145M} demonstrate that the results are independent of the grid resolution for the turbulent heating model. Here, we considered black hole spin, $a=0.9375$.

First, we ran the GRMHD simulations without Coulomb interactions and radiative cooling until $t=10\,000\,t_{\mathrm{g}}$, where $t_{\mathrm{g}} = GM_{\mathrm{BH}}/c^3$. The simulations mostly reached a quasi-stationary MHD state at this time. Then, we applied radiative cooling and Coulomb interactions with different mass accretion rates. Here, based on the accretion rate estimated from \cite{2021ApJ...910L..13E} and the total observed flux $S_{\mathrm{230}}=0.5~\mathrm{Jy}$ \citep{2019ApJ...875L...4E,2024A&A...681A..79E}, we chose three different mass accretion rates normalized to the Eddington rate, $\dot{m}=\dot{M}_{\mathrm{BH}}/\dot{M}_{\mathrm{Edd}}=1\times10^{-5}$, $5\times10^{-6}$, and $1\times10^{-6}$ at the horizon for the turbulent heating models as well as $\dot{m}= 5\times10^{-6}$ for the reconnection heating model as a reference. We carried out our simulations up to $t=15\,000\,t_{\mathrm{g}}$.

\subsection{General relativistic radiative transfer calculations}
To calculate black hole shadow images from GRMHD simulations, we used the GRRT code \texttt{BHOSS} \citep{2012A&A...545A..13Y,2020IAUS..342....9Y, Younsi2023}, which solves covariant radiative transfer equations via the ray-tracing method. Here, we considered electron synchrotron radiation as a mechanism for calculating the shadow.
Additionally, we employed a hybrid distribution composed of thermal and variable $\kappa$ components for our GRRT calculations. The thermal distribution used in this study follows the Maxwell-J\"uttner distribution given by Eq.~\eqref{eq1}. The $\kappa$ distribution can simultaneously represent both thermal eDFs and extended power-law characteristics by adjusting the $\kappa$ parameter, given in Eq.~\eqref{eq2}. 
The Maxwell-J\"uttner distribution is expressed as follows:
\begin{equation} \label{eq1}
\frac{d n_{\mathrm{e}}}{d \gamma_{\mathrm{e}}}=\frac{n_{\mathrm{e}}}{\Theta_{\mathrm{e}}} \frac{\gamma_{\mathrm{e}} \sqrt{\gamma_{\mathrm{e}}^{2}-1}}{K_{2}\left(1 / \Theta_{\mathrm{e}}\right)} \exp \left(-\frac{\gamma_{\mathrm{e}}}{\Theta_{\mathrm{e}}}\right)\,,
\end{equation}
where $n_{\mathrm{e}}$ is the electron number density, $\gamma_{\mathrm{e}}$ is the electron Lorentz factor, $K_{2}$ is the Bessel function of the second kind, and $\Theta_{\mathrm{e}}$ is the dimensionless electron temperature \citep[e.g.,][]{2021MNRAS.506..741M}. 

The relativistic nonthermal $\kappa$ distribution \citep{2006PPCF...48..203X} is expressed as follows:
\begin{equation} \label{eq2}
\frac{d n_{\mathrm{e}}}{d \gamma_{\mathrm{e}}}=N \gamma_{\mathrm{e}} \sqrt{\gamma_{\mathrm{e}}^{2}-1}\left(1+\frac{\gamma_{\mathrm{e}}-1}{\kappa w}\right)^{-(\kappa+1)},
\end{equation}
where $N$ is the normalization factor \citep{2016ApJ...822...34P,2018A&A...612A..34D} and $\kappa$ is related to the slope of the power-law distribution, $s=\kappa-1$.
When $\gamma_{\mathrm{e}}$ is large, particles satisfy $d n_{\mathrm{e}} / d \gamma_{\mathrm{e}} \propto \gamma_{\mathrm{e}}^{-s}$, and the nonthermal $\kappa$ distribution approximates the power-law distribution.
The parameter $w$ specifies the width of the $\kappa$ distribution.
Considering the contribution of both thermal and magnetic energies to heating and accelerating electrons \citep{2019A&A...632A...2D, 2022NatAs...6..103C,2022A&A...660A.107F}, the specific expression of $w$ is
\begin{equation} \label{eq3}
w:=\frac{\kappa-3}{\kappa} \Theta_{\mathrm{e}}+\frac{\varepsilon}{2}\left[1+\tanh \left(r-r_{\mathrm{inj}}\right)\right] \frac{\kappa-3}{6 \kappa} \frac{m_{\mathrm{p}}}{m_{\mathrm{e}}} \sigma\,,
\end{equation}
where $r_{\mathrm{inj}}$ is the injection radius, $m_{\mathrm{e}}$ is the electron mass, $m_{\mathrm{p}}$ is the proton mass, $\sigma=b^2/\rho$ is the magnetization, $b^2$ is the square of the four-magnetic field, $\rho$ is the fluid rest-mass density, and $\varepsilon$ is a tunable parameter for the region with a radius larger than $r_{\mathrm{inj}}$. The energy is dominated by thermal energy with a limit of $\sigma \ll 1$, while the magnetic energy contributes to highly magnetized regions.  
We set $\varepsilon=0.5$ to account for the contribution of magnetic energy to the GRRT images and the spatial distribution of $w$. The jet stagnation surface is a potential injection site and defines the injection radius.
The stagnation surface is located at $u^{r} = 0$, where the potential injection radius is usually between $5$ and $10~r_{\mathrm{g}}$ \citep[e.g.,][]{2018ApJ...868..146N}.
We thus assumed $r_{\mathrm{inj}} = 10\,r_{\mathrm{g}}$ in this study.

The $\kappa$ value is variable in different locations, and it is defined to be parametrically dependent on magnetization, $\sigma$, and plasma beta, $\beta=p_{\mathrm{g}}/p_{\mathrm{m}}$, where $p_{\mathrm{g}}$ is the fluid pressure and $p_{\mathrm{m}}=b^2/2$ is the magnetic pressure. For the $\texttt{PIC-CS}$ model of \cite{2018ApJ...862...80B}, the $\kappa$ function can be expressed as
\begin{equation}
\kappa:=2.8+0.7 \sigma^{-1/2} + 3.7\,\sigma^{-0.19} \, \tanh \left(23.4\, \sigma^{0.26} \,\beta\right) \,,
\end{equation}
which was empirically obtained from particle-in-cell (PIC) simulations of magnetic reconnection in a Harris current sheet. For the $\texttt{PIC-TURB}$ model of \cite{Meringolo2023}, $\kappa$ function is expressed as
\begin{equation} \label{eqM23}
\kappa:=2.8+0.2 \sigma^{-1/2} + 1.6\,\sigma^{-6/10} \, \tanh \left(2.25\, \sigma^{1/3} \,\beta\right) \,,
\end{equation}
which was obtained from PIC simulations of decaying plasma turbulence (see also \citet{Imbrogno2024,2025ApJ...990L..33I} for results from PIC simulations of turbulent plasma). To more self-consistently account for the nonthermal electron distribution for the reconnection heating model, we used the $\texttt{PIC-CS}$ model for $\kappa$ and the $\texttt{PIC-TURB}$ model for the turbulent heating case. 

Following \citep{2022ApJ...930L..16E}, we assumed that the proportion of nonthermal electrons depends on $\sigma$ and $\beta$. The emission coefficients $c_{\nu}$ (emissivity and absorptivity) combine the thermal \citep{2011ApJ...737...21L} and $\kappa$ coefficients \citep{2016ApJ...822...34P} through
\begin{equation}
c_{\nu, \text { tot }}=(1-\eta) c_{\nu, \text { thermal }}+\eta c_{\nu, \, \kappa},
\end{equation}
where the nonthermal efficiency is expressed as
\begin{equation} \label{eq6}
\eta(\epsilon, \beta, \sigma)=\epsilon\left[1-e^{-\beta^{-2}}\right]\left[1-e^{-\left(\sigma / \sigma_{\min }\right)^2}\right].
\end{equation}
Here, $\eta \rightarrow 0$ on the disk, and $\eta \rightarrow \epsilon$ in the jet. Because the emission at highly magnetized regions ($\sigma > \sigma_{\mathrm{cut}}=1$) were removed, the nonthermal electrons are mostly confined to the jet sheath. We set $\sigma_{\mathrm{min}} = 0.0 1$ and $\epsilon=0.5$.
In this study, we directly calculated the electron temperature from the two-temperature GRMHD simulations \citep{2021MNRAS.506..741M,2023MNRAS.518..405D}.

For the GRRT calculation, we modeled M87$^{*}$ as a target source, with a mass of $M_{\mathrm{BH}} = 6.5 \times 10^9 \, M_{\odot}$ at a distance of $D = 16.8 \, \mathrm{Mpc}$ \citep{2019ApJ...875L...6E}.
The field of view (FoV) was set to 760 $\mathrm{\mu as}$ in both directions, with a resolution of $1520 \times 1520$ pixels. 
The GRRT calculations were performed over the time range $t \in [12\,000\,t_{\mathrm{g}}, 15\,000\,t_{\mathrm{g}}]$, with a $10\,t_{\mathrm{g}}$ cadence, at 230~GHz and an inclination angle of $163^{\circ}$.
We varied the mass accretion rates ($\dot{m}$) in the radiative cooling GRMHD simulations as follows: $1\times10^{-6} \, \dot{M}_{\mathrm{Edd}}$, $5\times10^{-6} \, \dot{M}_{\mathrm{Edd}}$, and $1\times10^{-5} \, \dot{M}_{\mathrm{Edd}}$, where $\dot{M}_{\mathrm{Edd}}$ is the Eddington accretion rate defined as
\begin{equation}
\dot{M}_{\mathrm{Edd}} = \frac{L_{\mathrm{Edd}}}{\zeta c^{2}} = 1.4 \times 10^{17} \, \frac{M_{\mathrm{BH}}}{M_{\odot}} \, \mathrm{gs^{-1}}.
\end{equation}
Here, $L_{\mathrm{Edd}} = 4\pi G M_{\mathrm{BH}} c m_{\mathrm{p}}/\sigma_{\mathrm{T}}$ is the Eddington luminosity, and $\sigma_{\mathrm{T}}$ is the Thomson cross section. By setting the efficiency to $\zeta = 1$ and adopting the black hole mass for M87$^{*}$ as $M_{\mathrm{BH}} = 6.5 \times 10^9 \, M_{\odot}$, the Eddington accretion rate becomes $\dot{M}_{\mathrm{Edd}} \approx 9.1 \times 10^{26} \, \mathrm{gs^{-1}} \approx 14 \, M_{\odot} \mathrm{yr^{-1}}$.

To exclude regions with strong magnetization, a ceiling in magnetization was set at $\sigma_{\mathrm{cut}}=1$ for all models. To assess the impact of this choice on the shadow images, we also explore different values of $\sigma_{\mathrm{cut}} = 2$, 5, 10, and 25 in Appendix~\ref{AppendixA}. 

\section{Results}
\label{results}

\subsection{Evolution of mass accretion rate and magnetic flux}

\begin{figure*}
\centering
    \includegraphics[width=\linewidth]{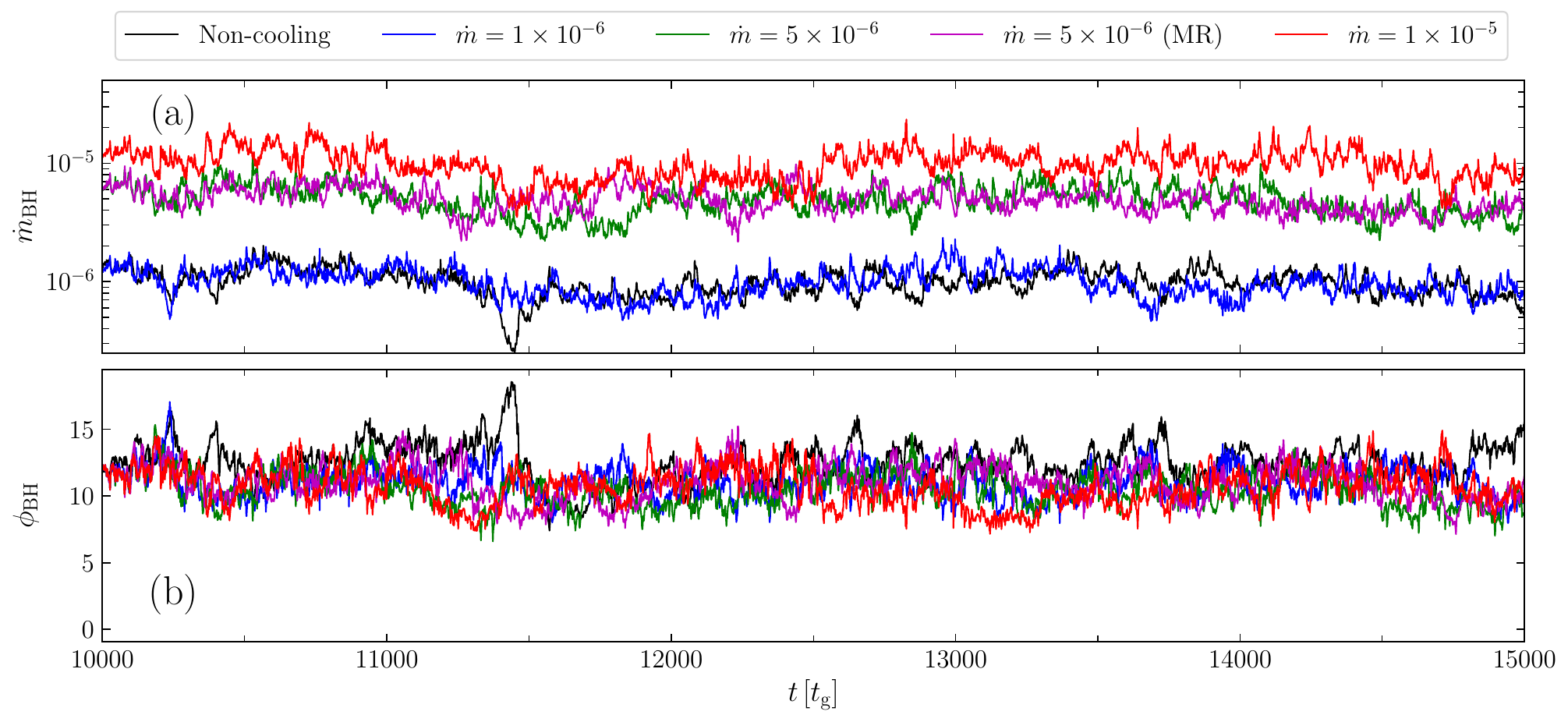} 
    \caption{Top: Accretion rates measured at the event horizon. Bottom: Normalized magnetic flux at the horizon. The different colored curves correspond to the various electron heating prescriptions, radiative cooling, and time-averaged accretion rates. The black curve corresponds to the model without cooling ($\dot{m} = 1\times10^{-6}$). The blue, green, and red curves depict the turbulent heating model with cooling for $\dot{m} = 1\times10^{-6},\,5\times10^{-6},$ and $ 1\times10^{-5}$, respectively. The magenta curve represents reconnection heating with cooling ($\dot{m} = 5\times10^{-6}$).
    }
    \label{figure1}
\end{figure*}

To analyze the temporal evolution of the simulation models, Fig.~\ref{figure1} shows the normalized mass accretion rate, $\dot{m}_\mathrm{BH}\equiv \dot{M}_{\mathrm{BH}}/\dot{M}_{\mathrm{Edd}}$, measured at the event horizon in Eddington units and the normalized magnetic flux, $\phi_{\mathrm{BH}} \equiv \Phi_{\mathrm{BH}} \,/\sqrt{\dot{M}_{\mathrm{BH}}}$, at the horizon \citep[see e.g.,][for the definition of the magnetic flux]{2019ApJS..243...26P}.
As shown in Fig.~\ref{figure1}, the mass accretion rate profile and the normalized magnetic flux settle to steady states at $t \gtrsim 10\,000\,t_{\mathrm{g}}$, with the small oscillations in time due to flux eruption events. The averaged values ($t=12\,000\,t_{\mathrm{g}}\,-\,15\,000 \,t_{\mathrm{g}}$) for the non-cooling case ($\texttt{NC}$), cooling with time-averaged mass accretion rates $\dot{m}=1\times10^{-6}$ ($\texttt{C\_KA1e-6}$), $5\times10^{-6}$ ($\texttt{C\_KA5e-6}$), $5\times10^{-6}$ ($\texttt{C\_MR5e-6}$), and $1\times10^{-5}$ ($\texttt{C\_KA1e-5}$) are $\phi_{\mathrm{BH}} \sim$ 12.5, 11.0, 10.4, 11.1, and 10.6, respectively. 
The decrease in $\phi_{\mathrm{BH}}$ with radiative cooling is consistent with the result found in \cite{2025ApJ...981L..11S}. Our simulations indicate that the decrease in $\phi_{\mathrm{BH}}$ results from the decrease in $\Phi_{\mathrm{BH}}$ and the increase in $\sqrt{\dot{M}_{\mathrm{BH}}}$.
In radiative cooling, the local magnetic field near the horizon becomes weaker due to the lower MRI growth rate.
From $t=12\,000\,t_{\mathrm{g}}$ to $15\,000 \,t_{\mathrm{g}}$, the ratios of the standard deviation to the average value of the normalized magnetic flux -- used to characterize variability -- are 0.091, 0.102, 0.113, 0.110, and 0.132 for the $\texttt{NC}$, $\texttt{C\_KA1e-6}$, $\texttt{C\_KA5e-6}$, $\texttt{C\_MR5e-6}$, and $\texttt{C\_KA1e-5}$ simulations, respectively. The variability in the normalized magnetic flux becomes greater with radiative cooling. 
Our simulations considered the accretion flow at low accretion rates. Under these conditions, the radiative-cooling processes are not strong enough to collapse the geometrically thick torus into a thin-disc structure. Consequently, the temporal evolution and general trends of the mass accretion rate and magnetic flux rate are mostly similar.

\subsection{Density distribution}
\label{3.2}

\begin{figure*}
\centering
    \includegraphics[width=\linewidth]{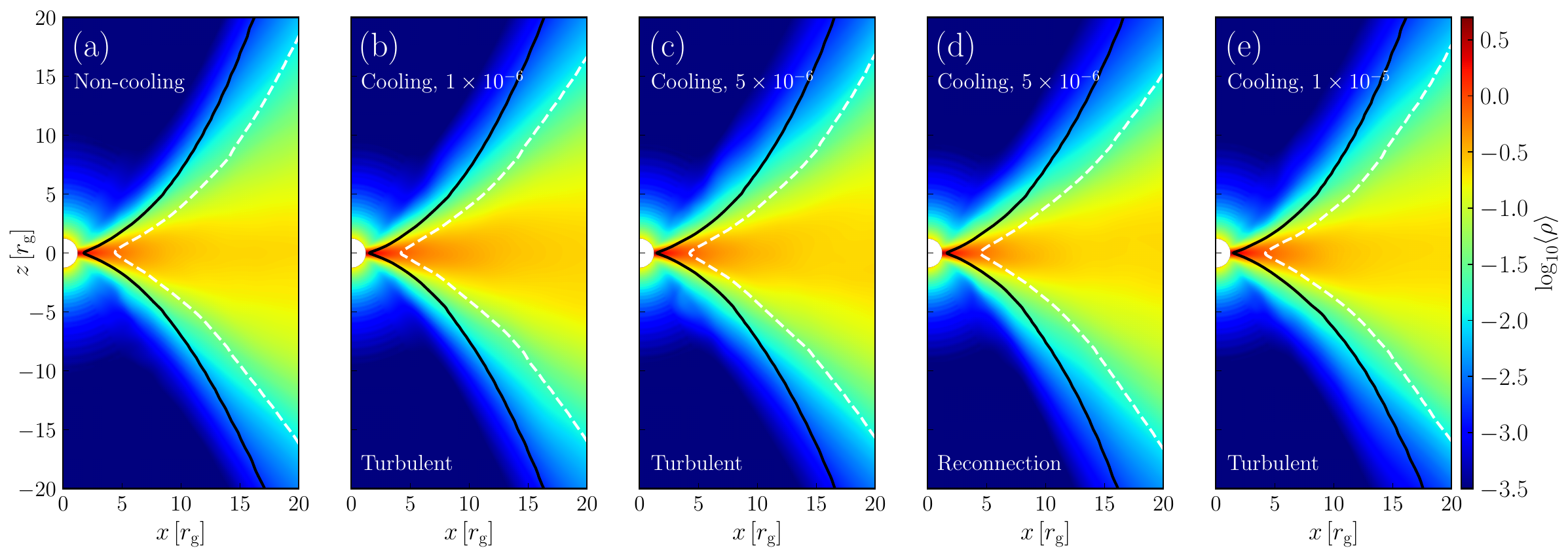}
    \caption{Logarithmic density distribution averaged in time and azimuth over the interval $t = 12\,000\,t_{\mathrm{g}}$ to $15\,000\,t_{\mathrm{g}}$.
    From left to right: Without cooling (a); turbulent heating with cooling at $\dot{m} = 1 \times 10^{-6}$ (b), $\dot{m} = 5 \times 10^{-6}$ (c), and $\dot{m} = 1 \times 10^{-5}$  (e); and reconnection heating with cooling at $\dot{m} = 5 \times 10^{-6}$ (d). 
    The dashed white and solid black curves represent magnetization for $\sigma = 0.1$ and 1, respectively.
    }
    \label{figure2}
\end{figure*}
 
Figure~\ref{figure2} shows the time- and azimuthally averaged density distribution over the interval $t = 12\,000\,t_{\mathrm{g}}$ to $15\,000\,t_{\mathrm{g}}$ for the electron heating prescriptions and radiative cooling conditions, along with the corresponding time-averaged normalized accretion rates.
Specifically, panels (a), (b), (c), and (e) compare the density distribution: (a) without radiative cooling, (b), (c), and (e) with turbulent heating and radiative cooling at normalized mass accretion rates of $\dot{m} = 1\times10^{-6}$, $5\times10^{-6}$, and $1\times10^{-5}$, respectively. Panels (c) and (d) compare the density profile with radiative cooling at $\dot{m} = 5\times10^{-6}$ under turbulent heating (c) and reconnection heating (d). The dashed white and solid black curves represent $\sigma =0.1$ and 1, respectively.
All cases generally show a relatively dense disk around the equatorial plane, a low-density off-equatorial region, and a sparse funnel region around the bipolar directions. The disk structure is related to the mass accretion rate when radiative cooling is considered. As shown in panels (a), (b), (c), and (e), a thinner disk forms with a higher accretion rate, due to the increase in radiative cooling efficiency \citep{2025ApJ...981L..11S, Dihingia_2025}. 
Compared with panels (c) and (d), under the same mass accretion rate $\dot{m} = 5\times10^{-6}$, there is no significant difference in the density profile caused by the electron heating prescriptions (i.e., turbulent heating and reconnection heating). 
The low-density region contributes to the disk winds, and the sparse funnel region contributes to the relativistic Poynting-dominated jet \citep{2019ApJ...882....2V, 2021MNRAS.505.3596D}. Due to the higher density region around the equatorial plane, the efficiency of Coulomb interactions and electron radiative cooling increase \citep{2023MNRAS.518..405D}. Hence, we expect the electron temperature near the inner disk to be affected (see Sect.~\ref{3.3} for more details).

\subsection{Temperature distribution}
\label{3.3}

\begin{figure*}
\centering
    \includegraphics[width=\linewidth]{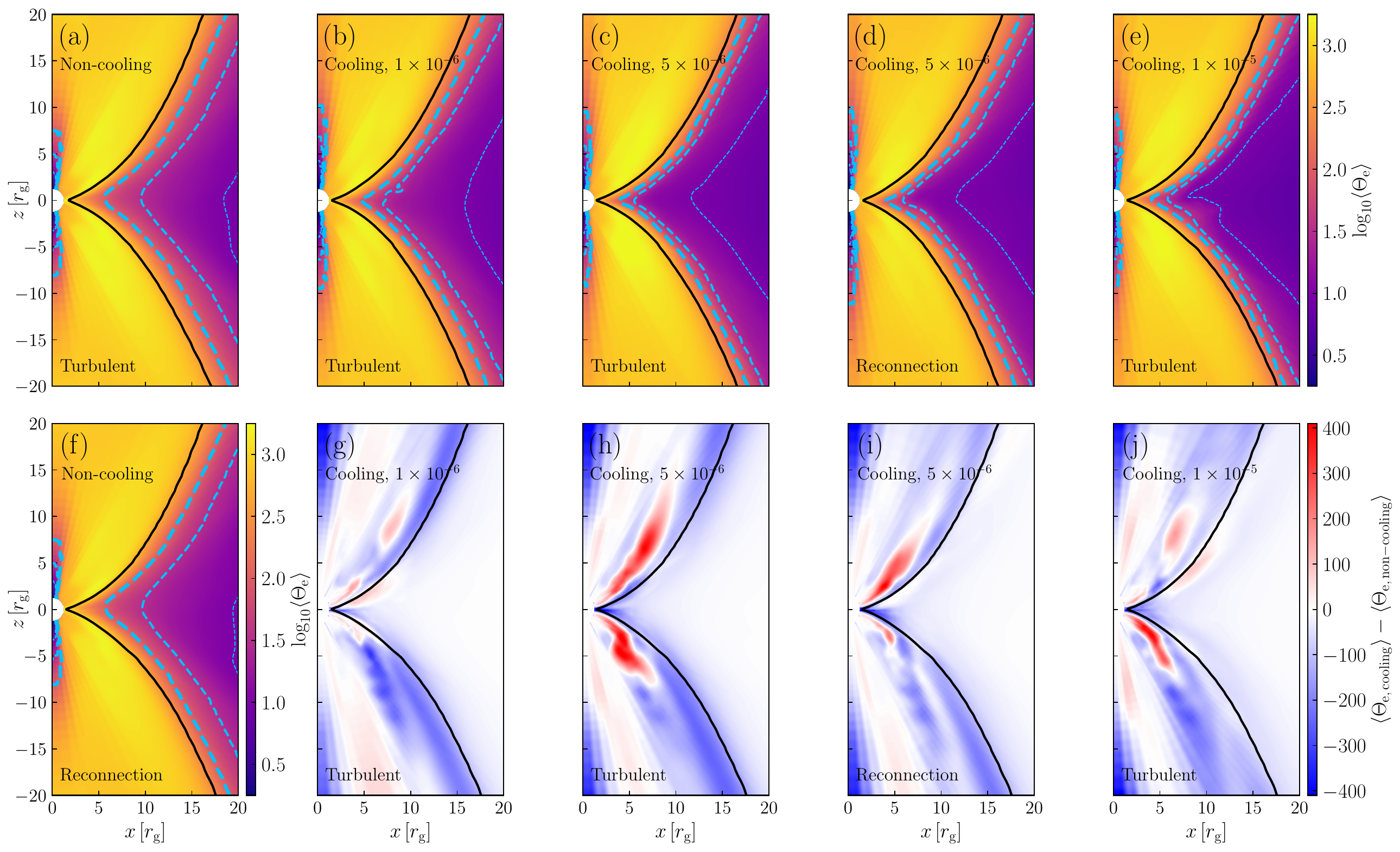}
    \caption{Panels $(\mathrm{a})-(\mathrm{f})$: Logarithm of the dimensionless electron temperature averaged in time and azimuth over the interval $t = 12\,000\,t_{\mathrm{g}}$ to $15\,000\,t_{\mathrm{g}}$. Panels $(\mathrm{g})-(\mathrm{j})$: Differences in the linear scale, calculated by subtracting the dimensionless electron temperature in the corresponding non-cooling case.
    The solid black curves represent $\sigma = 1$. The dashed sky-blue thin to thick curves represent $\Theta_\mathrm{e} =$ 10, 32, and 100, respectively.
    }
    \label{figure3}
\end{figure*}

\begin{figure}
\centering
    \includegraphics[width=\linewidth]{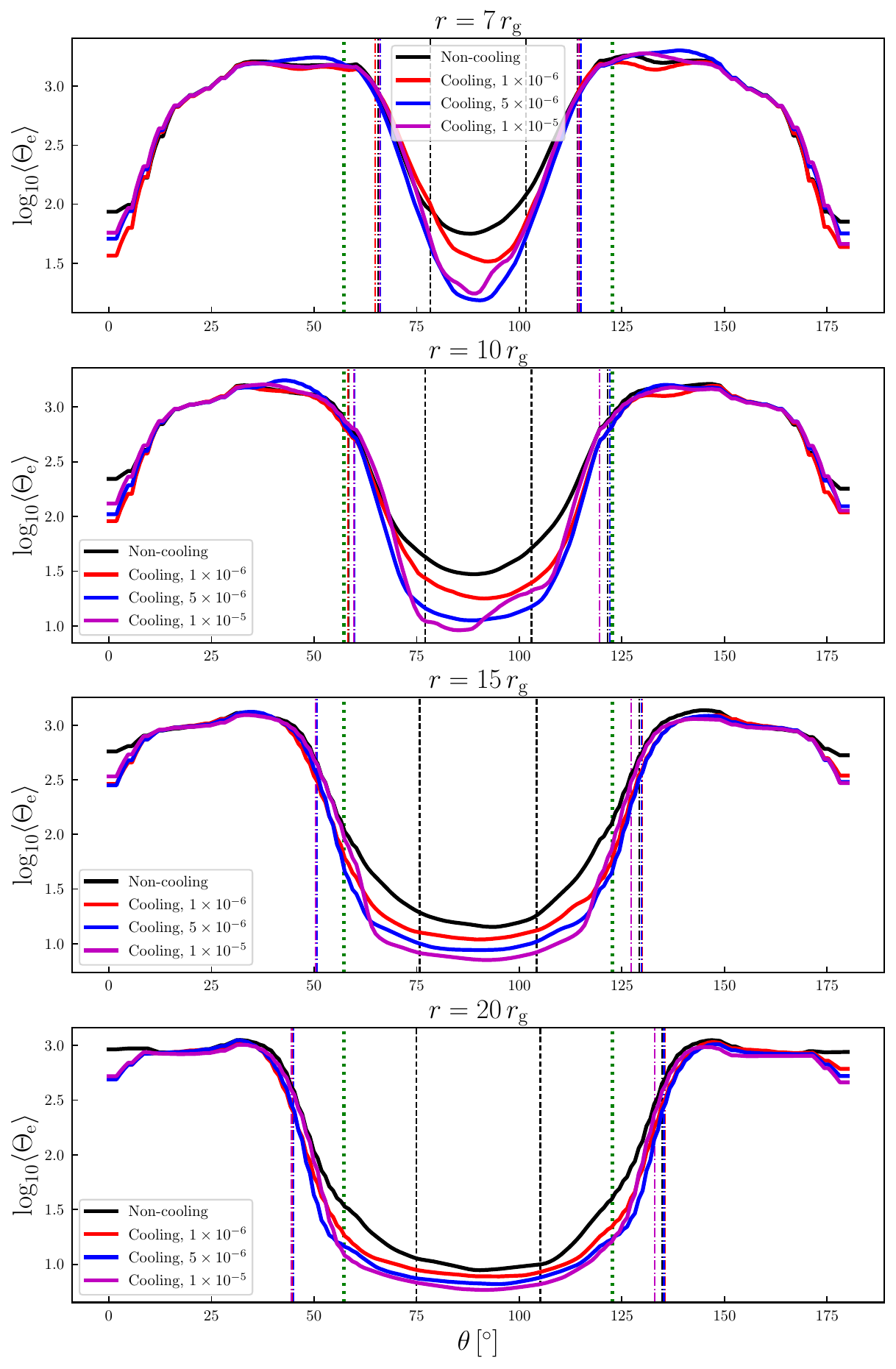}
    \caption{Angular distribution of time- and azimuthally averaged dimensionless electron temperatures at the given radii, plotted on a logarithmic scale. From top to bottom: Radial increase from $7 \, r_{\mathrm{g}}$ to $20 \, r_{\mathrm{g}}$. The different colored curves correspond to the non-cooling and radiative cooling models at the normalized mass accretion rates. The black curve represents the turbulent heating model without cooling. The red, blue, and magenta curves correspond to the models with cooling at $\dot{m} = 1\times10^{-6}$, $5\times10^{-6}$, and $1\times10^{-5}$, respectively. The dash-dotted lines in the same color represent $\sigma=1$ for each case. The vertical dashed lines in black on both sides correspond to the disk region of the non-cooling case (see Sect.~\ref{3.3} for more details).
    The dotted lines in green mark the boundary where the images are decomposed (see Sect.~\ref{3.4} for more details). 
    }
    \label{figure4}
\end{figure}

The electron temperature is an important quantity for modeling the electromagnetic radiation of an accreting black hole. In this section, we discuss the dependence of the dimensionless electron temperature on radiative cooling and mass accretion rates.
 
Figure~\ref {figure3} shows the time- and azimuthally averaged distribution of the logarithmic dimensionless electron temperature $\Theta_{\mathrm{e}} = k_{\mathrm{B}}T_{\mathrm{e}}/m_{\mathrm{e}}c^{2}$ \citep[e.g.,][]{2021MNRAS.506..741M} 
without (panels $(\mathrm{a})$ and $(\mathrm{f})$) and with (panels $(\mathrm{b})-(\mathrm{e})$) radiative cooling at different mass accretion rates. The differences are shown in a linear scale in panels $(\mathrm{g})-(\mathrm{j})$ and are compared with the corresponding non-cooling case. The red shaded region indicates the increase in density compared to the non-cooling case, while the blue shaded region highlights the decrease in density compared to the non-cooling case. A smooth transition appears from red to blue via white. 
The solid black curves represent $\sigma = 1$, and the dashed sky-blue thin to thick curves represent $\Theta_\mathrm{e} =$ 10, 32, and 100, respectively.
Figure~\ref{figure4} shows the angular distribution of the time- and azimuthally averaged dimensionless electron temperatures at the given radii.
The different colored curves correspond to the non-cooling and radiative cooling models at different accretion rates. The black curves represents the turbulent heating model without cooling; and the red, blue, and magenta curves represents cooling at $\dot{m} = 1\times10^{-6}$, $5\times10^{-6}$, and $1\times10^{-5}$, respectively. The dash-dotted lines in the same color represent $\sigma = 1$ for each case. The vertical dashed lines in black on both sides correspond to the disk region in the non-cooling case. The lines for other cases are omitted for simplicity, as they are located at a similar position. Here, disk angular thickness at a given radius $r$ is $(h/r)_{r}=\left[\iint_{\theta, \, \varphi}(\theta-\pi / 2)^2 ~\rho \mathrm{~d} A_{\theta \varphi} ~/ \iint_{\theta, \, \varphi} ~\rho \mathrm{~d} A_{\theta \varphi}\right]^{1 / 2}$, where $\mathrm{d} A_{\theta \varphi}=\sqrt{-g} \mathrm{~d} \theta \mathrm{~d} \varphi$ is an area element in $\theta-\varphi$ plane, and $g$ is the metric determinant. The integrals are over all $\theta, \, \varphi$ on a sphere of radius $r$ \citep[see e.g.,][for more details about the disk angular thickness]{2012MNRAS.423L..55T}. 
The dotted green lines mark the boundary where the images are decomposed (see Sect.~\ref{3.4} for more details).

From Fig.~\ref{figure3}, we see that the electron temperature at the inner disk falls sharply for all cooling cases around the equatorial plane. This dramatic drop can be confirmed quantitatively from Fig.~\ref{figure4} at least within $10\,r_{\mathrm{g}}$.  
In addition, in the slightly farther outer disk region ($\sim 20\,r_{\mathrm{g}}$), the electron temperature also drops, with an increase in the mass accretion rate. Interestingly, the temperature profiles look similar in both the $\texttt{C\_KA5e-6}$ and $\texttt{C\_MR5e-6}$ cases, even though the underlying heating functions are different. This may be because they evolve from the same fluid, so there is no significant difference induced by the MRI, which was excited by random perturbations at the beginning of the simulation. 
In addition, temperature differences remain in the disk and jets (see Appendix~\ref{AppendixC} for more details).
In Fig.~\ref{figure4}, the disk region for the non-cooling case is outlined by dashed black lines. At a smaller radius, we see a smaller angular thickness. This is because, at a smaller radius, the disk near the equatorial plane is compressed vertically by magnetic fields at the funnel region. 
The temperature in the jet regions (outside the midplane) slightly decreases when radiative cooling is considered, compared to the non-cooling case.

In summary, when we consider radiative cooling, the electron temperature in the inner disk around the equatorial plane significantly decreases. Furthermore, the temperature slightly decreases farther from the equatorial plane. This may profoundly impact the shadow images of the black holes, which we discuss in Sect. \ref{3.4}.

\subsection{Image decomposition}
\label{3.4}

\begin{figure*}
\centering
    \includegraphics[width=0.85\linewidth]{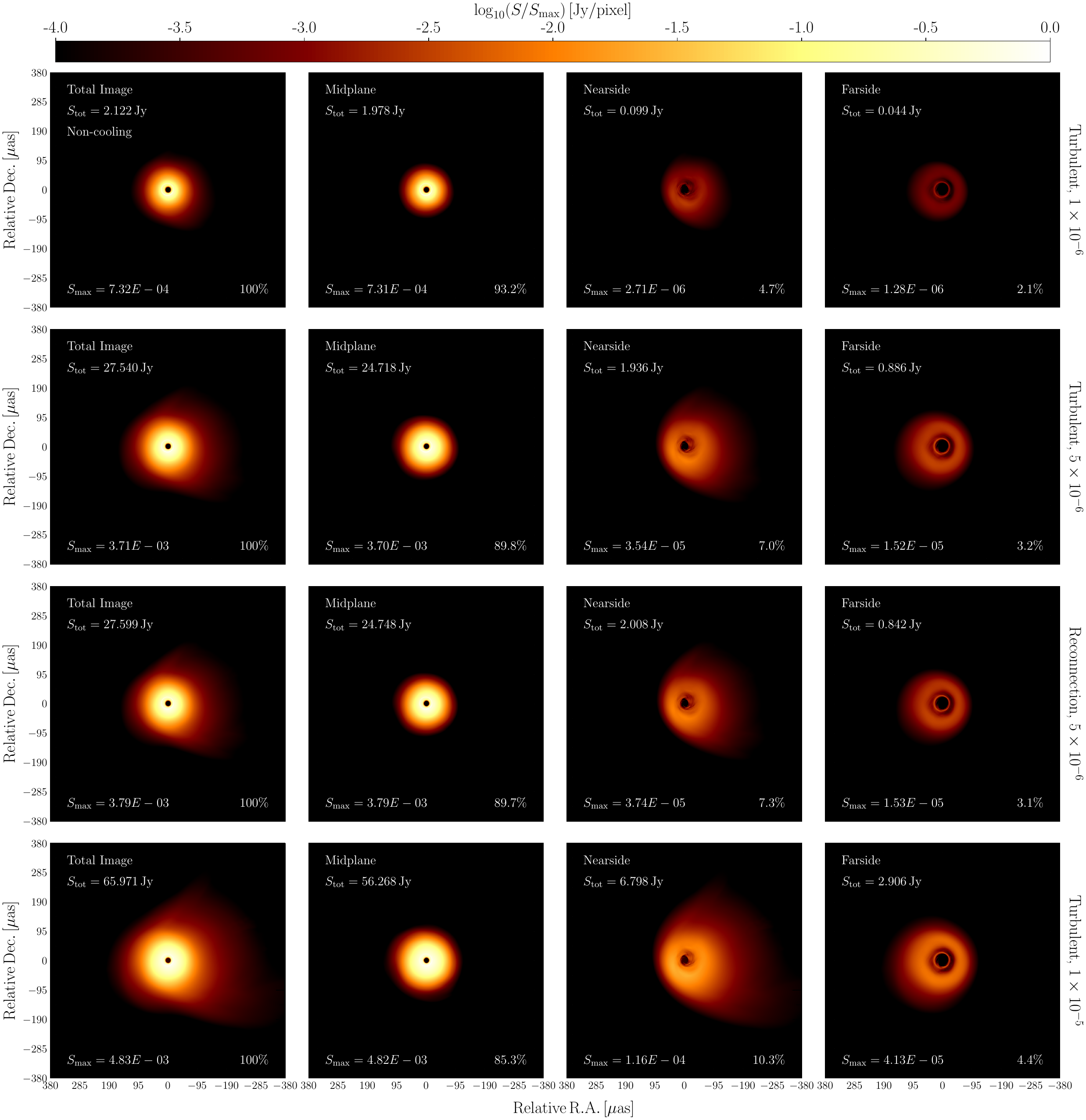}
    \caption{Time-averaged GRRT decomposed images from MAD simulations from $t = 12\,000\,t_{\mathrm{g}}$ to $15\,000\,t_{\mathrm{g}}$, assuming a black hole spin of $a = 0.9375$, observed at $230$ GHz with an inclination angle of $163^\circ$.
    From top to bottom: Accretion rates of $\dot{M}_{\mathrm{BH}}/\dot{M}_\mathrm{Edd} =1\times10^{-6}$, $5\times10^{-6}$, $5\times10^{-6}$, and $1\times10^{-5}$, respectively. The electron heating prescriptions are used for turbulent heating, turbulent heating, reconnection heating, and turbulent heating, respectively.
    From left to right: Emissions from every region, including the full, midplane, nearside jet, and farside jet regions.
    The eDF is modeled as a hybrid of thermal and variable $\kappa$ components, with $\varepsilon=0.5$ in $\kappa$ width, $w$. Cooling is not included.}
    \label{figure5}
\end{figure*}

\begin{figure*}
\centering
    \includegraphics[width=0.85\linewidth]{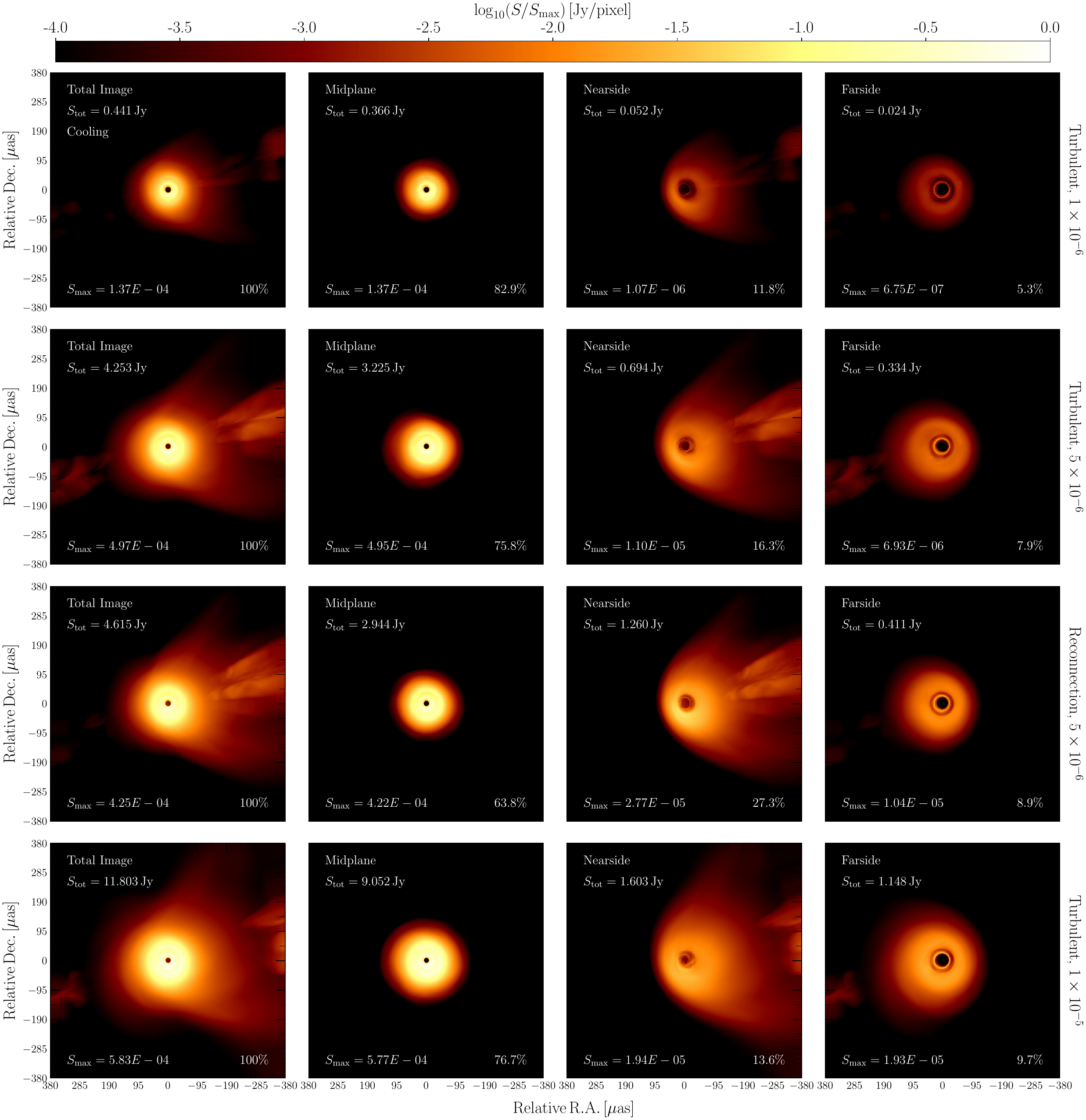}
    \caption{As in Fig.~\ref{figure5} but with radiative cooling included. 
    }
    \label{figure6}
\end{figure*}

Decomposed images estimate the fraction of the total emission contributed by different regions. This allows us to better understand the emission sources and their morphology in the image. By analyzing the decomposed images, we can gain insights into the underlying physical processes that are either responsible for or impact the observed image.
Following previous studies \citep{2019ApJ...875L...5E, 2024A&A...687A..88Z}, we divided the entire region into three parts: the midplane, the nearside jet, and the farside jet. Specifically, the polar angles per region range as follows: the farside jet spans from $0^{\circ}$ to $57.3^{\circ}$ ($1.0$~rad); the midplane lies between $57.3^{\circ}$ ($1.0$~rad) and $122.7^{\circ}$ ($2.14$~rad); and the nearside jet extends from $122.7^{\circ}$ ($2.14$~rad) to $180^{\circ}$.
Emissivity in our GRRT calculations are set at zero in regions outside the specific polar angles mentioned above.

Figure~\ref{figure5} shows the time-averaged decomposed images without radiative cooling, along with their fractional emission contribution to the total image over the interval $t\in [12\,000\,t_{\mathrm{g}},\,15\,000\,t_{\mathrm{g}}]$, at 230~GHz. The panels represent different heating prescriptions and various mass accretion rates with the eDF modeled as a hybrid of thermal and variable $\kappa$ components. 
The percentage marked on the bottom right of each image represents the fraction of the total emission contributed by this region relative to the whole image. 
Morphologically, we can confirm that the extended structure is seen in cases with different heating prescriptions and mass accretion rates from the nearside jet.  
In the turbulent heating scenario, the emission fraction from the nearside jet increases from $4.7 \, \%$ to $10.3 \, \%$ as the mass accretion rates increase. Quantitatively, the increase in mass accretion rate enhances the optical thickness in the midplane (see Sect.~\ref{3.5} for more details). 
Consequently, the contribution from optically thin, extended (nearside) jet emission increases with the accretion rate in the non-cooling case. 
Comparing the non-cooling cases for turbulent and reconnection heating at the same accretion rate $\dot{M}_{\mathrm{BH}}/\dot{M}_{\mathrm{Edd}} = 5 \times 10^{-6}$, the decomposed images are similar. This is because the temperatures are comparable in the absence of cooling (see Appendix~\ref{AppendixC} for more details).

Similar to Fig.~\ref{figure5}, Fig.~\ref{figure6} shows the decomposed images at different mass accretion rates, but including radiative cooling. Taking the non-cooling images as the reference, the effects of radiative cooling on the images can be studied.
First and foremost, compared to the non-cooling case, including radiative cooling leads to a sharp decrease in the total flux for every region at the same mass accretion rate due to the reduced electron temperature. 
Furthermore, the images with radiative cooling clearly exhibit more extended jet structures, and the nearside jet emission contribution increases. Thus, radiative cooling leads to a dim disk, more extended and brighter jets, and reduced total flux. These are consistent with radiative cooling effects on electron temperature, as illustrated in Sect.~\ref{3.3}: radiative cooling results in a cooler disk as well as a slight decrease in electron temperature in the jet sheath. 
Notably, under turbulent heating, as the mass accretion rates increase, the fraction of emission contributed by the nearside jet first increases from $11.8 \, \%$ to $16.3 \, \%$ and then decreases to $13.6 \, \%$. This is because the viewing angle for the nearside jet is smaller for the scenario at a mass accretion rate of $\dot{M}_{\mathrm{BH}}/\dot{M}_{\mathrm{Edd}} = 1 \times 10^{-5}$ when we consider $\sigma_{\mathrm{cut}}=1$ (e.g., see the dash-dotted magenta lines at 15~$r_{\mathrm{g}}$ and at 20~$r_{\mathrm{g}}$ in Fig.~\ref{figure4}). They are located further to the left than in the other cases, indicating that larger regions were excluded. Meanwhile, the viewing angle of the farside jet is similar across cases with different mass accretion rates. Hence, the farside jet contributes relatively more emission as disk becomes cooler with increasing mass accretion rate. Furthermore, as shown in Fig.~\ref{figure10}, when we consider $\sigma_{\mathrm{cut}}=2$ for the case with a mass accretion rate of $\dot{M}_{\mathrm{BH}}/\dot{M}_{\mathrm{Edd}} = 1 \times 10^{-5}$ (for which the viewing angle of the nearside jet is similar to that in the other cases with $\sigma_{\mathrm{cut}}=1$), the fraction of emission from the nearside jet increases to the level seen in the case with $\dot{M}_{\mathrm{BH}}/\dot{M}_{\mathrm{Edd}} = 5 \times 10^{-6}$. Nonetheless, the difference is very small. This may be because the Coulomb interactions dominate over radiative cooling in the inner dense disk ($\sim 7 \, r_{\mathrm{g}}$) for the case with $\dot{M}_{\mathrm{BH}}/\dot{M}_{\mathrm{Edd}} = 1 \times 10^{-5}$. It leads to higher electron temperatures in the inner dense disk (see Fig.~\ref{figure4} for more details) and more emission from the midplane. Therefore, the increase in the fraction of emission from the nearside jet slows down.
When comparing the radiative cooling case with turbulent and reconnection heating at the same mass accretion rate, $\dot{M}_{\mathrm{BH}}/\dot{M}_{\mathrm{Edd}} = 5 \times 10^{-6}$, we find that case $\texttt{C\_KA5e-6}$ exhibits more emissions from the midplane. By contrast, case $\texttt{C\_MR5e-6}$ exhibits more emissions from the nearside and farside jets. These differences using various electron heating prescriptions are significantly not shown in the non-cooling cases (see Appendix~\ref{AppendixC} for further details).
\subsection{Spectral energy distribution}
\label{3.5}

\begin{figure}
\centering
    \includegraphics[width=\linewidth]{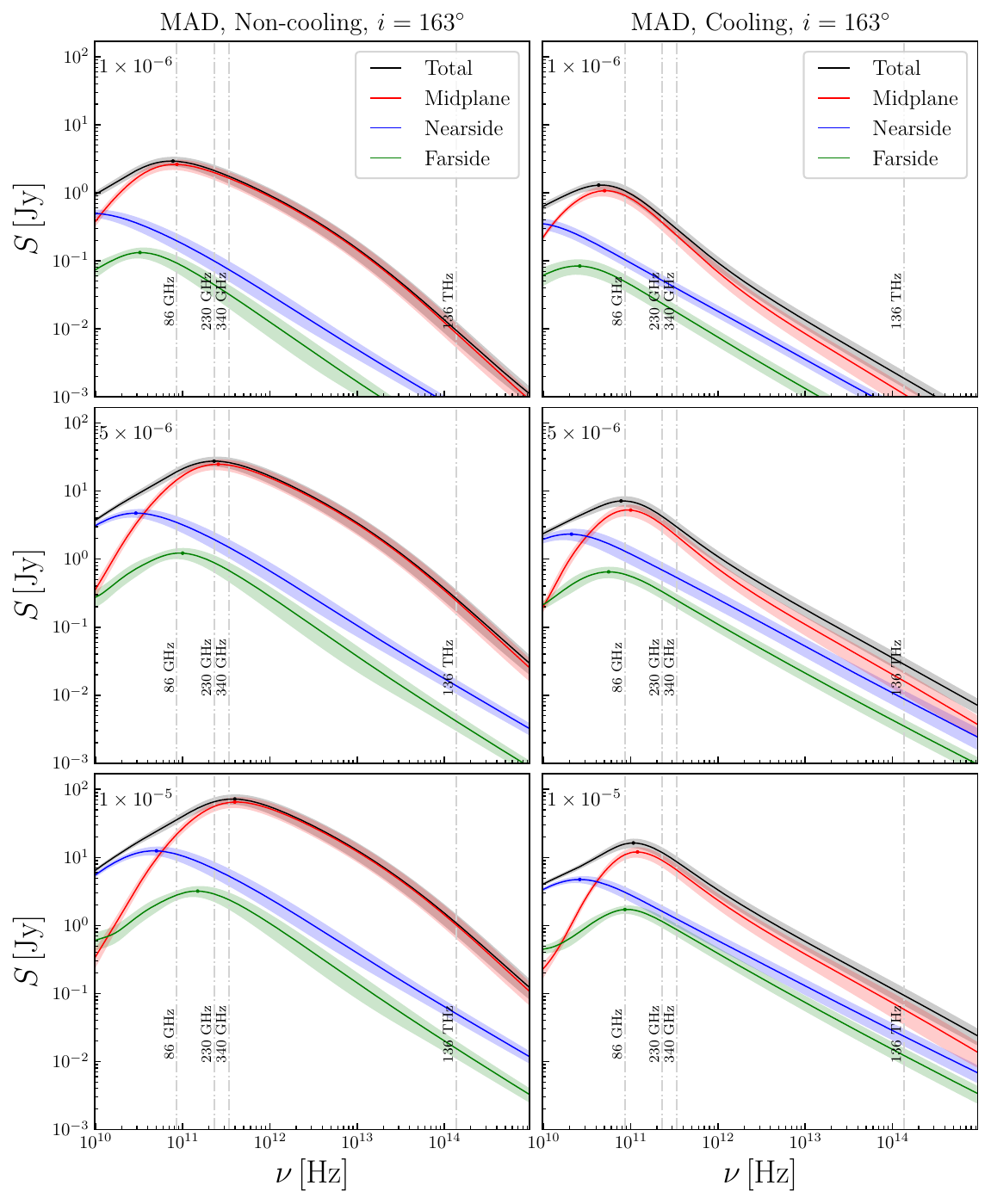}
    \caption{SED curves of different regions under turbulent heating at various accretion rates, without cooling (left) and with radiative cooling (right).
    From top to bottom: Time-averaged mass accretion rates of $\dot{M}_{\mathrm{BH}}/\dot{M}_\mathrm{Edd} =1\times10^{-6}$, $5\times10^{-6}$, and $1\times10^{-5}$.
    All curves adopt the hybrid thermal and variable $\kappa$ eDF.  
    The solid curves represent average values. The points correspond to the peaks. The shaded regions denote the standard deviation relative to the average values.  The vertical dash-dotted lines correspond to 86~GHz, 230~GHz, 340~GHz, and 136~THz. 
    }
    \label{figure7}
\end{figure}

Even though none of the models is precisely tuned to fit the M87$^{*}$ data, we show the SEDs of different regions at various accretion rates without and with cooling in Fig.~\ref{figure7} to investigate how radiative cooling affects flux variation with frequencies at different mass accretion rates.
The solid curves represent the average flux from $t = 12\,000\,t_{\mathrm{g}}$ to $t = 15\,000 \, t_{\mathrm{g}}$, the points correspond to the peaks, and the shaded regions denote the standard deviation resulting from the time variation relative to the average values. The black, red, blue, and green curves represent emission from the whole region, the midplane, the nearside jet, and the farside jet, respectively. 
To compare the effects of radiative cooling, non-cooling cases with various mass accretion rates are included in the left panel of Fig.~\ref{figure7}.

First, it is useful to understand SED behavior as a whole.
As shown in Fig.~\ref{figure7}, SED curve initially increases with increasing frequency, followed by a decrease after reaching a turnover frequency. The turnover frequencies for the non-cooling cases with $\dot{M}_{\mathrm{BH}}/\dot{M}_\mathrm{Edd} =1\times10^{-6}$, $5\times10^{-6}$, and $1\times10^{-5}$ are 78~GHz, 230~GHz, and 397~GHz, respectively. For the radiative cooling cases at the same accretion rates, the turnover frequencies are 43~GHz, 78~GHz, and 108~GHz, respectively. The peak frequency is related to the electron temperature, the magnetic field strength, and the optical depth \citep[e.g.,][]{1998MNRAS.301..435Z}. Notably, radiative cooling reduces the temperature. The optical depth is related to the mass accretion rate and electron distribution. As a result, at the same accretion rate, the peak (turnover frequency) in the cooling case shifts to a lower frequency compared to the non-cooling case. Regardless of whether the case is non-cooling or cooling, the SED with a higher accretion rate reaches its peak at a higher frequency. Thus, the peak position is sensitive to accretion rates and radiative cooling. 

Second, based on the position of the 230~GHz flux relative to the peak (turnover frequency) in the SED, synchrotron radiation is more self-absorbed at higher accretion rates in the non-cooling cases. This means that the emission at 230~GHz becomes optically thick at higher accretion rates. In the radiative cooling cases, however, the emission at 230~GHz remains in the optically thin regime, and the turnover frequencies maintain relatively similar positions.
For the non-cooling cases, this self-absorption mainly occurs in the midplane, while the jet emission is exponentially cut off at 230~GHz. This explains why more emission originates from the jets as the accretion rate increases, as shown in Sect.~\ref{3.4}.

Third, increasing the accretion rates leads to higher fluxes at higher frequencies. For a given accretion rate, radiative cooling reduces the flux at high frequencies compared to the non-cooling case. At low frequencies and at the same accretion rate, the non-cooling and radiative cooling cases behave similarly. 

\subsection{Time variability}

\begin{figure}
\centering
    \includegraphics[width=\linewidth]{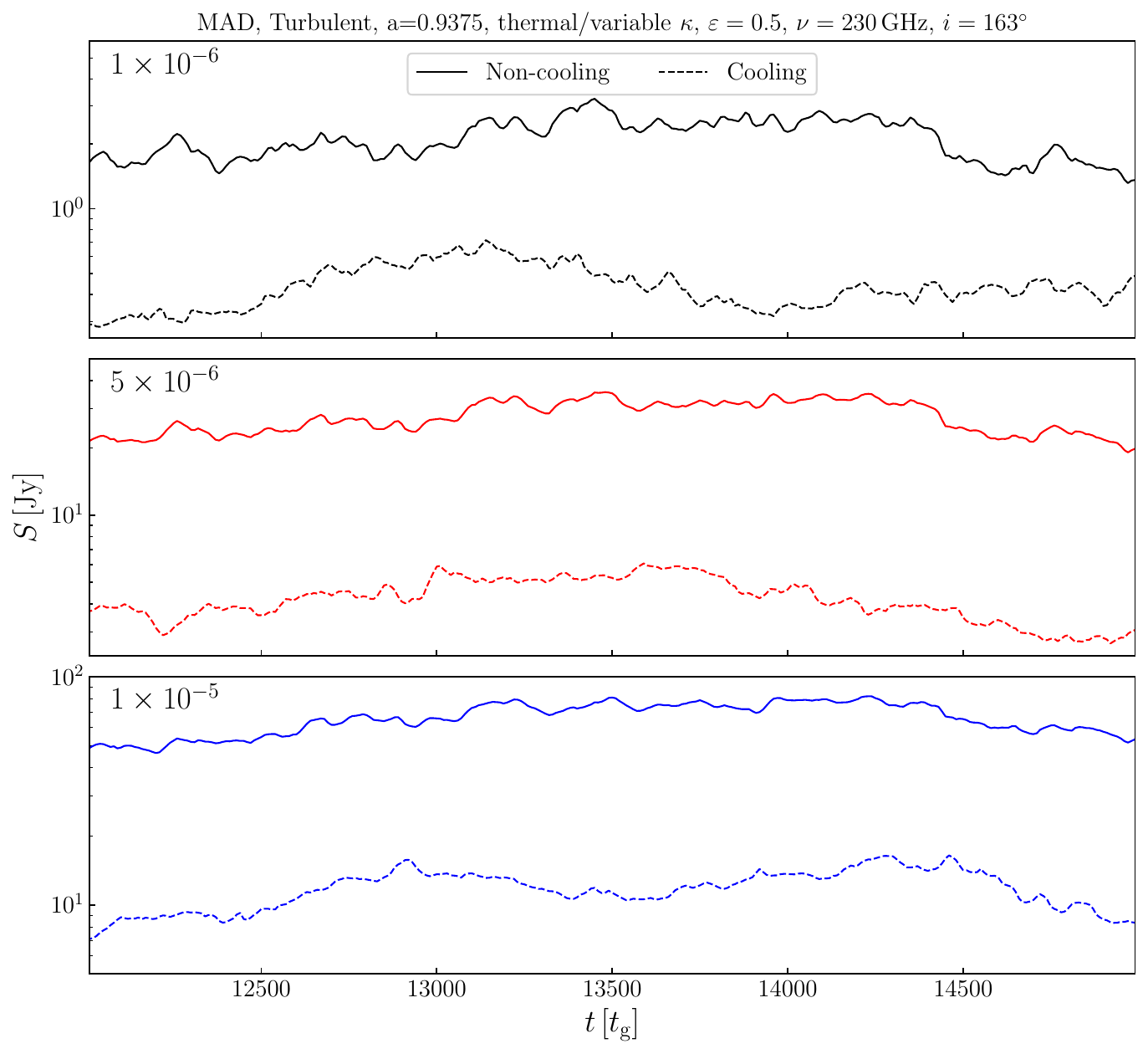}
    \caption{Light curves of the 230~GHz flux at a $163^\circ$ inclination angle and spin $a=0.9375$. The curves are plotted for accretion rates of $\dot{M}_{\mathrm{BH}}/\dot{M}_\mathrm{Edd} =1 \times 10^{-6}$ (top), $5 \times 10^{-6}$ (middle), and $1 \times 10^{-5}$ (bottom). The solid curves represent the non-cooling cases, and the dashed curves represent the radiative cooling cases. All curves assume turbulent heating and adopt the eDF modeled as a hybrid of thermal and variable $\kappa$ components.
    }
    \label{figure8}
\end{figure}

\begin{figure*}
\centering
    \includegraphics[width=\linewidth]{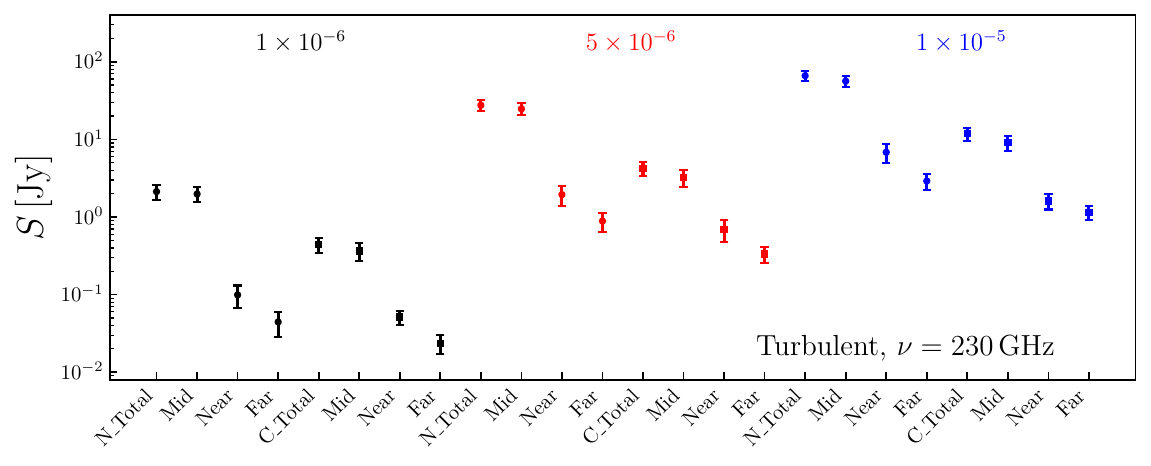}
    \caption{Total flux variation in turbulent heating in the non-cooling (dots) and radiative cooling (squares) cases at 230~GHz at a $163^\circ$ inclination angle and spin $a=0.9375$. The different colors correspond to the different accretion rates: $\dot{M}_{\mathrm{BH}}/\dot{M}_\mathrm{Edd}=1 \times 10^{-6}$ (black), $5 \times 10^{-6}$ (red), and $1 \times 10^{-5}$ (blue). The labels on the x-axis denote the emissions from each region: the full region is depicted first (Total), followed by the midplane (Mid), the nearside jet (Near), and the farside jet (Far). The colored dots and squares represent the time-averaged values, with the error bars indicating the standard deviation about the mean. 
    }
    \label{figure9}
\end{figure*}

\begin{table}
\caption{Modulation index for the different cases.}
\label{table1}
\centering
\begin{tabular}{ccccccc}
\toprule
$\dot{M}_{\mathrm{BH}}/\dot{M}_{\mathrm{Edd}}$ & Regions & $\sigma/\mu$ & $\sigma/\mu_{\mathrm{tot}}$ & $\sigma/\mu$ & $\sigma/\mu_{\mathrm{tot}}$ \\
$[10^{-6}]$ & & \multicolumn{2}{c}{Non-cooling} & \multicolumn{2}{c}{Cooling} \\
\midrule
$1$ & Total & 0.21 & 0.21 & 0.23 & 0.23 \\
$1$ & Midplane & 0.22 & 0.21 & 0.25 & 0.21 \\
$1$ & Nearside & 0.32 & 0.01 & 0.21 & 0.02 \\
$1$ & Farside & 0.36 & 0.01 & 0.27 & 0.01 \\
$5$ & Total & 0.17 & 0.17 & 0.21 & 0.21 \\
$5$ & Midplane & 0.18 & 0.16 & 0.24 & 0.18 \\
$5$ & Nearside & 0.29 & 0.02 & 0.31 & 0.05 \\
$5$ & Farside & 0.29 & 0.01 & 0.23 & 0.02 \\
$10$ & Total & 0.15 & 0.15 & 0.19 & 0.19 \\
$10$ & Midplane & 0.16 & 0.14 & 0.21 & 0.16 \\
$10$ & Nearside & 0.27 & 0.03 & 0.22 & 0.03 \\
$10$ & Farside & 0.24 & 0.01 & 0.20 & 0.02 \\
\bottomrule
\end{tabular}
\tablefoot{The ratios of the standard deviation ($\sigma$) relative to the averaged value for each case ($\mu$) and to the total case ($\mu_{\mathrm{tot}}$) are shown in Fig.~\ref{figure9}. All cases use turbulence heating \citep[][]{2019PNAS..116..771K}.}
\end{table}

Figure~\ref{figure8} shows the 230~GHz light curves for the models with turbulent heating at various mass accretion rates both without and with radiative cooling. 
Generally, the average flux increases with the accretion rate. For a given accretion rate, the flux in the non-cooling case is higher than in the radiative cooling case. 
Additionally, the variability in both the non-cooling and radiative cooling cases decreases as the accretion rate increases.

To quantitatively compare the models at different accretion rates, without and with cooling, Fig.~\ref{figure9} shows the time-averaged flux at 230~GHz and its standard deviation. 
We fixed the accretion rates rather than the flux, so the average value is different from case to case. Overall, the average flux increases with the accretion rate for both non-cooling and cooling cases. At a given accretion rate, the non-cooling case exhibits higher flux than the radiative cooling case.

To understand the time variability for each case, we list the ratio of standard deviation relative to the average value in Table~\ref{table1}. For the non-cooling cases with emission from the entire region (total), the results show that the modulation index (the standard deviation divided by the average value) decreases from 0.21 to 0.15 as the mass accretion rate increases from $\dot{M}_{\mathrm{BH}}/\dot{M}_\mathrm{Edd}=1 \times 10^{-6}$ to $1 \times 10^{-5}$.   
This trend is the same for the midplane, nearside, and farside regions. It indicates that for non-cooling cases, the time variability of all regions decreases with increasing accretion rate. 
For radiative cooling cases, we see a similar dependence of time variability on the accretion rate for the total region and the midplane. However, no clear dependence of time variability on the accretion rate is evident for the nearside and farside jet regions in the cooling cases. 

To determine the origin of the time variability in the total region, we also list the ratio of the standard deviation relative to the average value of the total case in Table~\ref{table1}. Our results show that, the time variability of the total region is primarily driven by the midplane. Furthermore, in the cooling cases, the jet regions also partially contribute to the variability. The proportion of emission from the jet regions is higher in the cooling than in the non-cooling cases (see Figs.~\ref{figure5} Fig.~\ref{figure6} for these proportions). 

\section{Summary and discussion}
\label{discussion}

In our previous study \citep{2024A&A...687A..88Z}, we ignored energy exchange impacts from Coulomb coupling and radiative cooling on black hole shadow and extended jet images. To address this issue, we adopted the electron heating prescriptions of \cite{2021MNRAS.506..741M} and added both Coulomb interactions and radiative cooling, following \cite{2023MNRAS.518..405D}, in our GRMHD simulations. In our GRRT calculations, we adopted a hybrid thermal and variable nonthermal $\kappa$ eDF, \citep[e.g.,][]{2022ApJ...930L..16E,2022NatAs...6..103C,2023ApJ...959L...3D,2024A&A...687A..88Z}, with $\varepsilon=0.5$. We also considered the non-cooling cases as a reference for exploring radiative cooling impacts on density distribution, electron temperature distribution, shadow images, jet morphology, SEDs, and total flux variation at 230~GHz at an inclination angle of $i=163^\circ$. Below we list our conclusions point-by-point.

\begin{enumerate}
    \item We find that Coulomb coupling and radiative cooling significantly modify the electron temperature distribution, leading to a cooler accretion disk, a slightly cooler jet sheath, and changes in black hole shadow morphology.

    \item Our results show that radiative cooling produces a dimmer ring, more extended and brighter jets, and a reduced total flux emission.

    \item Our results show that radiative cooling mediates heating effects. At an accretion rate of $\dot{M}_{\mathrm{BH}}/\dot{M}_\mathrm{Edd} = 5 \times 10^{-6}$,  the inclusion of radiative cooling results in a dimmer disk but brighter jets under reconnection heating compared to the turbulent heating case. Notably, the total flux remains higher under reconnection heating than under turbulent heating in both the non-cooling and radiative cooling scenarios.
    
    \item We find that  cooling reduces synchrotron self-absorption, shifting the turnover frequency to lower values at the same accretion rate.
    
    \item Our results show that time variability primarily originates from the midplane in both the cooling and non-cooling cases, and higher accretion rates reduce time variability.

    \item We find that although our simulations are scaled to M87$^{*}$, our results are generally applicable to other low-luminosity active galactic nuclei (LLAGNs) in the MAD state, where radiative cooling universally leads to a cooler disk, brighter jets, reduced total flux, and smoother light curves, independent of black hole mass and distance.
\end{enumerate}
 
We note that, in this study, we only considered the $\sigma_{\mathrm{cut}} = 1$, zeroing out 230~GHz emission from the regions with a higher magnetization. Several studies have examined the impact of different $\sigma_{\mathrm{cut}}$ assumptions. For example, the results are sensitive to the choice of $\sigma_{\mathrm{cut}}$ at frequencies $\nu > 230$~GHz \citep{2019MNRAS.486.2873C}. Including contributions from regions with stronger magnetization enhances the jet \citep{2019MNRAS.486.2873C, 2024A&A...687A..88Z}. Additionally, a lower $\sigma_{\mathrm{cut}}$ value requires a higher accretion rate to satisfy the observed flux density, resulting in more efficient radiative cooling and a greater Faraday rotation depth \citep{2025MNRAS.537.2496C}. Hence, with radiative cooling, different treatments of $\sigma_{\mathrm{cut}}$ yield varying jet structures and polarization fractions, as these depend on the accretion rate (follow Appendix \ref{AppendixA} for more details). 

In addition, we used radiative cooling and heating to avoid overestimating electron temperature. We find that radiative cooling reduces the flux at high frequencies compared to the non-cooling case. 
However, future GRMHD simulations should incorporate cooling functions more self-consistently to account for different nonthermal distributions. This includes adding cooling terms for nonthermal electrons due to bremsstrahlung, synchrotron radiation, and inverse Compton scattering. These nonthermal electrons would radiate more efficiently than thermal electrons and might possibly cool the plasma further. Moreover, it would be interesting to find the proper mass accretion rates (likely between $1 \times 10^{-6}\,\dot{M}_\mathrm{Edd}$ and $5 \times 10^{-6}\,\dot{M}_\mathrm{Edd}$) for different models targeting the same flux, and to assess which model best fits the observation. For instance,  \cite{2022NatAs...6..103C} successfully reproduce the broadband spectrum of M87 from radio to near-infrared bands without considering radiative cooling. To reproduce the broadband spectrum while accounting for radiative cooling, it is necessary to include more nonthermal electrons. This ensures sufficient emission in simulations in the near-infrared and optical bands. This raises important questions about which electron heating prescriptions can accelerate the electrons and where this electron energization occurs. 

Lastly, \cite{2024A&A...687A..88Z} show that nonthermal electrons result in more extended and brighter jets. Furthermore, as shown in this paper, radiative cooling still produces a faint large-scale jet emission on horizon scales, which appears brighter in GRRT images at 230~GHz. The total intensity of this emission varies with different treatments of $\sigma_{\mathrm{cut}}$. The presence of such horizon-scale jet emission, as well as the explicit consideration of $\sigma_{\mathrm{cut}}$, should be rigorously evaluated against observational data. Although currently below the dynamic range of EHT observations, both the horizon-scale shadow and the extended jet images may be resolved simultaneously at 230~GHz by the next-generation arrays, such as the ngEHT \citep{2023Galax..11...61J, 2023Galax..11....5R, 2025LRR....28....4A}. These arrays would have sufficient dynamic range to detect the features reported in this study.

\begin{acknowledgements}
    This research is supported by the National Key R\&D Program of China (2023YFE0101200), the National Natural Science Foundation of China (Grant No. 12273022, 12511540053), and the Shanghai Municipality orientation program of Basic Research for International Scientists (Grant No. 22JC1410600). MZ is supported by the Doctoral Student Program of the Young S$\&$T Talents Cultivation Project, CAST, and by the T.D. Lee scholarship. IKD acknowledges the TDLI postdoctoral fellowship for financial support. 
    ACO acknowledges to DGAPA-UNAM (grant IN110522), the Ciencia B\'{a}sica y de Frontera 2023–2024 program of SECIHTI M\'{e}xico (projects CBF2023-2024-1102 and 257435), and the European Horizon Europe Staff Exchange (SE) programme HORIZON-MSCA2021-SE-01 under Grant No. NewFunFiCO-101086251.
    The simulations were performed on the Astro cluster at Tsung-Dao Lee Institute, Pi~2.0, and the Siyuan-1 cluster in the Center for High Performance Computing at Shanghai Jiao Tong University.
    This work has made use of NASA's Astrophysics Data System (ADS). 
\end{acknowledgements}

%
   \bibliographystyle{aa} 
   \bibliography{example} 

@ARTICLE{2025ApJ...990L..33I,
       author = {{Imbrogno}, Mario and {Meringolo}, Claudio and {Cruz-Osorio}, Alejandro and {Rezzolla}, Luciano and {Cerutti}, Beno{\^\i}t and {Servidio}, Sergio},
        title = "{Turbulence and Magnetic Reconnection in Relativistic Multispecies Plasmas}",
      journal = {\apjl},
         year = 2025,
        month = sep,
       volume = {990},
       number = {2},
          eid = {L33},
        pages = {L33},
          doi = {10.3847/2041-8213/adfb4c},
archivePrefix = {arXiv},
       eprint = {2505.09700},
 primaryClass = {physics.plasm-ph},
       adsurl = {https://ui.adsabs.harvard.edu/abs/2025ApJ...990L..33I}
}

@article{Imbrogno2024,
       author = {{Imbrogno}, Mario and {Meringolo}, Claudio and {Servidio}, Sergio and {Cruz-Osorio}, Alejandro and {Cerutti}, Beno{\^\i}t and {Pegoraro}, Francesco},
        title = "{Long-lived Equilibria in Kinetic Astrophysical Plasma Turbulence}",
      journal = {Astrophys. J. Lett.},
         year = 2024,
        month = sep,
       volume = {972},
       number = {1},
          eid = {L5},
        pages = {L5},
          doi = {10.3847/2041-8213/ad6b9d},
archivePrefix = {arXiv},
       eprint = {2408.02656},
 primaryClass = {physics.plasm-ph},
       adsurl = {https://ui.adsabs.harvard.edu/abs/2024ApJ...972L...5I}
}

@ARTICLE{2019ApJS..243...26P,
       author = {{Porth}, Oliver and {Chatterjee}, Koushik and {Narayan}, Ramesh and {Gammie}, Charles F. and {Mizuno}, Yosuke and {Anninos}, Peter and {Baker}, John G. and {Bugli}, Matteo and {Chan}, Chi-kwan and {Davelaar}, Jordy and {Del Zanna}, Luca and {Etienne}, Zachariah B. and {Fragile}, P. Chris and {Kelly}, Bernard J. and {Liska}, Matthew and {Markoff}, Sera and {McKinney}, Jonathan C. and {Mishra}, Bhupendra and {Noble}, Scott C. and {Olivares}, H{\'e}ctor and {Prather}, Ben and {Rezzolla}, Luciano and {Ryan}, Benjamin R. and {Stone}, James M. and {Tomei}, Niccol{\`o} and {White}, Christopher J. and {Younsi}, Ziri and {Akiyama}, Kazunori and {Alberdi}, Antxon and {Alef}, Walter and {Asada}, Keiichi and {Azulay}, Rebecca and {Baczko}, Anne-Kathrin and {Ball}, David and {Balokovi{\'c}}, Mislav and {Barrett}, John and {Bintley}, Dan and {Blackburn}, Lindy and {Boland}, Wilfred and {Bouman}, Katherine L. and {Bower}, Geoffrey C. and {Bremer}, Michael and {Brinkerink}, Christiaan D. and {Brissenden}, Roger and {Britzen}, Silke and {Broderick}, Avery E. and {Broguiere}, Dominique and {Bronzwaer}, Thomas and {Byun}, Do-Young and {Carlstrom}, John E. and {Chael}, Andrew and {Chatterjee}, Shami and {Chen}, Ming-Tang and {Chen}, Yongjun and {Cho}, Ilje and {Christian}, Pierre and {Conway}, John E. and {Cordes}, James M. and {Geoffrey} and {Crew}, B. and {Cui}, Yuzhu and {De Laurentis}, Mariafelicia and {Deane}, Roger and {Dempsey}, Jessica and {Desvignes}, Gregory and {Doeleman}, Sheperd S. and {Eatough}, Ralph P. and {Falcke}, Heino and {Fish}, Vincent L. and {Fomalont}, Ed and {Fraga-Encinas}, Raquel and {Freeman}, Bill and {Friberg}, Per and {Fromm}, Christian M. and {G{\'o}mez}, Jos{\'e} L. and {Galison}, Peter and {Garc{\'\i}a}, Roberto and {Gentaz}, Olivier and {Georgiev}, Boris and {Goddi}, Ciriaco and {Gold}, Roman and {Gu}, Minfeng and {Gurwell}, Mark and {Hada}, Kazuhiro and {Hecht}, Michael H. and {Hesper}, Ronald and {Ho}, Luis C. and {Ho}, Paul and {Honma}, Mareki and {Huang}, Chih-Wei L. and {Huang}, Lei and {Hughes}, David H. and {Ikeda}, Shiro and {Inoue}, Makoto and {Issaoun}, Sara and {James}, David J. and {Jannuzi}, Buell T. and {Janssen}, Michael and {Jeter}, Britton and {Jiang}, Wu and {Johnson}, Michael D. and {Jorstad}, Svetlana and {Jung}, Taehyun and {Karami}, Mansour and {Karuppusamy}, Ramesh and {Kawashima}, Tomohisa and {Keating}, Garrett K. and {Kettenis}, Mark and {Kim}, Jae-Young and {Kim}, Junhan and {Kim}, Jongsoo and {Kino}, Motoki and {Koay}, Jun Yi and {Patrick} and {Koch}, M. and {Koyama}, Shoko and {Kramer}, Michael and {Kramer}, Carsten and {Krichbaum}, Thomas P. and {Kuo}, Cheng-Yu and {Lauer}, Tod R. and {Lee}, Sang-Sung and {Li}, Yan-Rong and {Li}, Zhiyuan and {Lindqvist}, Michael and {Liu}, Kuo and {Liuzzo}, Elisabetta and {Lo}, Wen-Ping and {Lobanov}, Andrei P. and {Loinard}, Laurent and {Lonsdale}, Colin and {Lu}, Ru-Sen and {MacDonald}, Nicholas R. and {Mao}, Jirong and {Marrone}, Daniel P. and {Marscher}, Alan P. and {Mart{\'\i}-Vidal}, Iv{\'a}n and {Matsushita}, Satoki and {Matthews}, Lynn D. and {Medeiros}, Lia and {Menten}, Karl M. and {Mizuno}, Izumi and {Moran}, James M. and {Moriyama}, Kotaro and {Moscibrodzka}, Monika and {M{\"u}ller}, Cornelia and {Nagai}, Hiroshi and {Nagar}, Neil M. and {Nakamura}, Masanori and {Narayanan}, Gopal and {Natarajan}, Iniyan and {Neri}, Roberto and {Ni}, Chunchong and {Noutsos}, Aristeidis and {Okino}, Hiroki and {Oyama}, Tomoaki and {{\"O}zel}, Feryal and {Palumbo}, Daniel C.~M. and {Patel}, Nimesh and {Pen}, Ue-Li and {Pesce}, Dominic W. and {Pi{\'e}tu}, Vincent and {Plambeck}, Richard and {PopStefanija}, Aleksandar and {Preciado-L{\'o}pez}, Jorge A. and {Psaltis}, Dimitrios and {Pu}, Hung-Yi and {Ramakrishnan}, Venkatessh and {Rao}, Ramprasad and {Rawlings}, Mark G. and {Raymond}, Alexander W. and {Ripperda}, Bart and {Roelofs}, Freek and {Rogers}, Alan and {Ros}, Eduardo and {Rose}, Mel and {Roshanineshat}, Arash and {Rottmann}, Helge and {Roy}, Alan L. and {Ruszczyk}, Chet and {Rygl}, Kazi L.~J. and {S{\'a}nchez}, Salvador and {S{\'a}nchez-Arguelles}, David and {Sasada}, Mahito and {Savolainen}, Tuomas and {Schloerb}, F. Peter and {Schuster}, Karl-Friedrich and {Shao}, Lijing and {Shen}, Zhiqiang and {Small}, Des and {Sohn}, Bong Won and {SooHoo}, Jason and {Tazaki}, Fumie and {Tiede}, Paul and {Tilanus}, Remo P.~J. and {Titus}, Michael and {Toma}, Kenji and {Torne}, Pablo and {Trent}, Tyler and {Trippe}, Sascha},
        title = "{The Event Horizon General Relativistic Magnetohydrodynamic Code Comparison Project}",
      journal = {\apjs},
         year = 2019,
        month = aug,
       volume = {243},
       number = {2},
          eid = {26},
        pages = {26},
          doi = {10.3847/1538-4365/ab29fd},
archivePrefix = {arXiv},
       eprint = {1904.04923},
 primaryClass = {astro-ph.HE},
       adsurl = {https://ui.adsabs.harvard.edu/abs/2019ApJS..243...26P}
}

@ARTICLE{2019ApJ...882....2V,
       author = {{Vourellis}, Christos and {Fendt}, Christian and {Qian}, Qian and {Noble}, Scott C.},
        title = "{GR-MHD Disk Winds and Jets from Black Holes and Resistive Accretion Disks}",
      journal = {\apj},
         year = 2019,
        month = sep,
       volume = {882},
       number = {1},
          eid = {2},
        pages = {2},
          doi = {10.3847/1538-4357/ab32e2},
archivePrefix = {arXiv},
       eprint = {1907.10622},
 primaryClass = {astro-ph.HE},
       adsurl = {https://ui.adsabs.harvard.edu/abs/2019ApJ...882....2V}
}

@ARTICLE{2021MNRAS.505.3596D,
       author = {{Dihingia}, Indu K. and {Vaidya}, Bhargav and {Fendt}, Christian},
        title = "{Jets, disc-winds, and oscillations in general relativistic, magnetically driven flows around black hole}",
      journal = {\mnras},
         year = 2021,
        month = aug,
       volume = {505},
       number = {3},
        pages = {3596-3615},
          doi = {10.1093/mnras/stab1512},
archivePrefix = {arXiv},
       eprint = {2105.11468},
 primaryClass = {astro-ph.HE},
       adsurl = {https://ui.adsabs.harvard.edu/abs/2021MNRAS.505.3596D}
}

@ARTICLE{Dihingia_2025,
       author = {{Dihingia}, Indu K. and {Mizuno}, Yosuke and {Fromm}, Christian M. and {Younsi}, Ziri},
        title = "{Impact of radiative cooling on the magnetised geometrically thin  accretion disc around Kerr black hole}",
      journal = {\jcap},
         year = 2025,
        month = jan,
       volume = {2025},
       number = {01},
        pages = {152},
          doi = {10.1088/1475-7516/2025/01/152},
archivePrefix = {arXiv},
       eprint = {2305.09698},
 primaryClass = {astro-ph.HE},
       adsurl = {https://dx.doi.org/10.1088/1475-7516/2025/01/152}
}

@ARTICLE{2023MNRAS.518..405D,
       author = {{Dihingia}, Indu K. and {Mizuno}, Yosuke and {Fromm}, Christian M. and {Rezzolla}, Luciano},
        title = "{Temperature properties in magnetized and radiatively cooled two-temperature accretion flows on to a black hole}",
      journal = {\mnras},
         year = 2023,
        month = jan,
       volume = {518},
       number = {1},
        pages = {405-417},
          doi = {10.1093/mnras/stac3165},
archivePrefix = {arXiv},
       eprint = {2206.13184},
 primaryClass = {astro-ph.HE},
       adsurl = {https://ui.adsabs.harvard.edu/abs/2023MNRAS.518..405D}
}

@ARTICLE{2012MNRAS.423L..55T,
       author = {{Tchekhovskoy}, Alexander and {McKinney}, Jonathan C.},
        title = "{Prograde and retrograde black holes: whose jet is more powerful?}",
      journal = {\mnras},
         year = 2012,
        month = jun,
       volume = {423},
       number = {1},
        pages = {L55-L59},
          doi = {10.1111/j.1745-3933.2012.01256.x},
archivePrefix = {arXiv},
       eprint = {1201.4385},
 primaryClass = {astro-ph.HE},
       adsurl = {https://ui.adsabs.harvard.edu/abs/2012MNRAS.423L..55T}
}

@ARTICLE{2021MNRAS.506..741M,
       author = {{Mizuno}, Yosuke and {Fromm}, Christian M. and {Younsi}, Ziri and {Porth}, Oliver and {Olivares}, Hector and {Rezzolla}, Luciano},
        title = "{Comparison of the ion-to-electron temperature ratio prescription: GRMHD simulations with electron thermodynamics}",
      journal = {\mnras},
         year = 2021,
        month = sep,
       volume = {506},
       number = {1},
        pages = {741-758},
          doi = {10.1093/mnras/stab1753},
archivePrefix = {arXiv},
       eprint = {2106.09272},
 primaryClass = {astro-ph.HE},
       adsurl = {https://ui.adsabs.harvard.edu/abs/2021MNRAS.506..741M}
}

@ARTICLE{2019ApJ...875L...5E,
       author = {{Event Horizon Telescope Collaboration} and {Akiyama}, Kazunori and {Alberdi}, Antxon and {Alef}, Walter and {Asada}, Keiichi and {Azulay}, Rebecca and {Baczko}, Anne-Kathrin and {Ball}, David and {Balokovi{\'c}}, Mislav and {Barrett}, John and {Bintley}, Dan and {Blackburn}, Lindy and {Boland}, Wilfred and {Bouman}, Katherine L. and {Bower}, Geoffrey C. and {Bremer}, Michael and {Brinkerink}, Christiaan D. and {Brissenden}, Roger and {Britzen}, Silke and {Broderick}, Avery E. and {Broguiere}, Dominique and {Bronzwaer}, Thomas and {Byun}, Do-Young and {Carlstrom}, John E. and {Chael}, Andrew and {Chan}, Chi-kwan and {Chatterjee}, Shami and {Chatterjee}, Koushik and {Chen}, Ming-Tang and {Chen}, Yongjun and {Cho}, Ilje and {Christian}, Pierre and {Conway}, John E. and {Cordes}, James M. and {Crew}, Geoffrey B. and {Cui}, Yuzhu and {Davelaar}, Jordy and {De Laurentis}, Mariafelicia and {Deane}, Roger and {Dempsey}, Jessica and {Desvignes}, Gregory and {Dexter}, Jason and {Doeleman}, Sheperd S. and {Eatough}, Ralph P. and {Falcke}, Heino and {Fish}, Vincent L. and {Fomalont}, Ed and {Fraga-Encinas}, Raquel and {Friberg}, Per and {Fromm}, Christian M. and {G{\'o}mez}, Jos{\'e} L. and {Galison}, Peter and {Gammie}, Charles F. and {Garc{\'\i}a}, Roberto and {Gentaz}, Olivier and {Georgiev}, Boris and {Goddi}, Ciriaco and {Gold}, Roman and {Gu}, Minfeng and {Gurwell}, Mark and {Hada}, Kazuhiro and {Hecht}, Michael H. and {Hesper}, Ronald and {Ho}, Luis C. and {Ho}, Paul and {Honma}, Mareki and {Huang}, Chih-Wei L. and {Huang}, Lei and {Hughes}, David H. and {Ikeda}, Shiro and {Inoue}, Makoto and {Issaoun}, Sara and {James}, David J. and {Jannuzi}, Buell T. and {Janssen}, Michael and {Jeter}, Britton and {Jiang}, Wu and {Johnson}, Michael D. and {Jorstad}, Svetlana and {Jung}, Taehyun and {Karami}, Mansour and {Karuppusamy}, Ramesh and {Kawashima}, Tomohisa and {Keating}, Garrett K. and {Kettenis}, Mark and {Kim}, Jae-Young and {Kim}, Junhan and {Kim}, Jongsoo and {Kino}, Motoki and {Koay}, Jun Yi and {Koch}, Patrick M. and {Koyama}, Shoko and {Kramer}, Michael and {Kramer}, Carsten and {Krichbaum}, Thomas P. and {Kuo}, Cheng-Yu and {Lauer}, Tod R. and {Lee}, Sang-Sung and {Li}, Yan-Rong and {Li}, Zhiyuan and {Lindqvist}, Michael and {Liu}, Kuo and {Liuzzo}, Elisabetta and {Lo}, Wen-Ping and {Lobanov}, Andrei P. and {Loinard}, Laurent and {Lonsdale}, Colin and {Lu}, Ru-Sen and {MacDonald}, Nicholas R. and {Mao}, Jirong and {Markoff}, Sera and {Marrone}, Daniel P. and {Marscher}, Alan P. and {Mart{\'\i}-Vidal}, Iv{\'a}n and {Matsushita}, Satoki and {Matthews}, Lynn D. and {Medeiros}, Lia and {Menten}, Karl M. and {Mizuno}, Yosuke and {Mizuno}, Izumi and {Moran}, James M. and {Moriyama}, Kotaro and {Moscibrodzka}, Monika and {Mul{\ensuremath{\ddot{}}}ler}, Cornelia and {Nagai}, Hiroshi and {Nagar}, Neil M. and {Nakamura}, Masanori and {Narayan}, Ramesh and {Narayanan}, Gopal and {Natarajan}, Iniyan and {Neri}, Roberto and {Ni}, Chunchong and {Noutsos}, Aristeidis and {Okino}, Hiroki and {Olivares}, H{\'e}ctor and {Oyama}, Tomoaki and {{\"O}zel}, Feryal and {Palumbo}, Daniel C.~M. and {Patel}, Nimesh and {Pen}, Ue-Li and {Pesce}, Dominic W. and {Pi{\'e}tu}, Vincent and {Plambeck}, Richard and {PopStefanija}, Aleksandar and {Porth}, Oliver and {Prather}, Ben and {Preciado-L{\'o}pez}, Jorge A. and {Psaltis}, Dimitrios and {Pu}, Hung-Yi and {Ramakrishnan}, Venkatessh and {Rao}, Ramprasad and {Rawlings}, Mark G. and {Raymond}, Alexander W. and {Rezzolla}, Luciano and {Ripperda}, Bart and {Roelofs}, Freek and {Rogers}, Alan and {Ros}, Eduardo and {Rose}, Mel and {Roshanineshat}, Arash and {Rottmann}, Helge and {Roy}, Alan L. and {Ruszczyk}, Chet and {Ryan}, Benjamin R. and {Rygl}, Kazi L.~J. and {S{\'a}nchez}, Salvador and {S{\'a}nchez-Arguelles}, David and {Sasada}, Mahito and {Savolainen}, Tuomas and {Schloerb}, F. Peter and {Schuster}, Karl-Friedrich and {Shao}, Lijing and {Shen}, Zhiqiang and {Small}, Des and {Sohn}, Bong Won and {SooHoo}, Jason and {Tazaki}, Fumie and {Tiede}, Paul and {Tilanus}, Remo P.~J. and {Titus}, Michael and {Toma}, Kenji and {Torne}, Pablo and {Trent}, Tyler and {Trippe}, Sascha and {Tsuda}, Shuichiro and {van Bemmel}, Ilse and {van Langevelde}, Huib Jan and {van Rossum}, Daniel R. and {Wagner}, Jan and {Wardle}, John and {Weintroub}, Jonathan and {Wex}, Norbert and {Wharton}, Robert and {Wielgus}, Maciek and {Wong}, George N. and {Wu}, Qingwen and {Young}, Andr{\'e} and {Young}, Ken and {Younsi}, Ziri and {Yuan}, Feng and {Yuan}, Ye-Fei and {Zensus}, J. Anton and {Zhao}, Guangyao and {Zhao}, Shan-Shan and {Zhu}, Ziyan and {Anczarski}, Jadyn and {Baganoff}, Frederick K. and {Eckart}, Andreas and {Farah}, Joseph R. and {Haggard}, Daryl and {Meyer-Zhao}, Zheng and {Michalik}, Daniel and {Nadolski}, Andrew and {Neilsen}, Joseph and {Nishioka}, Hiroaki and {Nowak}, Michael A. and {Pradel}, Nicolas and {Primiani}, Rurik A. and {Souccar}, Kamal and {Vertatschitsch}, Laura and {Yamaguchi}, Paul and {Zhang}, Shuo},
        title = "{First M87 Event Horizon Telescope Results. V. Physical Origin of the Asymmetric Ring}",
      journal = {\apjl},
         year = 2019,
        month = apr,
       volume = {875},
       number = {1},
          eid = {L5},
        pages = {L5},
          doi = {10.3847/2041-8213/ab0f43},
archivePrefix = {arXiv},
       eprint = {1906.11242},
 primaryClass = {astro-ph.GA},
       adsurl = {https://ui.adsabs.harvard.edu/abs/2019ApJ...875L...5E}
}

@ARTICLE{2024A&A...687A..88Z,
       author = {{Zhang}, Mingyuan and {Mizuno}, Yosuke and {Fromm}, Christian M. and {Younsi}, Ziri and {Cruz-Osorio}, Alejandro},
        title = "{Impacts of nonthermal emission on the images of a black hole shadow and extended jets in two-temperature GRMHD simulations}",
      journal = {\aap},
         year = 2024,
        month = jul,
       volume = {687},
          eid = {A88},
        pages = {A88},
          doi = {10.1051/0004-6361/202449497},
archivePrefix = {arXiv},
       eprint = {2404.04033},
 primaryClass = {astro-ph.HE},
       adsurl = {https://ui.adsabs.harvard.edu/abs/2024A&A...687A..88Z}
}

@ARTICLE{2022A&A...660A.107F,
       author = {{Fromm}, Christian M. and {Cruz-Osorio}, Alejandro and {Mizuno}, Yosuke and {Nathanail}, Antonios and {Younsi}, Ziri and {Porth}, Oliver and {Olivares}, Hector and {Davelaar}, Jordy and {Falcke}, Heino and {Kramer}, Michael and {Rezzolla}, Luciano},
        title = "{Impact of non-thermal particles on the spectral and structural properties of M87}",
      journal = {\aap},
         year = 2022,
        month = apr,
       volume = {660},
          eid = {A107},
        pages = {A107},
          doi = {10.1051/0004-6361/202142295},
archivePrefix = {arXiv},
       eprint = {2111.02518},
 primaryClass = {astro-ph.HE},
       adsurl = {https://ui.adsabs.harvard.edu/abs/2022A&A...660A.107F}
}

@ARTICLE{1998MNRAS.301..435Z,
       author = {{Zdziarski}, Andrzej A. and {Poutanen}, Juri and {Mikolajewska}, Joanna and {Gierlinski}, Marek and {Ebisawa}, Ken and {Johnson}, W. Neil},
        title = "{Broad-band X-ray/gamma-ray spectra and binary parameters of GX 339-4 and their astrophysical implications}",
      journal = {\mnras},
         year = 1998,
        month = dec,
       volume = {301},
       number = {2},
        pages = {435-450},
          doi = {10.1046/j.1365-8711.1998.02021.x},
archivePrefix = {arXiv},
       eprint = {astro-ph/9807300},
 primaryClass = {astro-ph},
       adsurl = {https://ui.adsabs.harvard.edu/abs/1998MNRAS.301..435Z}
}

@ARTICLE{2023ApJ...959L...3D,
       author = {{Davelaar}, J. and {Ripperda}, B. and {Sironi}, L. and {Philippov}, A.~A. and {Olivares}, H. and {Porth}, O. and {Berg}, B. van den and {Bronzwaer}, T. and {Chatterjee}, K. and {Liska}, M.},
        title = "{Synchrotron Polarization Signatures of Surface Waves in Supermassive Black Hole Jets}",
      journal = {\apjl},
         year = 2023,
        month = dec,
       volume = {959},
       number = {1},
          eid = {L3},
        pages = {L3},
          doi = {10.3847/2041-8213/ad0b79},
archivePrefix = {arXiv},
       eprint = {2309.07963},
 primaryClass = {astro-ph.HE},
       adsurl = {https://ui.adsabs.harvard.edu/abs/2023ApJ...959L...3D}
}

@ARTICLE{2022ApJ...930L..16E,
       author = {{Event Horizon Telescope Collaboration} and {Akiyama}, Kazunori and {Alberdi}, Antxon and {Alef}, Walter and {Algaba}, Juan Carlos and {Anantua}, Richard and {Asada}, Keiichi and {Azulay}, Rebecca and {Bach}, Uwe and {Baczko}, Anne-Kathrin and {Ball}, David and {Balokovi{\'c}}, Mislav and {Barrett}, John and {Baub{\"o}ck}, Michi and {Benson}, Bradford A. and {Bintley}, Dan and {Blackburn}, Lindy and {Blundell}, Raymond and {Bouman}, Katherine L. and {Bower}, Geoffrey C. and {Boyce}, Hope and {Bremer}, Michael and {Brinkerink}, Christiaan D. and {Brissenden}, Roger and {Britzen}, Silke and {Broderick}, Avery E. and {Broguiere}, Dominique and {Bronzwaer}, Thomas and {Bustamante}, Sandra and {Byun}, Do-Young and {Carlstrom}, John E. and {Ceccobello}, Chiara and {Chael}, Andrew and {Chan}, Chi-kwan and {Chatterjee}, Koushik and {Chatterjee}, Shami and {Chen}, Ming-Tang and {Chen}, Yongjun and {Cheng}, Xiaopeng and {Cho}, Ilje and {Christian}, Pierre and {Conroy}, Nicholas S. and {Conway}, John E. and {Cordes}, James M. and {Crawford}, Thomas M. and {Crew}, Geoffrey B. and {Cruz-Osorio}, Alejandro and {Cui}, Yuzhu and {Davelaar}, Jordy and {De Laurentis}, Mariafelicia and {Deane}, Roger and {Dempsey}, Jessica and {Desvignes}, Gregory and {Dexter}, Jason and {Dhruv}, Vedant and {Doeleman}, Sheperd S. and {Dougal}, Sean and {Dzib}, Sergio A. and {Eatough}, Ralph P. and {Emami}, Razieh and {Falcke}, Heino and {Farah}, Joseph and {Fish}, Vincent L. and {Fomalont}, Ed and {Ford}, H. Alyson and {Fraga-Encinas}, Raquel and {Freeman}, William T. and {Friberg}, Per and {Fromm}, Christian M. and {Fuentes}, Antonio and {Galison}, Peter and {Gammie}, Charles F. and {Garc{\'\i}a}, Roberto and {Gentaz}, Olivier and {Georgiev}, Boris and {Goddi}, Ciriaco and {Gold}, Roman and {G{\'o}mez-Ruiz}, Arturo I. and {G{\'o}mez}, Jos{\'e} L. and {Gu}, Minfeng and {Gurwell}, Mark and {Hada}, Kazuhiro and {Haggard}, Daryl and {Haworth}, Kari and {Hecht}, Michael H. and {Hesper}, Ronald and {Heumann}, Dirk and {Ho}, Luis C. and {Ho}, Paul and {Honma}, Mareki and {Huang}, Chih-Wei L. and {Huang}, Lei and {Hughes}, David H. and {Ikeda}, Shiro and {Violette Impellizzeri}, C.~M. and {Inoue}, Makoto and {Issaoun}, Sara and {James}, David J. and {Jannuzi}, Buell T. and {Janssen}, Michael and {Jeter}, Britton and {Jiang}, Wu and {Jim{\'e}nez-Rosales}, Alejandra and {Johnson}, Michael D. and {Jorstad}, Svetlana and {Joshi}, Abhishek V. and {Jung}, Taehyun and {Karami}, Mansour and {Karuppusamy}, Ramesh and {Kawashima}, Tomohisa and {Keating}, Garrett K. and {Kettenis}, Mark and {Kim}, Dong-Jin and {Kim}, Jae-Young and {Kim}, Jongsoo and {Kim}, Junhan and {Kino}, Motoki and {Koay}, Jun Yi and {Kocherlakota}, Prashant and {Kofuji}, Yutaro and {Koch}, Patrick M. and {Koyama}, Shoko and {Kramer}, Carsten and {Kramer}, Michael and {Krichbaum}, Thomas P. and {Kuo}, Cheng-Yu and {La Bella}, Noemi and {Lauer}, Tod R. and {Lee}, Daeyoung and {Lee}, Sang-Sung and {Leung}, Po Kin and {Levis}, Aviad and {Li}, Zhiyuan and {Lico}, Rocco and {Lindahl}, Greg and {Lindqvist}, Michael and {Lisakov}, Mikhail and {Liu}, Jun and {Liu}, Kuo and {Liuzzo}, Elisabetta and {Lo}, Wen-Ping and {Lobanov}, Andrei P. and {Loinard}, Laurent and {Lonsdale}, Colin J. and {Lu}, Ru-Sen and {Mao}, Jirong and {Marchili}, Nicola and {Markoff}, Sera and {Marrone}, Daniel P. and {Marscher}, Alan P. and {Mart{\'\i}-Vidal}, Iv{\'a}n and {Matsushita}, Satoki and {Matthews}, Lynn D. and {Medeiros}, Lia and {Menten}, Karl M. and {Michalik}, Daniel and {Mizuno}, Izumi and {Mizuno}, Yosuke and {Moran}, James M. and {Moriyama}, Kotaro and {Moscibrodzka}, Monika and {M{\"u}ller}, Cornelia and {Mus}, Alejandro and {Musoke}, Gibwa and {Myserlis}, Ioannis and {Nadolski}, Andrew and {Nagai}, Hiroshi and {Nagar}, Neil M. and {Nakamura}, Masanori and {Narayan}, Ramesh and {Narayanan}, Gopal and {Natarajan}, Iniyan and {Nathanail}, Antonios and {Navarro Fuentes}, Santiago and {Neilsen}, Joey and {Neri}, Roberto and {Ni}, Chunchong and {Noutsos}, Aristeidis and {Nowak}, Michael A. and {Oh}, Junghwan and {Okino}, Hiroki and {Olivares}, H{\'e}ctor and {Ortiz-Le{\'o}n}, Gisela N. and {Oyama}, Tomoaki and {{\"O}zel}, Feryal and {Palumbo}, Daniel C.~M. and {Filippos Paraschos}, Georgios and {Park}, Jongho and {Parsons}, Harriet and {Patel}, Nimesh and {Pen}, Ue-Li and {Pesce}, Dominic W. and {Pi{\'e}tu}, Vincent and {Plambeck}, Richard and {PopStefanija}, Aleksandar and {Porth}, Oliver and {P{\"o}tzl}, Felix M. and {Prather}, Ben and {Preciado-L{\'o}pez}, Jorge A. and {Psaltis}, Dimitrios},
        title = "{First Sagittarius A* Event Horizon Telescope Results. V. Testing Astrophysical Models of the Galactic Center Black Hole}",
      journal = {\apjl},
         year = 2022,
        month = may,
       volume = {930},
       number = {2},
          eid = {L16},
        pages = {L16},
          doi = {10.3847/2041-8213/ac6672},
       adsurl = {https://ui.adsabs.harvard.edu/abs/2022ApJ...930L..16E}
}

@ARTICLE{2017ComAC...4....1P,
       author = {{Porth}, Oliver and {Olivares}, Hector and {Mizuno}, Yosuke and {Younsi}, Ziri and {Rezzolla}, Luciano and {Moscibrodzka}, Monika and {Falcke}, Heino and {Kramer}, Michael},
        title = "{The black hole accretion code}",
      journal = {Computational Astrophysics and Cosmology},
         year = 2017,
        month = may,
       volume = {4},
       number = {1},
          eid = {1},
        pages = {1},
          doi = {10.1186/s40668-017-0020-2},
archivePrefix = {arXiv},
       eprint = {1611.09720},
 primaryClass = {gr-qc},
       adsurl = {https://ui.adsabs.harvard.edu/abs/2017ComAC...4....1P}
}

@ARTICLE{2019A&A...629A..61O,
       author = {{Olivares}, Hector and {Porth}, Oliver and {Davelaar}, Jordy and {Most}, Elias R. and {Fromm}, Christian M. and {Mizuno}, Yosuke and {Younsi}, Ziri and {Rezzolla}, Luciano},
        title = "{Constrained transport and adaptive mesh refinement in the Black Hole Accretion Code}",
      journal = {\aap},
         year = 2019,
        month = sep,
       volume = {629},
          eid = {A61},
        pages = {A61},
          doi = {10.1051/0004-6361/201935559},
archivePrefix = {arXiv},
       eprint = {1906.10795},
 primaryClass = {astro-ph.HE},
       adsurl = {https://ui.adsabs.harvard.edu/abs/2019A&A...629A..61O}
}

@ARTICLE{2015MNRAS.454.1848R,
       author = {{Ressler}, S.~M. and {Tchekhovskoy}, A. and {Quataert}, E. and {Chandra}, M. and {Gammie}, C.~F.},
        title = "{Electron thermodynamics in GRMHD simulations of low-luminosity black hole accretion}",
      journal = {\mnras},
         year = 2015,
        month = dec,
       volume = {454},
       number = {2},
        pages = {1848-1870},
          doi = {10.1093/mnras/stv2084},
archivePrefix = {arXiv},
       eprint = {1509.04717},
 primaryClass = {astro-ph.HE},
       adsurl = {https://ui.adsabs.harvard.edu/abs/2015MNRAS.454.1848R}
}

@ARTICLE{2019PNAS..116..771K,
       author = {{Kawazura}, Yohei and {Barnes}, Michael and {Schekochihin}, Alexander A.},
        title = "{Thermal disequilibration of ions and electrons by collisionless plasma turbulence}",
      journal = {Proceedings of the National Academy of Science},
         year = 2019,
        month = jan,
       volume = {116},
       number = {3},
        pages = {771-776},
          doi = {10.1073/pnas.1812491116},
archivePrefix = {arXiv},
       eprint = {1807.07702},
 primaryClass = {physics.plasm-ph},
       adsurl = {https://ui.adsabs.harvard.edu/abs/2019PNAS..116..771K}
}

@ARTICLE{2017ApJ...850...29R,
       author = {{Rowan}, Michael E. and {Sironi}, Lorenzo and {Narayan}, Ramesh},
        title = "{Electron and Proton Heating in Transrelativistic Magnetic Reconnection}",
      journal = {\apj},
         year = 2017,
        month = nov,
       volume = {850},
       number = {1},
          eid = {29},
        pages = {29},
          doi = {10.3847/1538-4357/aa9380},
archivePrefix = {arXiv},
       eprint = {1708.04627},
 primaryClass = {astro-ph.HE},
       adsurl = {https://ui.adsabs.harvard.edu/abs/2017ApJ...850...29R}
}

@ARTICLE{2012A&A...545A..13Y,
       author = {{Younsi}, Z. and {Wu}, K. and {Fuerst}, S.~V.},
        title = "{General relativistic radiative transfer: formulation and emission from structured tori around black holes}",
      journal = {\aap},
         year = 2012,
        month = sep,
       volume = {545},
          eid = {A13},
        pages = {A13},
          doi = {10.1051/0004-6361/201219599},
archivePrefix = {arXiv},
       eprint = {1207.4234},
 primaryClass = {astro-ph.HE},
       adsurl = {https://ui.adsabs.harvard.edu/abs/2012A&A...545A..13Y}
}

@INPROCEEDINGS{2020IAUS..342....9Y,
       author = {{Younsi}, Ziri and {Porth}, Oliver and {Mizuno}, Yosuke and {Fromm}, Christian M. and {Olivares}, Hector},
        title = "{Modelling the polarised emission from black holes on event horizon-scales}",
    booktitle = {Perseus in Sicily: From Black Hole to Cluster Outskirts},
         year = 2020,
       editor = {{Asada}, Keiichi and {de Gouveia Dal Pino}, Elisabete and {Giroletti}, Marcello and {Nagai}, Hiroshi and {Nemmen}, Rodrigo},
       volume = {342},
        month = jan,
        pages = {9-12},
          doi = {10.1017/S1743921318007263},
archivePrefix = {arXiv},
       eprint = {1907.09196},
 primaryClass = {astro-ph.HE},
       adsurl = {https://ui.adsabs.harvard.edu/abs/2020IAUS..342....9Y}
}

@ARTICLE{Younsi2023,
       author = {{Younsi}, Ziri and {Psaltis}, Dimitrios and {{\"O}zel}, Feryal},
        title = "{Black Hole Images as Tests of General Relativity: Effects of Spacetime Geometry}",
      journal = {\apj},
         year = 2023,
        month = jan,
       volume = {942},
       number = {1},
          eid = {47},
        pages = {47},
          doi = {10.3847/1538-4357/aca58a},
archivePrefix = {arXiv},
       eprint = {2111.01752},
 primaryClass = {astro-ph.HE},
       adsurl = {https://ui.adsabs.harvard.edu/abs/2023ApJ...942...47Y}
}

@ARTICLE{2006PPCF...48..203X,
       author = {{Xiao}, Fuliang},
        title = "{Modelling energetic particles by a relativistic kappa-loss-cone distribution function in plasmas}",
      journal = {Plasma Physics and Controlled Fusion},
         year = 2006,
        month = feb,
       volume = {48},
       number = {2},
        pages = {203-213},
          doi = {10.1088/0741-3335/48/2/003},
       adsurl = {https://ui.adsabs.harvard.edu/abs/2006PPCF...48..203X}
}

@ARTICLE{2016ApJ...822...34P,
       author = {{Pandya}, Alex and {Zhang}, Zhaowei and {Chandra}, Mani and {Gammie}, Charles F.},
        title = "{Polarized Synchrotron Emissivities and Absorptivities for Relativistic Thermal, Power-law, and Kappa Distribution Functions}",
      journal = {\apj},
         year = 2016,
        month = may,
       volume = {822},
       number = {1},
          eid = {34},
        pages = {34},
          doi = {10.3847/0004-637X/822/1/34},
archivePrefix = {arXiv},
       eprint = {1602.08749},
 primaryClass = {astro-ph.HE},
       adsurl = {https://ui.adsabs.harvard.edu/abs/2016ApJ...822...34P}
}

@ARTICLE{2019A&A...632A...2D,
       author = {{Davelaar}, Jordy and {Olivares}, Hector and {Porth}, Oliver and {Bronzwaer}, Thomas and {Janssen}, Michael and {Roelofs}, Freek and {Mizuno}, Yosuke and {Fromm}, Christian M. and {Falcke}, Heino and {Rezzolla}, Luciano},
        title = "{Modeling non-thermal emission from the jet-launching region of M 87 with adaptive mesh refinement}",
      journal = {\aap},
         year = 2019,
        month = dec,
       volume = {632},
          eid = {A2},
        pages = {A2},
          doi = {10.1051/0004-6361/201936150},
archivePrefix = {arXiv},
       eprint = {1906.10065},
 primaryClass = {astro-ph.HE},
       adsurl = {https://ui.adsabs.harvard.edu/abs/2019A&A...632A...2D}
}

@ARTICLE{2022NatAs...6..103C,
       author = {{Cruz-Osorio}, Alejandro and {Fromm}, Christian M. and {Mizuno}, Yosuke and {Nathanail}, Antonios and {Younsi}, Ziri and {Porth}, Oliver and {Davelaar}, Jordy and {Falcke}, Heino and {Kramer}, Michael and {Rezzolla}, Luciano},
        title = "{State-of-the-art energetic and morphological modelling of the launching site of the M87 jet}",
      journal = {Nature Astronomy},
         year = 2022,
        month = jan,
       volume = {6},
        pages = {103-108},
          doi = {10.1038/s41550-021-01506-w},
archivePrefix = {arXiv},
       eprint = {2111.02517},
 primaryClass = {astro-ph.HE},
       adsurl = {https://ui.adsabs.harvard.edu/abs/2022NatAs...6..103C}
}

@ARTICLE{2018ApJ...868..146N,
       author = {{Nakamura}, Masanori and {Asada}, Keiichi and {Hada}, Kazuhiro and {Pu}, Hung-Yi and {Noble}, Scott and {Tseng}, Chihyin and {Toma}, Kenji and {Kino}, Motoki and {Nagai}, Hiroshi and {Takahashi}, Kazuya and {Algaba}, Juan-Carlos and {Orienti}, Monica and {Akiyama}, Kazunori and {Doi}, Akihiro and {Giovannini}, Gabriele and {Giroletti}, Marcello and {Honma}, Mareki and {Koyama}, Shoko and {Lico}, Rocco and {Niinuma}, Kotaro and {Tazaki}, Fumie},
        title = "{Parabolic Jets from the Spinning Black Hole in M87}",
      journal = {\apj},
         year = 2018,
        month = dec,
       volume = {868},
       number = {2},
          eid = {146},
        pages = {146},
          doi = {10.3847/1538-4357/aaeb2d},
archivePrefix = {arXiv},
       eprint = {1810.09963},
 primaryClass = {astro-ph.HE},
       adsurl = {https://ui.adsabs.harvard.edu/abs/2018ApJ...868..146N}
}

@ARTICLE{2018ApJ...862...80B,
       author = {{Ball}, David and {Sironi}, Lorenzo and {{\"O}zel}, Feryal},
        title = "{Electron and Proton Acceleration in Trans-relativistic Magnetic Reconnection: Dependence on Plasma Beta and Magnetization}",
      journal = {\apj},
         year = 2018,
        month = jul,
       volume = {862},
       number = {1},
          eid = {80},
        pages = {80},
          doi = {10.3847/1538-4357/aac820},
archivePrefix = {arXiv},
       eprint = {1803.05556},
 primaryClass = {astro-ph.HE},
       adsurl = {https://ui.adsabs.harvard.edu/abs/2018ApJ...862...80B}
}

@ARTICLE{Meringolo2023,
       author = {{Meringolo}, Claudio and {Cruz-Osorio}, Alejandro and {Rezzolla}, Luciano and {Servidio}, Sergio},
        title = "{Microphysical Plasma Relations from Special-relativistic Turbulence}",
      journal = {Astrophys. J.},
         year = 2023,
        month = feb,
       volume = {944},
       number = {2},
          eid = {122},
        pages = {122},
          doi = {10.3847/1538-4357/acaefe},
archivePrefix = {arXiv},
       eprint = {2301.02669},
 primaryClass = {astro-ph.HE},
       adsurl = {https://ui.adsabs.harvard.edu/abs/2023ApJ...944..122M}
}

@ARTICLE{2019ApJ...875L...6E,
       author = {{Event Horizon Telescope Collaboration} and {Akiyama}, Kazunori and {Alberdi}, Antxon and {Alef}, Walter and {Asada}, Keiichi and {Azulay}, Rebecca and {Baczko}, Anne-Kathrin and {Ball}, David and {Balokovi{\'c}}, Mislav and {Barrett}, John and {Bintley}, Dan and {Blackburn}, Lindy and {Boland}, Wilfred and {Bouman}, Katherine L. and {Bower}, Geoffrey C. and {Bremer}, Michael and {Brinkerink}, Christiaan D. and {Brissenden}, Roger and {Britzen}, Silke and {Broderick}, Avery E. and {Broguiere}, Dominique and {Bronzwaer}, Thomas and {Byun}, Do-Young and {Carlstrom}, John E. and {Chael}, Andrew and {Chan}, Chi-kwan and {Chatterjee}, Shami and {Chatterjee}, Koushik and {Chen}, Ming-Tang and {Chen}, Yongjun and {Cho}, Ilje and {Christian}, Pierre and {Conway}, John E. and {Cordes}, James M. and {Crew}, Geoffrey B. and {Cui}, Yuzhu and {Davelaar}, Jordy and {De Laurentis}, Mariafelicia and {Deane}, Roger and {Dempsey}, Jessica and {Desvignes}, Gregory and {Dexter}, Jason and {Doeleman}, Sheperd S. and {Eatough}, Ralph P. and {Falcke}, Heino and {Fish}, Vincent L. and {Fomalont}, Ed and {Fraga-Encinas}, Raquel and {Friberg}, Per and {Fromm}, Christian M. and {G{\'o}mez}, Jos{\'e} L. and {Galison}, Peter and {Gammie}, Charles F. and {Garc{\'\i}a}, Roberto and {Gentaz}, Olivier and {Georgiev}, Boris and {Goddi}, Ciriaco and {Gold}, Roman and {Gu}, Minfeng and {Gurwell}, Mark and {Hada}, Kazuhiro and {Hecht}, Michael H. and {Hesper}, Ronald and {Ho}, Luis C. and {Ho}, Paul and {Honma}, Mareki and {Huang}, Chih-Wei L. and {Huang}, Lei and {Hughes}, David H. and {Ikeda}, Shiro and {Inoue}, Makoto and {Issaoun}, Sara and {James}, David J. and {Jannuzi}, Buell T. and {Janssen}, Michael and {Jeter}, Britton and {Jiang}, Wu and {Johnson}, Michael D. and {Jorstad}, Svetlana and {Jung}, Taehyun and {Karami}, Mansour and {Karuppusamy}, Ramesh and {Kawashima}, Tomohisa and {Keating}, Garrett K. and {Kettenis}, Mark and {Kim}, Jae-Young and {Kim}, Junhan and {Kim}, Jongsoo and {Kino}, Motoki and {Koay}, Jun Yi and {Koch}, Patrick M. and {Koyama}, Shoko and {Kramer}, Michael and {Kramer}, Carsten and {Krichbaum}, Thomas P. and {Kuo}, Cheng-Yu and {Lauer}, Tod R. and {Lee}, Sang-Sung and {Li}, Yan-Rong and {Li}, Zhiyuan and {Lindqvist}, Michael and {Liu}, Kuo and {Liuzzo}, Elisabetta and {Lo}, Wen-Ping and {Lobanov}, Andrei P. and {Loinard}, Laurent and {Lonsdale}, Colin and {Lu}, Ru-Sen and {MacDonald}, Nicholas R. and {Mao}, Jirong and {Markoff}, Sera and {Marrone}, Daniel P. and {Marscher}, Alan P. and {Mart{\'\i}-Vidal}, Iv{\'a}n and {Matsushita}, Satoki and {Matthews}, Lynn D. and {Medeiros}, Lia and {Menten}, Karl M. and {Mizuno}, Yosuke and {Mizuno}, Izumi and {Moran}, James M. and {Moriyama}, Kotaro and {Moscibrodzka}, Monika and {M{\"u}ller}, Cornelia and {Nagai}, Hiroshi and {Nagar}, Neil M. and {Nakamura}, Masanori and {Narayan}, Ramesh and {Narayanan}, Gopal and {Natarajan}, Iniyan and {Neri}, Roberto and {Ni}, Chunchong and {Noutsos}, Aristeidis and {Okino}, Hiroki and {Olivares}, H{\'e}ctor and {Oyama}, Tomoaki and {{\"O}zel}, Feryal and {Palumbo}, Daniel C.~M. and {Patel}, Nimesh and {Pen}, Ue-Li and {Pesce}, Dominic W. and {Pi{\'e}tu}, Vincent and {Plambeck}, Richard and {PopStefanija}, Aleksandar and {Porth}, Oliver and {Prather}, Ben and {Preciado-L{\'o}pez}, Jorge A. and {Psaltis}, Dimitrios and {Pu}, Hung-Yi and {Ramakrishnan}, Venkatessh and {Rao}, Ramprasad and {Rawlings}, Mark G. and {Raymond}, Alexander W. and {Rezzolla}, Luciano and {Ripperda}, Bart and {Roelofs}, Freek and {Rogers}, Alan and {Ros}, Eduardo and {Rose}, Mel and {Roshanineshat}, Arash and {Rottmann}, Helge and {Roy}, Alan L. and {Ruszczyk}, Chet and {Ryan}, Benjamin R. and {Rygl}, Kazi L.~J. and {S{\'a}nchez}, Salvador and {S{\'a}nchez-Arguelles}, David and {Sasada}, Mahito and {Savolainen}, Tuomas and {Schloerb}, F. Peter and {Schuster}, Karl-Friedrich and {Shao}, Lijing and {Shen}, Zhiqiang and {Small}, Des and {Sohn}, Bong Won and {SooHoo}, Jason and {Tazaki}, Fumie and {Tiede}, Paul and {Tilanus}, Remo P.~J. and {Titus}, Michael and {Toma}, Kenji and {Torne}, Pablo and {Trent}, Tyler and {Trippe}, Sascha and {Tsuda}, Shuichiro and {van Bemmel}, Ilse and {van Langevelde}, Huib Jan and {van Rossum}, Daniel R. and {Wagner}, Jan and {Wardle}, John and {Weintroub}, Jonathan and {Wex}, Norbert and {Wharton}, Robert and {Wielgus}, Maciek and {Wong}, George N. and {Wu}, Qingwen and {Young}, Andr{\'e} and {Young}, Ken and {Younsi}, Ziri and {Yuan}, Feng and {Yuan}, Ye-Fei and {Zensus}, J. Anton and {Zhao}, Guangyao and {Zhao}, Shan-Shan and {Zhu}, Ziyan and {Farah}, Joseph R. and {Meyer-Zhao}, Zheng and {Michalik}, Daniel and {Nadolski}, Andrew and {Nishioka}, Hiroaki and {Pradel}, Nicolas and {Primiani}, Rurik A. and {Souccar}, Kamal and {Vertatschitsch}, Laura and {Yamaguchi}, Paul},
        title = "{First M87 Event Horizon Telescope Results. VI. The Shadow and Mass of the Central Black Hole}",
      journal = {\apjl},
         year = 2019,
        month = apr,
       volume = {875},
       number = {1},
          eid = {L6},
        pages = {L6},
          doi = {10.3847/2041-8213/ab1141},
archivePrefix = {arXiv},
       eprint = {1906.11243},
 primaryClass = {astro-ph.GA},
       adsurl = {https://ui.adsabs.harvard.edu/abs/2019ApJ...875L...6E}
}

@BOOK{1965pfig.book.....S,
       author = {{Spitzer}, Lyman},
        title = "{Physics of fully ionized gases}",
         year = 1965,
       adsurl = {https://ui.adsabs.harvard.edu/abs/1965pfig.book.....S}
}

@ARTICLE{1984ApJ...280..319C,
       author = {{Colpi}, M. and {Maraschi}, L. and {Treves}, A.},
        title = "{Two-temperature model of spherical accretion onto a black hole.}",
      journal = {\apj},
         year = 1984,
        month = may,
       volume = {280},
        pages = {319-327},
          doi = {10.1086/161998},
       adsurl = {https://ui.adsabs.harvard.edu/abs/1984ApJ...280..319C}
}

@INCOLLECTION{2025nfgs.book..327D,
       author = {{Dihingia}, Indu K. and {Fendt}, Christian},
        title = "{Thin Accretion Disks in GR-MHD Simulations}",
    booktitle = {New Frontiers in GRMHD Simulations},
         year = 2025,
       editor = {{Bambi}, Cosimo and {Mizuno}, Yosuke and {Shashank}, Swarnim and {Yuan}, Feng},
        pages = {327-360},
          doi = {10.1007/978-981-97-8522-3_10},
       adsurl = {https://ui.adsabs.harvard.edu/abs/2025nfgs.book..327D}
}

@ARTICLE{1996ApJ...465..312E,
       author = {{Esin}, Ann A. and {Narayan}, Ramesh and {Ostriker}, Eve and {Yi}, Insu},
        title = "{Hot One-Temperature Accretion Flows around Black Holes}",
      journal = {\apj},
         year = 1996,
        month = jul,
       volume = {465},
        pages = {312},
          doi = {10.1086/177421},
archivePrefix = {arXiv},
       eprint = {astro-ph/9601074},
 primaryClass = {astro-ph},
       adsurl = {https://ui.adsabs.harvard.edu/abs/1996ApJ...465..312E}
}

@ARTICLE{1995ApJ...452..710N,
       author = {{Narayan}, Ramesh and {Yi}, Insu},
        title = "{Advection-dominated Accretion: Underfed Black Holes and Neutron Stars}",
      journal = {\apj},
         year = 1995,
        month = oct,
       volume = {452},
        pages = {710},
          doi = {10.1086/176343},
archivePrefix = {arXiv},
       eprint = {astro-ph/9411059},
 primaryClass = {astro-ph},
       adsurl = {https://ui.adsabs.harvard.edu/abs/1995ApJ...452..710N}
}

@ARTICLE{2014ARA&A..52..529Y,
       author = {{Yuan}, Feng and {Narayan}, Ramesh},
        title = "{Hot Accretion Flows Around Black Holes}",
      journal = {\araa},
         year = 2014,
        month = aug,
       volume = {52},
        pages = {529-588},
          doi = {10.1146/annurev-astro-082812-141003},
archivePrefix = {arXiv},
       eprint = {1401.0586},
 primaryClass = {astro-ph.HE},
       adsurl = {https://ui.adsabs.harvard.edu/abs/2014ARA&A..52..529Y}
}

@ARTICLE{2019ARA&A..57..467B,
       author = {{Blandford}, Roger and {Meier}, David and {Readhead}, Anthony},
        title = "{Relativistic Jets from Active Galactic Nuclei}",
      journal = {\araa},
         year = 2019,
        month = aug,
       volume = {57},
        pages = {467-509},
          doi = {10.1146/annurev-astro-081817-051948},
archivePrefix = {arXiv},
       eprint = {1812.06025},
 primaryClass = {astro-ph.HE},
       adsurl = {https://ui.adsabs.harvard.edu/abs/2019ARA&A..57..467B}
}

@ARTICLE{1977ApJ...214..840I,
       author = {{Ichimaru}, S.},
        title = "{Bimodal behavior of accretion disks: theory and application to Cygnus X-1 transitions.}",
      journal = {\apj},
         year = 1977,
        month = jun,
       volume = {214},
        pages = {840-855},
          doi = {10.1086/155314},
       adsurl = {https://ui.adsabs.harvard.edu/abs/1977ApJ...214..840I}
}

@ARTICLE{1994ApJ...428L..13N,
       author = {{Narayan}, Ramesh and {Yi}, Insu},
        title = "{Advection-dominated Accretion: A Self-similar Solution}",
      journal = {\apjl},
         year = 1994,
        month = jun,
       volume = {428},
        pages = {L13},
          doi = {10.1086/187381},
archivePrefix = {arXiv},
       eprint = {astro-ph/9403052},
 primaryClass = {astro-ph},
       adsurl = {https://ui.adsabs.harvard.edu/abs/1994ApJ...428L..13N}
}

@ARTICLE{2019ApJ...875L...4E,
       author = {{Event Horizon Telescope Collaboration} and {Akiyama}, Kazunori and {Alberdi}, Antxon and {Alef}, Walter and {Asada}, Keiichi and {Azulay}, Rebecca and {Baczko}, Anne-Kathrin and {Ball}, David and {Balokovi{\'c}}, Mislav and {Barrett}, John and {Bintley}, Dan and {Blackburn}, Lindy and {Boland}, Wilfred and {Bouman}, Katherine L. and {Bower}, Geoffrey C. and {Bremer}, Michael and {Brinkerink}, Christiaan D. and {Brissenden}, Roger and {Britzen}, Silke and {Broderick}, Avery E. and {Broguiere}, Dominique and {Bronzwaer}, Thomas and {Byun}, Do-Young and {Carlstrom}, John E. and {Chael}, Andrew and {Chan}, Chi-kwan and {Chatterjee}, Shami and {Chatterjee}, Koushik and {Chen}, Ming-Tang and {Chen}, Yongjun and {Cho}, Ilje and {Christian}, Pierre and {Conway}, John E. and {Cordes}, James M. and {Crew}, Geoffrey B. and {Cui}, Yuzhu and {Davelaar}, Jordy and {De Laurentis}, Mariafelicia and {Deane}, Roger and {Dempsey}, Jessica and {Desvignes}, Gregory and {Dexter}, Jason and {Doeleman}, Sheperd S. and {Eatough}, Ralph P. and {Falcke}, Heino and {Fish}, Vincent L. and {Fomalont}, Ed and {Fraga-Encinas}, Raquel and {Freeman}, William T. and {Friberg}, Per and {Fromm}, Christian M. and {G{\'o}mez}, Jos{\'e} L. and {Galison}, Peter and {Gammie}, Charles F. and {Garc{\'\i}a}, Roberto and {Gentaz}, Olivier and {Georgiev}, Boris and {Goddi}, Ciriaco and {Gold}, Roman and {Gu}, Minfeng and {Gurwell}, Mark and {Hada}, Kazuhiro and {Hecht}, Michael H. and {Hesper}, Ronald and {Ho}, Luis C. and {Ho}, Paul and {Honma}, Mareki and {Huang}, Chih-Wei L. and {Huang}, Lei and {Hughes}, David H. and {Ikeda}, Shiro and {Inoue}, Makoto and {Issaoun}, Sara and {James}, David J. and {Jannuzi}, Buell T. and {Janssen}, Michael and {Jeter}, Britton and {Jiang}, Wu and {Johnson}, Michael D. and {Jorstad}, Svetlana and {Jung}, Taehyun and {Karami}, Mansour and {Karuppusamy}, Ramesh and {Kawashima}, Tomohisa and {Keating}, Garrett K. and {Kettenis}, Mark and {Kim}, Jae-Young and {Kim}, Junhan and {Kim}, Jongsoo and {Kino}, Motoki and {Koay}, Jun Yi and {Koch}, Patrick M. and {Koyama}, Shoko and {Kramer}, Michael and {Kramer}, Carsten and {Krichbaum}, Thomas P. and {Kuo}, Cheng-Yu and {Lauer}, Tod R. and {Lee}, Sang-Sung and {Li}, Yan-Rong and {Li}, Zhiyuan and {Lindqvist}, Michael and {Liu}, Kuo and {Liuzzo}, Elisabetta and {Lo}, Wen-Ping and {Lobanov}, Andrei P. and {Loinard}, Laurent and {Lonsdale}, Colin and {Lu}, Ru-Sen and {MacDonald}, Nicholas R. and {Mao}, Jirong and {Markoff}, Sera and {Marrone}, Daniel P. and {Marscher}, Alan P. and {Mart{\'\i}-Vidal}, Iv{\'a}n and {Matsushita}, Satoki and {Matthews}, Lynn D. and {Medeiros}, Lia and {Menten}, Karl M. and {Mizuno}, Yosuke and {Mizuno}, Izumi and {Moran}, James M. and {Moriyama}, Kotaro and {Moscibrodzka}, Monika and {M{\"u}ller}, Cornelia and {Nagai}, Hiroshi and {Nagar}, Neil M. and {Nakamura}, Masanori and {Narayan}, Ramesh and {Narayanan}, Gopal and {Natarajan}, Iniyan and {Neri}, Roberto and {Ni}, Chunchong and {Noutsos}, Aristeidis and {Okino}, Hiroki and {Olivares}, H{\'e}ctor and {Oyama}, Tomoaki and {{\"O}zel}, Feryal and {Palumbo}, Daniel C.~M. and {Patel}, Nimesh and {Pen}, Ue-Li and {Pesce}, Dominic W. and {Pi{\'e}tu}, Vincent and {Plambeck}, Richard and {PopStefanija}, Aleksandar and {Porth}, Oliver and {Prather}, Ben and {Preciado-L{\'o}pez}, Jorge A. and {Psaltis}, Dimitrios and {Pu}, Hung-Yi and {Ramakrishnan}, Venkatessh and {Rao}, Ramprasad and {Rawlings}, Mark G. and {Raymond}, Alexander W. and {Rezzolla}, Luciano and {Ripperda}, Bart and {Roelofs}, Freek and {Rogers}, Alan and {Ros}, Eduardo and {Rose}, Mel and {Roshanineshat}, Arash and {Rottmann}, Helge and {Roy}, Alan L. and {Ruszczyk}, Chet and {Ryan}, Benjamin R. and {Rygl}, Kazi L.~J. and {S{\'a}nchez}, Salvador and {S{\'a}nchez-Arguelles}, David and {Sasada}, Mahito and {Savolainen}, Tuomas and {Schloerb}, F. Peter and {Schuster}, Karl-Friedrich and {Shao}, Lijing and {Shen}, Zhiqiang and {Small}, Des and {Sohn}, Bong Won and {SooHoo}, Jason and {Tazaki}, Fumie and {Tiede}, Paul and {Tilanus}, Remo P.~J. and {Titus}, Michael and {Toma}, Kenji and {Torne}, Pablo and {Trent}, Tyler and {Trippe}, Sascha and {Tsuda}, Shuichiro and {van Bemmel}, Ilse and {van Langevelde}, Huib Jan and {van Rossum}, Daniel R. and {Wagner}, Jan and {Wardle}, John and {Weintroub}, Jonathan and {Wex}, Norbert and {Wharton}, Robert and {Wielgus}, Maciek and {Wong}, George N. and {Wu}, Qingwen and {Young}, Andr{\'e} and {Young}, Ken and {Younsi}, Ziri},
        title = "{First M87 Event Horizon Telescope Results. IV. Imaging the Central Supermassive Black Hole}",
      journal = {\apjl},
         year = 2019,
        month = apr,
       volume = {875},
       number = {1},
          eid = {L4},
        pages = {L4},
          doi = {10.3847/2041-8213/ab0e85},
archivePrefix = {arXiv},
       eprint = {1906.11241},
 primaryClass = {astro-ph.GA},
       adsurl = {https://ui.adsabs.harvard.edu/abs/2019ApJ...875L...4E}
}

@ARTICLE{1995ApJ...444..231N,
       author = {{Narayan}, Ramesh and {Yi}, Insu},
        title = "{Advection-dominated Accretion: Self-Similarity and Bipolar Outflows}",
      journal = {\apj},
         year = 1995,
        month = may,
       volume = {444},
        pages = {231},
          doi = {10.1086/175599},
archivePrefix = {arXiv},
       eprint = {astro-ph/9411058},
 primaryClass = {astro-ph},
       adsurl = {https://ui.adsabs.harvard.edu/abs/1995ApJ...444..231N}
}

@ARTICLE{2003ApJ...589..458D,
       author = {{De Villiers}, Jean-Pierre and {Hawley}, John F.},
        title = "{A Numerical Method for General Relativistic Magnetohydrodynamics}",
      journal = {\apj},
         year = 2003,
        month = may,
       volume = {589},
       number = {1},
        pages = {458-480},
          doi = {10.1086/373949},
archivePrefix = {arXiv},
       eprint = {astro-ph/0210518},
 primaryClass = {astro-ph},
       adsurl = {https://ui.adsabs.harvard.edu/abs/2003ApJ...589..458D}
}

@ARTICLE{2007CQGra..24S.259N,
       author = {{Noble}, Scott C. and {Leung}, Po Kin and {Gammie}, Charles F. and {Book}, Laura G.},
        title = "{Simulating the emission and outflows from accretion discs}",
      journal = {Classical and Quantum Gravity},
         year = 2007,
        month = jun,
       volume = {24},
       number = {12},
        pages = {S259-S274},
          doi = {10.1088/0264-9381/24/12/S17},
archivePrefix = {arXiv},
       eprint = {astro-ph/0701778},
 primaryClass = {astro-ph},
       adsurl = {https://ui.adsabs.harvard.edu/abs/2007CQGra..24S.259N}
}

@ARTICLE{2009ApJ...706..497M,
       author = {{Mo{\'s}cibrodzka}, Monika and {Gammie}, Charles F. and {Dolence}, Joshua C. and {Shiokawa}, Hotaka and {Leung}, Po Kin},
        title = "{Radiative Models of SGR A* from GRMHD Simulations}",
      journal = {\apj},
         year = 2009,
        month = nov,
       volume = {706},
       number = {1},
        pages = {497-507},
          doi = {10.1088/0004-637X/706/1/497},
archivePrefix = {arXiv},
       eprint = {0909.5431},
 primaryClass = {astro-ph.HE},
       adsurl = {https://ui.adsabs.harvard.edu/abs/2009ApJ...706..497M}
}

@ARTICLE{2014A&A...570A...7M,
       author = {{Mo{\'s}cibrodzka}, Monika and {Falcke}, Heino and {Shiokawa}, Hotaka and {Gammie}, Charles F.},
        title = "{Observational appearance of inefficient accretion flows and jets in 3D GRMHD simulations: Application to Sagittarius A*}",
      journal = {\aap},
         year = 2014,
        month = oct,
       volume = {570},
          eid = {A7},
        pages = {A7},
          doi = {10.1051/0004-6361/201424358},
archivePrefix = {arXiv},
       eprint = {1408.4743},
 primaryClass = {astro-ph.HE},
       adsurl = {https://ui.adsabs.harvard.edu/abs/2014A&A...570A...7M}
}

@ARTICLE{2016A&A...586A..38M,
       author = {{Mo{\'s}cibrodzka}, Monika and {Falcke}, Heino and {Shiokawa}, Hotaka},
        title = "{General relativistic magnetohydrodynamical simulations of the jet in M 87}",
      journal = {\aap},
         year = 2016,
        month = feb,
       volume = {586},
          eid = {A38},
        pages = {A38},
          doi = {10.1051/0004-6361/201526630},
archivePrefix = {arXiv},
       eprint = {1510.07243},
 primaryClass = {astro-ph.HE},
       adsurl = {https://ui.adsabs.harvard.edu/abs/2016A&A...586A..38M}
}

@ARTICLE{2010ApJ...717.1092D,
       author = {{Dexter}, Jason and {Agol}, Eric and {Fragile}, P. Chris and {McKinney}, Jonathan C.},
        title = "{The Submillimeter Bump in Sgr A* from Relativistic MHD Simulations}",
      journal = {\apj},
         year = 2010,
        month = jul,
       volume = {717},
       number = {2},
        pages = {1092-1104},
          doi = {10.1088/0004-637X/717/2/1092},
archivePrefix = {arXiv},
       eprint = {1005.4062},
 primaryClass = {astro-ph.HE},
       adsurl = {https://ui.adsabs.harvard.edu/abs/2010ApJ...717.1092D}
}

@ARTICLE{2012ApJ...755..133S,
       author = {{Shcherbakov}, Roman V. and {Penna}, Robert F. and {McKinney}, Jonathan C.},
        title = "{Sagittarius A* Accretion Flow and Black Hole Parameters from General Relativistic Dynamical and Polarized Radiative Modeling}",
      journal = {\apj},
         year = 2012,
        month = aug,
       volume = {755},
       number = {2},
          eid = {133},
        pages = {133},
          doi = {10.1088/0004-637X/755/2/133},
archivePrefix = {arXiv},
       eprint = {1007.4832},
 primaryClass = {astro-ph.HE},
       adsurl = {https://ui.adsabs.harvard.edu/abs/2012ApJ...755..133S}
}

@ARTICLE{2018A&A...612A..34D,
       author = {{Davelaar}, J. and {Mo{\'s}cibrodzka}, M. and {Bronzwaer}, T. and {Falcke}, H.},
        title = "{General relativistic magnetohydrodynamical {\ensuremath{\kappa}}-jet models for Sagittarius A*}",
      journal = {\aap},
         year = 2018,
        month = apr,
       volume = {612},
          eid = {A34},
        pages = {A34},
          doi = {10.1051/0004-6361/201732025},
archivePrefix = {arXiv},
       eprint = {1712.02266},
 primaryClass = {astro-ph.HE},
       adsurl = {https://ui.adsabs.harvard.edu/abs/2018A&A...612A..34D}
}

@ARTICLE{2018NatAs...2..585M,
       author = {{Mizuno}, Yosuke and {Younsi}, Ziri and {Fromm}, Christian M. and {Porth}, Oliver and {De Laurentis}, Mariafelicia and {Olivares}, Hector and {Falcke}, Heino and {Kramer}, Michael and {Rezzolla}, Luciano},
        title = "{The current ability to test theories of gravity with black hole shadows}",
      journal = {Nature Astronomy},
         year = 2018,
        month = apr,
       volume = {2},
        pages = {585-590},
          doi = {10.1038/s41550-018-0449-5},
archivePrefix = {arXiv},
       eprint = {1804.05812},
 primaryClass = {astro-ph.GA},
       adsurl = {https://ui.adsabs.harvard.edu/abs/2018NatAs...2..585M}
}

@ARTICLE{2003ApJ...589..444G,
       author = {{Gammie}, Charles F. and {McKinney}, Jonathan C. and {T{\'o}th}, G{\'a}bor},
        title = "{HARM: A Numerical Scheme for General Relativistic Magnetohydrodynamics}",
      journal = {\apj},
         year = 2003,
        month = may,
       volume = {589},
       number = {1},
        pages = {444-457},
          doi = {10.1086/374594},
archivePrefix = {arXiv},
       eprint = {astro-ph/0301509},
 primaryClass = {astro-ph},
       adsurl = {https://ui.adsabs.harvard.edu/abs/2003ApJ...589..444G}
}

@ARTICLE{2012MNRAS.426.3241N,
       author = {{Narayan}, Ramesh and {S{\"A} dowski}, Aleksander and {Penna}, Robert F. and {Kulkarni}, Akshay K.},
        title = "{GRMHD simulations of magnetized advection-dominated accretion on a non-spinning black hole: role of outflows}",
      journal = {\mnras},
         year = 2012,
        month = nov,
       volume = {426},
       number = {4},
        pages = {3241-3259},
          doi = {10.1111/j.1365-2966.2012.22002.x},
archivePrefix = {arXiv},
       eprint = {1206.1213},
 primaryClass = {astro-ph.HE},
       adsurl = {https://ui.adsabs.harvard.edu/abs/2012MNRAS.426.3241N}
}

@ARTICLE{2013MNRAS.436.3856S,
       author = {{S{\k{a}}dowski}, Aleksander and {Narayan}, Ramesh and {Penna}, Robert and {Zhu}, Yucong},
        title = "{Energy, momentum and mass outflows and feedback from thick accretion discs around rotating black holes}",
      journal = {\mnras},
         year = 2013,
        month = dec,
       volume = {436},
       number = {4},
        pages = {3856-3874},
          doi = {10.1093/mnras/stt1881},
archivePrefix = {arXiv},
       eprint = {1307.1143},
 primaryClass = {astro-ph.HE},
       adsurl = {https://ui.adsabs.harvard.edu/abs/2013MNRAS.436.3856S}
}

@ARTICLE{2003PASJ...55L..69N,
       author = {{Narayan}, Ramesh and {Igumenshchev}, Igor V. and {Abramowicz}, Marek A.},
        title = "{Magnetically Arrested Disk: an Energetically Efficient Accretion Flow}",
      journal = {\pasj},
         year = 2003,
        month = dec,
       volume = {55},
        pages = {L69-L72},
          doi = {10.1093/pasj/55.6.L69},
archivePrefix = {arXiv},
       eprint = {astro-ph/0305029},
 primaryClass = {astro-ph},
       adsurl = {https://ui.adsabs.harvard.edu/abs/2003PASJ...55L..69N}
}

@ARTICLE{2011MNRAS.418L..79T,
       author = {{Tchekhovskoy}, Alexander and {Narayan}, Ramesh and {McKinney}, Jonathan C.},
        title = "{Efficient generation of jets from magnetically arrested accretion on a rapidly spinning black hole}",
      journal = {\mnras},
         year = 2011,
        month = nov,
       volume = {418},
       number = {1},
        pages = {L79-L83},
          doi = {10.1111/j.1745-3933.2011.01147.x},
archivePrefix = {arXiv},
       eprint = {1108.0412},
 primaryClass = {astro-ph.HE},
       adsurl = {https://ui.adsabs.harvard.edu/abs/2011MNRAS.418L..79T}
}

@ARTICLE{2012MNRAS.423.3083M,
       author = {{McKinney}, Jonathan C. and {Tchekhovskoy}, Alexander and {Blandford}, Roger D.},
        title = "{General relativistic magnetohydrodynamic simulations of magnetically choked accretion flows around black holes}",
      journal = {\mnras},
         year = 2012,
        month = jul,
       volume = {423},
       number = {4},
        pages = {3083-3117},
          doi = {10.1111/j.1365-2966.2012.21074.x},
archivePrefix = {arXiv},
       eprint = {1201.4163},
 primaryClass = {astro-ph.HE},
       adsurl = {https://ui.adsabs.harvard.edu/abs/2012MNRAS.423.3083M}
}

@ARTICLE{2014MNRAS.439..503S,
       author = {{S{\k{a}}dowski}, Aleksander and {Narayan}, Ramesh and {McKinney}, Jonathan C. and {Tchekhovskoy}, Alexander},
        title = "{Numerical simulations of super-critical black hole accretion flows in general relativity}",
      journal = {\mnras},
         year = 2014,
        month = mar,
       volume = {439},
       number = {1},
        pages = {503-520},
          doi = {10.1093/mnras/stt2479},
archivePrefix = {arXiv},
       eprint = {1311.5900},
 primaryClass = {astro-ph.HE},
       adsurl = {https://ui.adsabs.harvard.edu/abs/2014MNRAS.439..503S}
}

@ARTICLE{2018ApJ...864..126R,
       author = {{Ryan}, Benjamin R. and {Ressler}, Sean M. and {Dolence}, Joshua C. and {Gammie}, Charles and {Quataert}, Eliot},
        title = "{Two-temperature GRRMHD Simulations of M87}",
      journal = {\apj},
         year = 2018,
        month = sep,
       volume = {864},
       number = {2},
          eid = {126},
        pages = {126},
          doi = {10.3847/1538-4357/aad73a},
archivePrefix = {arXiv},
       eprint = {1808.01958},
 primaryClass = {astro-ph.HE},
       adsurl = {https://ui.adsabs.harvard.edu/abs/2018ApJ...864..126R}
}

@ARTICLE{2019MNRAS.486.2873C,
       author = {{Chael}, Andrew and {Narayan}, Ramesh and {Johnson}, Michael D.},
        title = "{Two-temperature, Magnetically Arrested Disc simulations of the jet from the supermassive black hole in M87}",
      journal = {\mnras},
         year = 2019,
        month = jun,
       volume = {486},
       number = {2},
        pages = {2873-2895},
          doi = {10.1093/mnras/stz988},
archivePrefix = {arXiv},
       eprint = {1810.01983},
 primaryClass = {astro-ph.HE},
       adsurl = {https://ui.adsabs.harvard.edu/abs/2019MNRAS.486.2873C}
}

@ARTICLE{2018MNRAS.478.5209C,
       author = {{Chael}, Andrew and {Rowan}, Michael and {Narayan}, Ramesh and {Johnson}, Michael and {Sironi}, Lorenzo},
        title = "{The role of electron heating physics in images and variability of the Galactic Centre black hole Sagittarius A*}",
      journal = {\mnras},
         year = 2018,
        month = aug,
       volume = {478},
       number = {4},
        pages = {5209-5229},
          doi = {10.1093/mnras/sty1261},
archivePrefix = {arXiv},
       eprint = {1804.06416},
 primaryClass = {astro-ph.HE},
       adsurl = {https://ui.adsabs.harvard.edu/abs/2018MNRAS.478.5209C}
}

@ARTICLE{2017MNRAS.467.3604R,
       author = {{Ressler}, S.~M. and {Tchekhovskoy}, A. and {Quataert}, E. and {Gammie}, C.~F.},
        title = "{The disc-jet symbiosis emerges: modelling the emission of Sagittarius A* with electron thermodynamics}",
      journal = {\mnras},
         year = 2017,
        month = may,
       volume = {467},
       number = {3},
        pages = {3604-3619},
          doi = {10.1093/mnras/stx364},
archivePrefix = {arXiv},
       eprint = {1611.09365},
 primaryClass = {astro-ph.HE},
       adsurl = {https://ui.adsabs.harvard.edu/abs/2017MNRAS.467.3604R}
}

@ARTICLE{2020MNRAS.494.4168D,
       author = {{Dexter}, J. and {Jim{\'e}nez-Rosales}, A. and {Ressler}, S.~M. and {Tchekhovskoy}, A. and {Baub{\"o}ck}, M. and {de Zeeuw}, P.~T. and {Eisenhauer}, F. and {von Fellenberg}, S. and {Gao}, F. and {Genzel}, R. and {Gillessen}, S. and {Habibi}, M. and {Ott}, T. and {Stadler}, J. and {Straub}, O. and {Widmann}, F.},
        title = "{A parameter survey of Sgr A* radiative models from GRMHD simulations with self-consistent electron heating}",
      journal = {\mnras},
         year = 2020,
        month = may,
       volume = {494},
       number = {3},
        pages = {4168-4186},
          doi = {10.1093/mnras/staa922},
archivePrefix = {arXiv},
       eprint = {2004.00019},
 primaryClass = {astro-ph.HE},
       adsurl = {https://ui.adsabs.harvard.edu/abs/2020MNRAS.494.4168D}
}

@ARTICLE{2020MNRAS.499.3178Y,
       author = {{Yoon}, D. and {Chatterjee}, K. and {Markoff}, S.~B. and {van Eijnatten}, D. and {Younsi}, Z. and {Liska}, M. and {Tchekhovskoy}, A.},
        title = "{Spectral and imaging properties of Sgr A* from high-resolution 3D GRMHD simulations with radiative cooling}",
      journal = {\mnras},
         year = 2020,
        month = dec,
       volume = {499},
       number = {3},
        pages = {3178-3192},
          doi = {10.1093/mnras/staa3031},
archivePrefix = {arXiv},
       eprint = {2009.14227},
 primaryClass = {astro-ph.HE},
       adsurl = {https://ui.adsabs.harvard.edu/abs/2020MNRAS.499.3178Y}
}

@ARTICLE{2012MNRAS.426.1928D,
       author = {{Dibi}, S. and {Drappeau}, S. and {Fragile}, P.~C. and {Markoff}, S. and {Dexter}, J.},
        title = "{General relativistic magnetohydrodynamic simulations of accretion on to Sgr A*: how important are radiative losses?}",
      journal = {\mnras},
         year = 2012,
        month = nov,
       volume = {426},
       number = {3},
        pages = {1928-1939},
          doi = {10.1111/j.1365-2966.2012.21857.x},
archivePrefix = {arXiv},
       eprint = {1206.3976},
 primaryClass = {astro-ph.HE},
       adsurl = {https://ui.adsabs.harvard.edu/abs/2012MNRAS.426.1928D}
}

@ARTICLE{2021ApJ...910L..13E,
       author = {{Event Horizon Telescope Collaboration} and {Akiyama}, Kazunori and {Algaba}, Juan Carlos and {Alberdi}, Antxon and {Alef}, Walter and {Anantua}, Richard and {Asada}, Keiichi and {Azulay}, Rebecca and {Baczko}, Anne-Kathrin and {Ball}, David and {Balokovi{\'c}}, Mislav and {Barrett}, John and {Benson}, Bradford A. and {Bintley}, Dan and {Blackburn}, Lindy and {Blundell}, Raymond and {Boland}, Wilfred and {Bouman}, Katherine L. and {Bower}, Geoffrey C. and {Boyce}, Hope and {Bremer}, Michael and {Brinkerink}, Christiaan D. and {Brissenden}, Roger and {Britzen}, Silke and {Broderick}, Avery E. and {Broguiere}, Dominique and {Bronzwaer}, Thomas and {Byun}, Do-Young and {Carlstrom}, John E. and {Chael}, Andrew and {Chan}, Chi-kwan and {Chatterjee}, Shami and {Chatterjee}, Koushik and {Chen}, Ming-Tang and {Chen}, Yongjun and {Chesler}, Paul M. and {Cho}, Ilje and {Christian}, Pierre and {Conway}, John E. and {Cordes}, James M. and {Crawford}, Thomas M. and {Crew}, Geoffrey B. and {Cruz-Osorio}, Alejandro and {Cui}, Yuzhu and {Davelaar}, Jordy and {De Laurentis}, Mariafelicia and {Deane}, Roger and {Dempsey}, Jessica and {Desvignes}, Gregory and {Dexter}, Jason and {Doeleman}, Sheperd S. and {Eatough}, Ralph P. and {Falcke}, Heino and {Farah}, Joseph and {Fish}, Vincent L. and {Fomalont}, Ed and {Ford}, H. Alyson and {Fraga-Encinas}, Raquel and {Friberg}, Per and {Fromm}, Christian M. and {Fuentes}, Antonio and {Galison}, Peter and {Gammie}, Charles F. and {Garc{\'\i}a}, Roberto and {Gelles}, Zachary and {Gentaz}, Olivier and {Georgiev}, Boris and {Goddi}, Ciriaco and {Gold}, Roman and {G{\'o}mez}, Jos{\'e} L. and {G{\'o}mez-Ruiz}, Arturo I. and {Gu}, Minfeng and {Gurwell}, Mark and {Hada}, Kazuhiro and {Haggard}, Daryl and {Hecht}, Michael H. and {Hesper}, Ronald and {Himwich}, Elizabeth and {Ho}, Luis C. and {Ho}, Paul and {Honma}, Mareki and {Huang}, Chih-Wei L. and {Huang}, Lei and {Hughes}, David H. and {Ikeda}, Shiro and {Inoue}, Makoto and {Issaoun}, Sara and {James}, David J. and {Jannuzi}, Buell T. and {Janssen}, Michael and {Jeter}, Britton and {Jiang}, Wu and {Jimenez-Rosales}, Alejandra and {Johnson}, Michael D. and {Jorstad}, Svetlana and {Jung}, Taehyun and {Karami}, Mansour and {Karuppusamy}, Ramesh and {Kawashima}, Tomohisa and {Keating}, Garrett K. and {Kettenis}, Mark and {Kim}, Dong-Jin and {Kim}, Jae-Young and {Kim}, Jongsoo and {Kim}, Junhan and {Kino}, Motoki and {Koay}, Jun Yi and {Kofuji}, Yutaro and {Koch}, Patrick M. and {Koyama}, Shoko and {Kramer}, Michael and {Kramer}, Carsten and {Krichbaum}, Thomas P. and {Kuo}, Cheng-Yu and {Lauer}, Tod R. and {Lee}, Sang-Sung and {Levis}, Aviad and {Li}, Yan-Rong and {Li}, Zhiyuan and {Lindqvist}, Michael and {Lico}, Rocco and {Lindahl}, Greg and {Liu}, Jun and {Liu}, Kuo and {Liuzzo}, Elisabetta and {Lo}, Wen-Ping and {Lobanov}, Andrei P. and {Loinard}, Laurent and {Lonsdale}, Colin and {Lu}, Ru-Sen and {MacDonald}, Nicholas R. and {Mao}, Jirong and {Marchili}, Nicola and {Markoff}, Sera and {Marrone}, Daniel P. and {Marscher}, Alan P. and {Mart{\'\i}-Vidal}, Iv{\'a}n and {Matsushita}, Satoki and {Matthews}, Lynn D. and {Medeiros}, Lia and {Menten}, Karl M. and {Mizuno}, Izumi and {Mizuno}, Yosuke and {Moran}, James M. and {Moriyama}, Kotaro and {Moscibrodzka}, Monika and {M{\"u}ller}, Cornelia and {Musoke}, Gibwa and {Mus Mej{\'\i}as}, Alejandro and {Michalik}, Daniel and {Nadolski}, Andrew and {Nagai}, Hiroshi and {Nagar}, Neil M. and {Nakamura}, Masanori and {Narayan}, Ramesh and {Narayanan}, Gopal and {Natarajan}, Iniyan and {Nathanail}, Antonios and {Neilsen}, Joey and {Neri}, Roberto and {Ni}, Chunchong and {Noutsos}, Aristeidis and {Nowak}, Michael A. and {Okino}, Hiroki and {Olivares}, H{\'e}ctor and {Ortiz-Le{\'o}n}, Gisela N. and {Oyama}, Tomoaki and {{\"O}zel}, Feryal and {Palumbo}, Daniel C.~M. and {Park}, Jongho and {Patel}, Nimesh and {Pen}, Ue-Li and {Pesce}, Dominic W. and {Pi{\'e}tu}, Vincent and {Plambeck}, Richard and {PopStefanija}, Aleksandar and {Porth}, Oliver and {P{\"o}tzl}, Felix M. and {Prather}, Ben and {Preciado-L{\'o}pez}, Jorge A. and {Psaltis}, Dimitrios and {Pu}, Hung-Yi and {Ramakrishnan}, Venkatessh and {Rao}, Ramprasad and {Rawlings}, Mark G. and {Raymond}, Alexander W. and {Rezzolla}, Luciano and {Ricarte}, Angelo and {Ripperda}, Bart and {Roelofs}, Freek and {Rogers}, Alan and {Ros}, Eduardo and {Rose}, Mel and {Roshanineshat}, Arash and {Rottmann}, Helge and {Roy}, Alan L. and {Ruszczyk}, Chet and {Rygl}, Kazi L.~J. and {S{\'a}nchez}, Salvador and {S{\'a}nchez-Arguelles}, David},
        title = "{First M87 Event Horizon Telescope Results. VIII. Magnetic Field Structure near The Event Horizon}",
      journal = {\apjl},
         year = 2021,
        month = mar,
       volume = {910},
       number = {1},
          eid = {L13},
        pages = {L13},
          doi = {10.3847/2041-8213/abe4de},
archivePrefix = {arXiv},
       eprint = {2105.01173},
 primaryClass = {astro-ph.HE},
       adsurl = {https://ui.adsabs.harvard.edu/abs/2021ApJ...910L..13E}
}

@ARTICLE{2025ApJ...981L..11S,
       author = {{Singh}, Akshay and {B{\'e}gu{\'e}}, Damien and {Pe'er}, Asaf},
        title = "{Radiative Cooling Changes the Dynamics of Magnetically Arrested Disks}",
      journal = {\apjl},
         year = 2025,
        month = mar,
       volume = {981},
       number = {1},
          eid = {L11},
        pages = {L11},
          doi = {10.3847/2041-8213/adb749},
archivePrefix = {arXiv},
       eprint = {2412.11440},
 primaryClass = {astro-ph.HE},
       adsurl = {https://ui.adsabs.harvard.edu/abs/2025ApJ...981L..11S}
}

@ARTICLE{2026arXiv260509326S,
       author = {{Singh}, Akshay and {Begue}, Damien and {Pe'er}, Asaf},
        title = "{Characterizing the Scale Height and Filamentary Structure of Radiatively Cooled MADs}",
      journal = {ApJ, submitted},
         year = 2026,
        month = may,
          eid = {arXiv:2605.09326},
        pages = {arXiv:2605.09326},
archivePrefix = {arXiv},
       eprint = {2605.09326},
 primaryClass = {astro-ph.HE},
       adsurl = {https://ui.adsabs.harvard.edu/abs/2026arXiv260509326S}
}

@ARTICLE{2011ApJ...735....9M,
       author = {{Mo{\'s}cibrodzka}, M. and {Gammie}, C.~F. and {Dolence}, J.~C. and {Shiokawa}, H.},
        title = "{Pair Production in Low-luminosity Galactic Nuclei}",
      journal = {\apj},
         year = 2011,
        month = jul,
       volume = {735},
       number = {1},
          eid = {9},
        pages = {9},
          doi = {10.1088/0004-637X/735/1/9},
archivePrefix = {arXiv},
       eprint = {1104.2042},
 primaryClass = {astro-ph.HE},
       adsurl = {https://ui.adsabs.harvard.edu/abs/2011ApJ...735....9M}
}

@ARTICLE{2025MNRAS.537.2496C,
       author = {{Chael}, Andrew},
        title = "{Survey of radiative, two-temperature magnetically arrested simulations of the black hole M87* I: turbulent electron heating}",
      journal = {\mnras},
         year = 2025,
        month = mar,
       volume = {537},
       number = {3},
        pages = {2496-2515},
          doi = {10.1093/mnras/staf200},
archivePrefix = {arXiv},
       eprint = {2501.12448},
 primaryClass = {astro-ph.HE},
       adsurl = {https://ui.adsabs.harvard.edu/abs/2025MNRAS.537.2496C}
}

@ARTICLE{2025MNRAS.538..698S,
       author = {{Salas}, L.~D.~S. and {Liska}, M.~T.~P. and {Markoff}, S.~B. and {Chatterjee}, K. and {Musoke}, G. and {Porth}, O. and {Ripperda}, B. and {Yoon}, D. and {Mulaudzi}, W.},
        title = "{Two-temperature treatments in magnetically arrested disc GRMHD simulations more accurately predict light curves of Sagittarius A*}",
      journal = {\mnras},
         year = 2025,
        month = apr,
       volume = {538},
       number = {2},
        pages = {698-710},
          doi = {10.1093/mnras/staf240},
archivePrefix = {arXiv},
       eprint = {2411.09556},
 primaryClass = {astro-ph.HE},
       adsurl = {https://ui.adsabs.harvard.edu/abs/2025MNRAS.538..698S}
}

@ARTICLE{2016MNRAS.457.3801P,
       author = {{Prieto}, M.~A. and {Fern{\'a}ndez-Ontiveros}, J.~A. and {Markoff}, S. and {Espada}, D. and {Gonz{\'a}lez-Mart{\'\i}n}, O.},
        title = "{The central parsecs of M87: jet emission and an elusive accretion disc}",
      journal = {\mnras},
         year = 2016,
        month = apr,
       volume = {457},
       number = {4},
        pages = {3801-3816},
          doi = {10.1093/mnras/stw166},
archivePrefix = {arXiv},
       eprint = {1508.02302},
 primaryClass = {astro-ph.GA},
       adsurl = {https://ui.adsabs.harvard.edu/abs/2016MNRAS.457.3801P}
}

@ARTICLE{2010ApJ...708.1545D,
       author = {{Ding}, Jian and {Yuan}, Feng and {Liang}, Edison},
        title = "{Electron Heating and Acceleration by Magnetic Reconnection in Hot Accretion Flows}",
      journal = {\apj},
         year = 2010,
        month = jan,
       volume = {708},
       number = {2},
        pages = {1545-1550},
          doi = {10.1088/0004-637X/708/2/1545},
archivePrefix = {arXiv},
       eprint = {0911.4560},
 primaryClass = {astro-ph.HE},
       adsurl = {https://ui.adsabs.harvard.edu/abs/2010ApJ...708.1545D}
}

@ARTICLE{2013ApJ...773..118H,
       author = {{Hoshino}, Masahiro},
        title = "{Particle Acceleration during Magnetorotational Instability in a Collisionless Accretion Disk}",
      journal = {\apj},
         year = 2013,
        month = aug,
       volume = {773},
       number = {2},
          eid = {118},
        pages = {118},
          doi = {10.1088/0004-637X/773/2/118},
archivePrefix = {arXiv},
       eprint = {1306.6720},
 primaryClass = {astro-ph.HE},
       adsurl = {https://ui.adsabs.harvard.edu/abs/2013ApJ...773..118H}
}

@ARTICLE{1976ApJ...207..962F,
       author = {{Fishbone}, L.~G. and {Moncrief}, V.},
        title = "{Relativistic fluid disks in orbit around Kerr black holes.}",
      journal = {\apj},
         year = 1976,
        month = aug,
       volume = {207},
        pages = {962-976},
          doi = {10.1086/154565},
       adsurl = {https://ui.adsabs.harvard.edu/abs/1976ApJ...207..962F}
}

@ARTICLE{2023Galax..11....5R,
       author = {{Ricarte}, Angelo and {Johnson}, Michael D. and {Kovalev}, Yuri Y. and {Palumbo}, Daniel C.~M. and {Emami}, Razieh},
        title = "{How Spatially Resolved Polarimetry Informs Black Hole Accretion Flow Models}",
      journal = {Galaxies},
         year = 2023,
        month = jan,
       volume = {11},
       number = {1},
          eid = {5},
        pages = {5},
          doi = {10.3390/galaxies11010005},
archivePrefix = {arXiv},
       eprint = {2211.03907},
 primaryClass = {astro-ph.HE},
       adsurl = {https://ui.adsabs.harvard.edu/abs/2023Galax..11....5R}
}

\newpage
\begin{appendix}
\section{Exclusion of magnetized region}
\label{AppendixA}

Due to the density, pressure, and internal energy in simulations may reach the floor value in highly magnetized regions (where $\sigma \gg 1$), we chose a conservative cut-off value $\sigma_{\mathrm{cut}}=1$ in the previous results in this study to avoid the potential numerical error in highly magnetized regions.

\begin{figure*}[h]
\centering
        \centering
        \includegraphics[width=\linewidth]{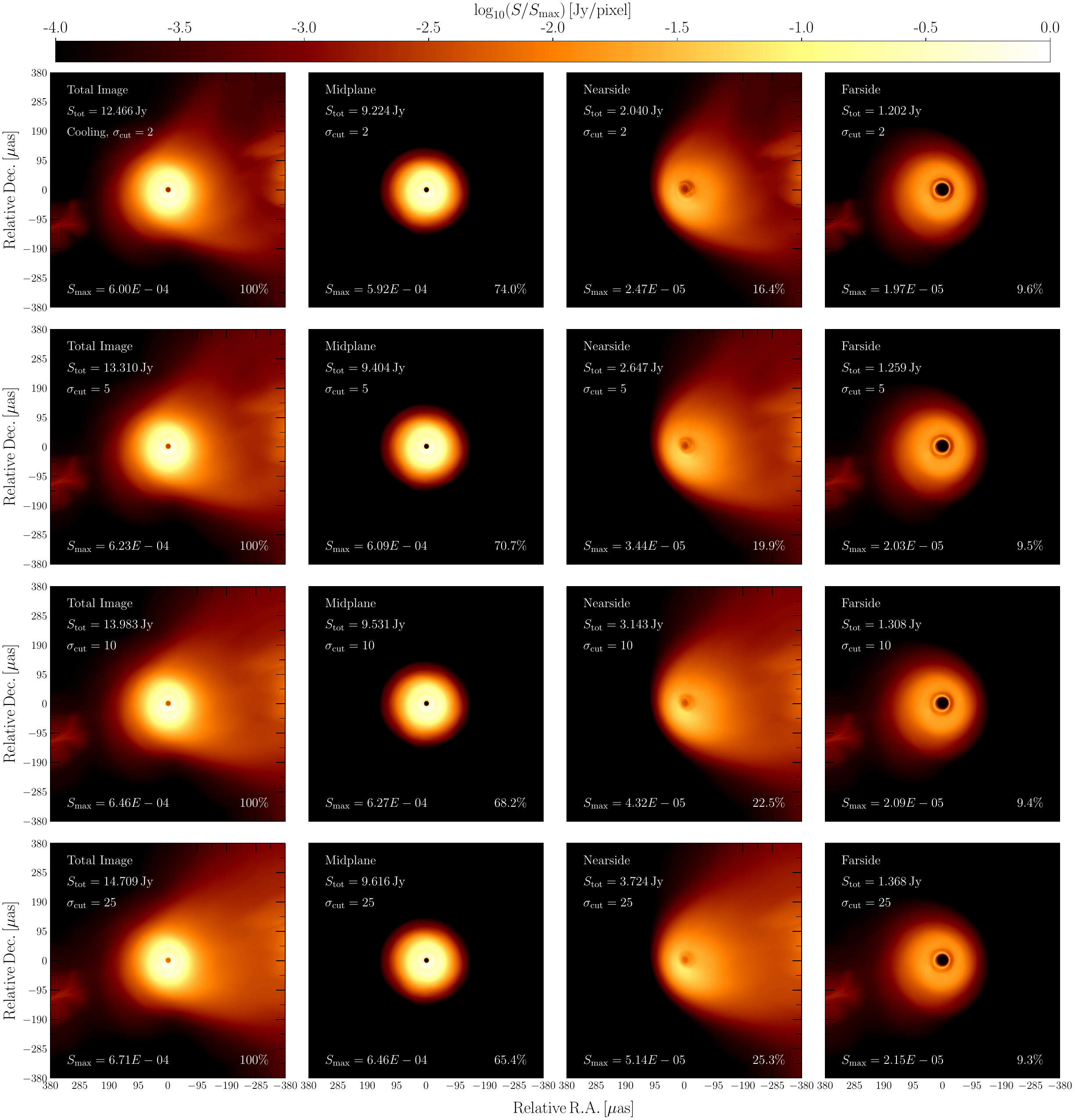}
        \caption{Same as Fig.~\ref{figure5} but using $\sigma_{\mathrm{cut}}=2$, 5, 10, and 25, respectively, with radiative cooling, in turbulent heating, and the mass accretion rate $\dot{M}_{\mathrm{BH}}/\dot{M}_\mathrm{Edd} = 1 \times 10^{-5}$. }
        \label{figure10}
\end{figure*}

Here, we investigate the effects of various $\sigma$ thresholds on flux and jet structure for a MAD state considering radiative cooling, under a mass accretion rate $\dot{M}_{\mathrm{BH}}/\dot{M}_{\mathrm{Edd}} = 1 \times 10^{-5}$. Figure~\ref{figure10} shows the time-averaged ($t=12\,000\,t_{\mathrm{g}}\,-\,15\,000 \,t_{\mathrm{g}}$) decomposed GRRT images with various $\sigma_{\mathrm{cut}}$ at 230\,GHz and an inclination angle of $163^{\circ}$, with hybrid thermal and variable $\kappa$ eDF. Combing with decomposed GRRT images with $\sigma_{\mathrm{cut}}=1$ shown in the last row in Fig.~\ref{figure6} and the images in Fig.~\ref{figure10}, the results show that with $\sigma_{\mathrm{cut}}$ increasing to 25, the ratio of extended jet emission to total emission increases from $16.4 \, \%$ to $25.3 \, \%$, compared to the case of $\sigma_{\mathrm {cut}}=1$. Meanwhile, the total flux increases from 12.5~Jy to 14.7~Jy. It is easy to confirm that the increase in flux mainly results from the brighter nearside jets, and the jet structure remains similar. 
Interestingly, the proportion of farside jet flux remains similar under the increase of $\sigma_{\mathrm{cut}}$. That means the flux of the farside jet, though increasing, is moving away from us and is therefore beyond our reach. The similar diminishing rate of increase on flux of farside jet with the increase of $\sigma_{\mathrm{cut}}$ can also be found in Fig.~10 in \cite{2024A&A...687A..88Z}. Therefore, consider radiative cooling, under a mass accretion rate $\dot{M}_{\mathrm{BH}}/\dot{M}_{\mathrm{Edd}} = 1 \times 10^{-5}$, there is no significant dependence of jet structure on $\sigma_{\mathrm{cut}}$ values, but the flux of nearside jet depends on $\sigma_{\mathrm{cut}}$ values. However, if one wants to find a proper mass accretion rate satisfying the total flux obtained from the observation, the calculated mass accretion rate and, thereafter, the radiative efficiency will be different for various choices of $\sigma_{\mathrm{cut}}$ value, which results in different theoretical predictions on jet emission, Faraday depth, and the polarization fraction.

\section{Difference maps of density 
}
\label{AppendixB}

\begin{figure}[h]
\centering
    \begin{minipage}[t]{1\textwidth}
        \centering
        \includegraphics[width=\linewidth]{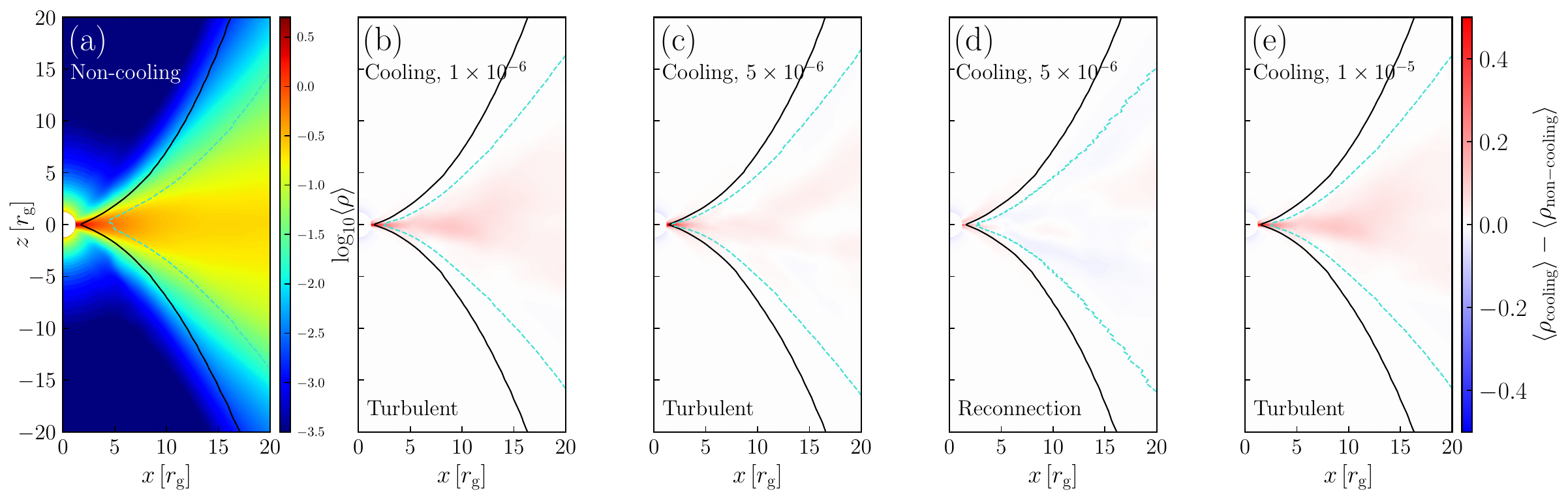}
        \caption{Same as Fig.~\ref{figure2} for panel (a), but panels (b) – (e) highlight the differences in linear scale by subtracting the density depicted in panel (a).  
        The solid black curves represent $\sigma = 1$.}
        \label{figure11}
    \end{minipage}
\end{figure}

To better visualize the differences in density in different heating prescriptions, the radiative cooling, and the mass accretion rates, we plotted the pixel-by-pixel difference maps as shown in Fig.~\ref{figure11}. We plotted the time- ($t=12\,000\,t_{\mathrm{g}}\,-\,15\,000 \,t_{\mathrm{g}}$) and azimuthally averaged distribution of logarithmic density without radiative cooling in panel (a), and the differences of density in linear scale in panels (b) $-$ (e) under various mass accretion rate and electron heating prescriptions with radiative cooling, compared with panel (a). The region shaded in red indicates the density exceeds that of the non-cooling case, while the region shaded in blue highlights the density is lower than that of the non-cooling case. There is a smooth transition from red to blue through white. The solid black curves represent $\sigma = 1$. As shown in panels (b), (c), and (e) in Fig.~\ref{figure11}, the density on the equatorial plane generally decreases, and a thinner disk is formed with a higher accretion rate, which is consistent with the results in Sec.~\ref{3.2}. Compared with panels (c) and (d), the density is higher in the former case.  

\section{Angular distribution of electron temperature and spectrum in reconnection heating}
\label{AppendixC}

To elucidate why there are more emissions from midplane in simulation $\texttt{C\_KA5e-6}$ than $\texttt{C\_MR5e-6}$ and more emissions from jets in latter case, and why the decomposed images become similar in both turbulent and reconnection cases without cooling as discussed in Sec.~\ref{3.4}, we present the angular distribution of the time- and azimuthally averaged dimensionless electron temperature for reconnection heating at specified radii, and the spectral energy distribution curves of different regions in reconnection heating. 

\begin{figure}[h]
\centering
    \begin{minipage}[t]{0.49\textwidth}
        \centering
        \includegraphics[width=\linewidth]{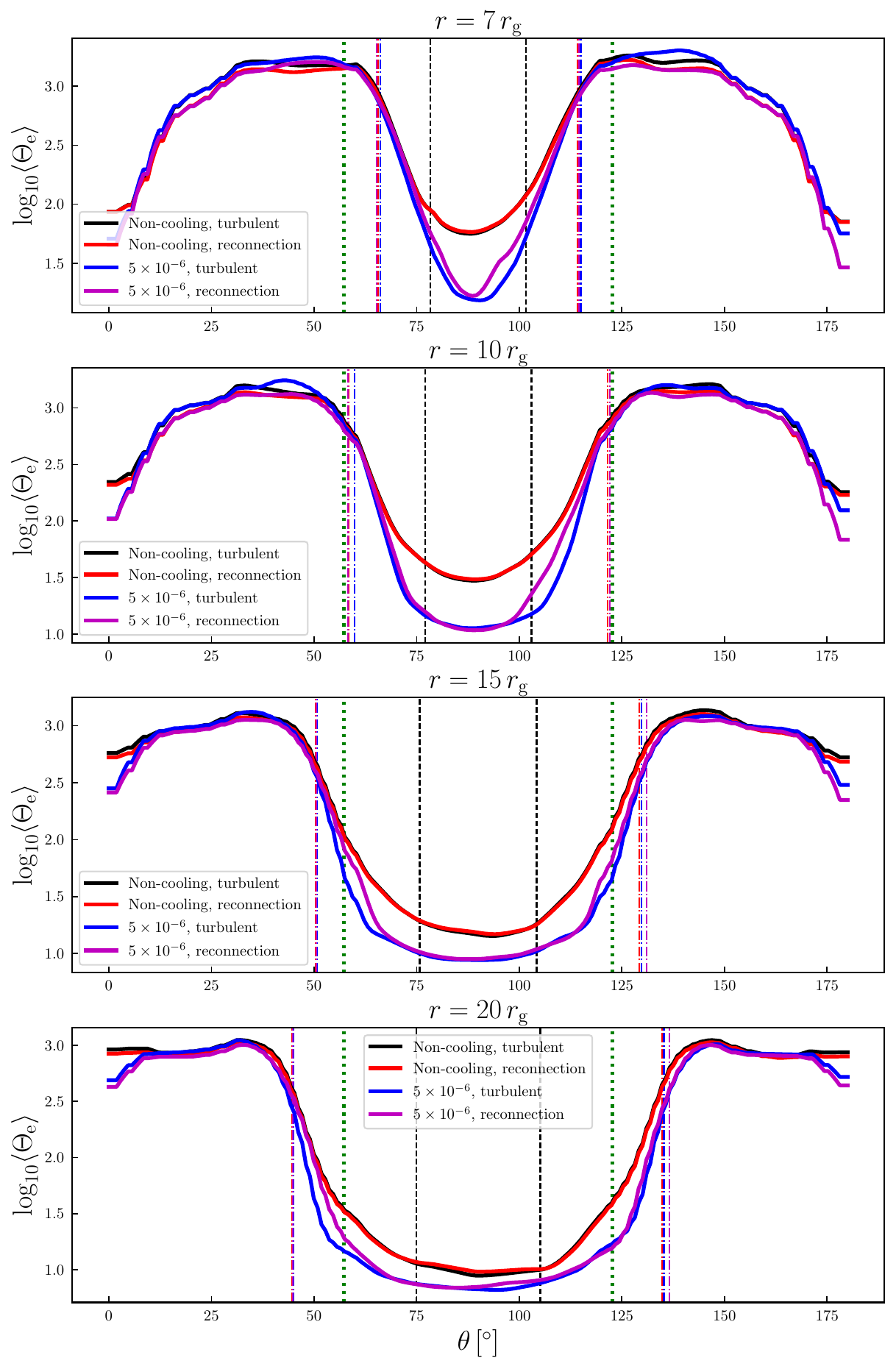}
        \caption{Same as Fig.~\ref{figure4} but the curves in various colors correspond to different cases: the turbulent heating model without cooling (black), with cooling of $\dot{M}_{\mathrm{BH}}/\dot{M}_\mathrm{Edd} = 5\times10^{-6}$ (blue), and the reconnetion heating model without cooling (red), with cooling of  $\dot{M}_{\mathrm{BH}}/\dot{M}_\mathrm{Edd} = 5\times10^{-6}$ (magenta).}
        \label{figure13}
    \end{minipage}
\end{figure}

The temperature distribution curves under turbulent heating are also included in Fig.~\ref{figure13} as a reference to compare the impacts of different electron heating prescriptions and radiative cooling. The curves in various colors correspond to different cases: the turbulent heating model without cooling (black), with cooling of $\dot{M}_{\mathrm{BH}}/\dot{M}_\mathrm{Edd} = 5\times10^{-6}$ (blue), and the reconnetion heating model without cooling (red), with cooling of  $\dot{M}_{\mathrm{BH}}/\dot{M}_\mathrm{Edd} = 5\times10^{-6}$ (magenta).

As shown in Fig.~\ref{figure13}, the temperature without cooling is similar in both the turbulent and reconnection heating. This is why under the same accretion rate $\dot{M}_{\mathrm{BH}}/\dot{M}_\mathrm{Edd} = 5\times10^{-6}$ without cooling, the total flux, emission contribution from each divided region, and the jet structure are similar in both cases. 

\begin{figure}
\vspace{7.24cm}
    \begin{minipage}[t]{0.49\textwidth}
        \centering
        \includegraphics[width=\linewidth]{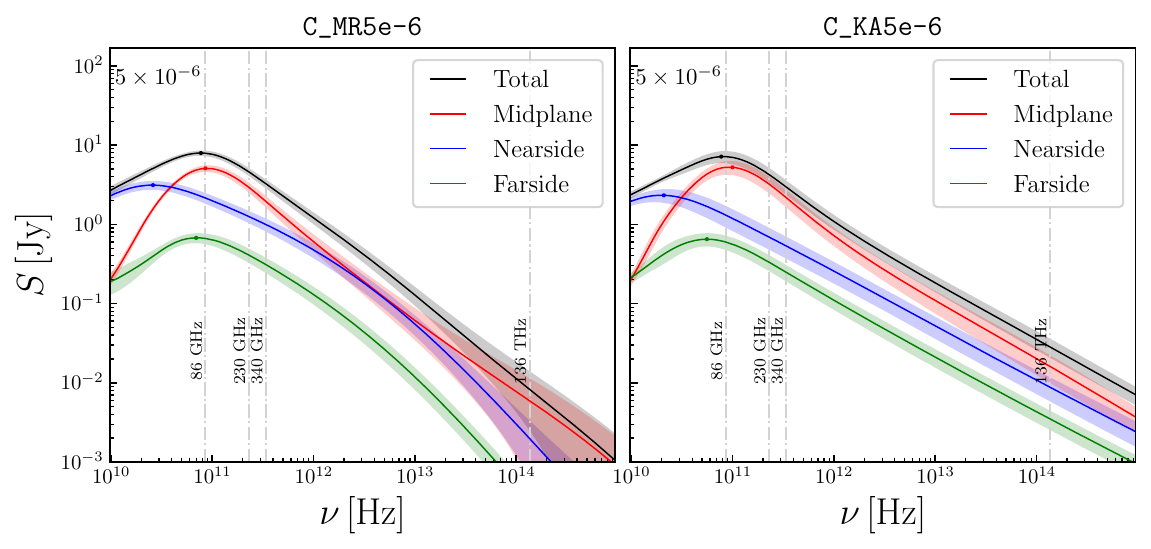}
        \caption{Same as Fig.~\ref{figure7} for simulation $\texttt{C\_KA5e-6}$ but the case $\texttt{C\_MR5e-6}$ is also added.}
        \label{figure14}
        \includegraphics[width=\linewidth]{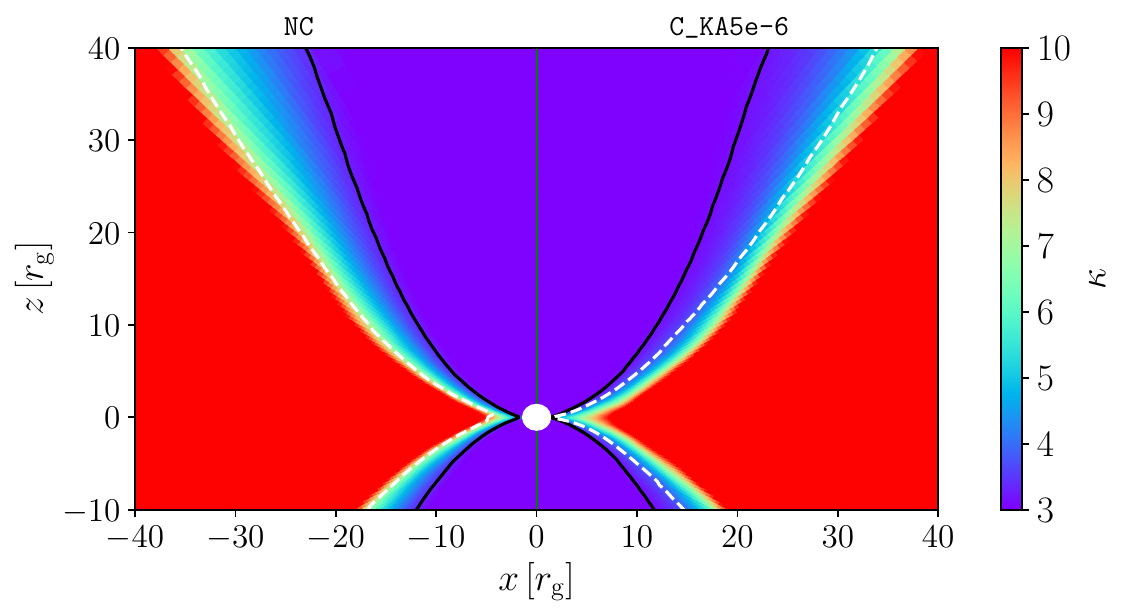}
        \caption{Time- and azimuthally averaged distributions of $\kappa$ for the $\texttt{PIC-TURB}$ obtained from decaying plasma turbulence [Eq.~\eqref{eqM23}] in cases $\texttt{NC}$ and $\texttt{C\_KA5e-6}$. The solid black and dashed white curves represent $\sigma=1$ and the Bernoulli parameter, $-hu_{\mathrm{t}}=1.02$, respectively.}
        \label{figure15}
    \end{minipage}
\end{figure}

To understand why there are more emissions from midplane for simulation $\texttt{C\_KA5e-6}$ than case $\texttt{C\_MR5e-6}$ and more emissions from jets in the latter, we need to consider the differences in density profile, temperature distribution, and the emissions from nonthermal electrons. Even though 
Fig.~\ref{figure13} indicates that the disk temperature in the $\texttt{C\_MR5e-6}$ case is higher than the $\texttt{C\_KA5e-6}$ case, the density is higher in the latter case (see panels (d) and (c) in Fig.~\ref{figure11}), which results in there being more emissions from the midplane in the case $\texttt{C\_KA5e-6}$. Besides, the spectral energy distribution curves under turbulent heating are also included in Fig.~\ref{figure14} as a reference to understand the impacts of nonthermal electrons in different heating prescriptions, at 230~GHz and on a broader range of frequencies. Compared with simulation $\texttt{C\_MR5e-6}$, Fig.~\ref{figure14} shows that the spectral curves of jets are flatter in case $\texttt{C\_KA5e-6}$ and the flux at higher frequency is also higher. Furthermore, the emissions from jets at 230~GHz are lower in case $\texttt{C\_KA5e-6}$. One reason is that the temperature of jets in case $\texttt{C\_MR5e-6}$ is higher than that in simulation $\texttt{C\_KA5e-6}$, as shown in Fig.~\ref{figure13}. The other reason is that the $\texttt{PIC-TURB}$ model of \cite{Meringolo2023} shows smaller values of $\kappa$ than the $\texttt{PIC-CS}$ model of  \cite{2018ApJ...862...80B} in the jet region \citep{2026ApJ..1001..227C}, which indicates a broader distribution of nonthermal electrons. As a result, as shown in Fig.~\ref{figure6} and Fig.~\ref{figure14}, there are more emissions from jets at higher frequencies rather than 230~GHz in case $\texttt{C\_KA5e-6}$, and more emissions are coming from jets in simulation $\texttt{C\_MR5e-6}$ at 230~GHz. Furthermore, we plotted the time- ($t = 12\,000\,t_{\mathrm{g}} - 15\,000\,t_{\mathrm{g}}$) and azimuthally averaged distribution of $\kappa$ in model $\texttt{PIC-TUBR}$ for simulations $\texttt{NC}$ and $\texttt{C\_KA5e-6}$ in Fig.~\ref{figure15}. As shown in Fig.~\ref{figure15}, the $\kappa$ values are smaller in the jet region in the cooling case $\texttt{C\_KA5e-6}$ than those in the non-cooling case $\texttt{NC}$. Hence, the nonthermal impacts in the $\texttt{PIC-TURB}$ model on the jets at 230~GHz and higher frequencies are more significant if the radiative cooling is considered (i.e., there are more emissions from jets in case $\texttt{C\_MR5e-6}$ than $\texttt{C\_KA5e-6}$ at 230~GHz as shown in Fig.~\ref{figure6}, due to the nonthermal emission contributing more at higher frequencies in the $\texttt{PIC-TURB}$ model with radiative cooling. And as shown in Fig.~\ref{figure5}, the jet emission is similar in both the $\texttt{PIC-TURB}$ model and $\texttt{PIC-CS}$ model without cooling). 


\end{appendix}

\end{document}